\documentclass{entcs} 
\usepackage{entcsmacro}

\usepackage{epsfig}  
\usepackage[utf8]{inputenc} 
\usepackage{graphicx}
\usepackage{pstricks} 
\usepackage{amssymb}
\usepackage{amsmath}
\usepackage[2cell,all]{xy}

\DeclareGraphicsExtensions{.eps,.bmp}
\definecolor{gray}{cmyk}{0,0,0,0.825}

\newcommand{\ket}[1]{\left\vert{#1}\right\rangle}

\newcommand{\ip}[2]{\left \langle {#1}|{#2}\right\rangle}
\newcommand{\HH}{\mathcal H}

\UseAllTwocells 
\CompileMatrices 

\def\lastname{B. Coecke, \'E. O. Paquette and S. Perdrix}

\begin{document}
\begin{frontmatter}
  \title{Bases in diagrammatic quantum protocols} 
  \author{Bob Coecke\thanksref{Bob}\thanksref{bob.coecke@comlab.ox.ac.uk}}
  \author{Simon Perdrix\thanksref{Sim}\thanksref{simon.perdrix@comlab.ox.ac.uk}}
  \address{Oxford University Computing Laboratory\\
    Wolfson Building, Parks Road, OX1 3QD Oxford, UK} 
  \author{\'Eric Oliver Paquette\thanksref{Eri}\thanksref{eopaquette@inexistant.net}}
  \address{D\'epartement d'Informatique et de Recherche Op\'erationelle \\ 
  LITQ, Universit\'e de Montr\'eal,   Montr\'eal, Canada}
    
    \thanks[Bob]{B.~C.~is supported by EPSRC Advanced Research Fellowship EP/D072786/1 entitled  \em The Structure of Quantum Information and its Ramifications for IT\em.  He thanks Bertfried Fauser for useful comments.}
    \thanks[bob.coecke@comlab.ox.ac.uk]{Email:
    \href{mailto:bob.coecke@comlab.ox.ac.uk} {\texttt{\normalshape
        bob.coecke@comlab.ox.ac.uk}}} 
        
    \thanks[Sim]{S.~P.~is employed on the EC-IST-FP6 STREP 033763 entitled \em Foundational Structures in Quantum Information and Computation \em -- acronym QICS.}
     \thanks[simon.perdrix@comlab.ox.ac.uk]{Email:
    \href{simon.perdrix@comlab.ox.ac.uk} {\texttt{\normalshape
        simon.perdrix@comlab.ox.ac.uk}}} 
        
    \thanks[Eri]{\'E.~O.~P.~thanks Oxford University Computing Laboratory, Bob Coecke and Mehrnoosh Sadrzadeh for their hospitality during his visit in which part of this work was realised. 
    This visit was supported by EPSRC Advanced Research Fellowship EP/D072786/1.
    He also thanks Gilles Brassard's chaire de recherche du Canada en informatique quantique for financing and Michel Boyer for discussions and feedback on some problems that are discussed in the present paper.}
    \thanks[eopaquette@inexistant.net]{Email:
    \href{eopaquette@inexistant.net} {\texttt{\normalshape
        couserid@codept.coinst.coedu}}}
\begin{abstract} 
This paper contains two new results:
\begin{enumerate}
\item We amend the notion of abstract basis in a  dagger symmetric monoidal category, as well as its corresponding graphical representation,  in order to accommodate non-self-dual  dagger compact structures; this is crucial for obtaining a \em planar \em diagrammatical representation of the induced dagger compact structure as well as for representing many complementary bases within one diagrammatic calculus.
\item We (crucially) rely on these basis  structures in a purely diagrammatic derivation of the \em quantum state transfer protocol\em; this derivation provides interesting insights 
in the distinct structural resources required for state-transfer and teleportation as models of quantum computing.
\end{enumerate}
\end{abstract}
\end{frontmatter}                
\maketitle

\section{Introduction}

Categorical axiomatisation of quantum computation and information, a research program initiated
by Abramsky and Coecke \cite{AC04}, enables rigorous, abstract, diagrammatic and automated design of quantum protocols.  They showed that dagger compact categories capture
essential structures of the quantum mechanical formalism including \em unitarity, (self-)adjointness,  trace, Bell-states, and Dirac calculus\em.  These then enable design of the quantum teleportation and related protocols \cite{AC04,Kindergarten}. In their work Abramsky and Coecke heavily relied on Kelly and Laplaza's earlier work on coherence for compact categories \cite{KL81}.   A particularly appealing feature of dagger compact categories, already present in Kelly's earlier work,  is that they come with an intuitive diagrammatic calculus, made precise by Selinger in \cite{Sel05}, extending the one due to Joyal and Street \cite{JS}. A more informal 
use of diagrammatic notation traces back to Penrose's work in the 1970's \cite{Penrose}.
Another notable contribution to the categorical axiomatisation of quantum computation and information program, also in \cite{Sel05},  is Selinger's  construction of \em mixed states \em and \em completely positive maps\em. 

Quantum teleportation involves measurement and operations that depend on measurement outcomes. The dagger compact structure can only handle them in a \em post-selected \em manner, that is, by conditioning on classical data.  One (ugly) solution is to represent  classical data syntactically, by `indices' which relate control operations to measurement outcomes \cite{Kindergarten}.  If one wants to represent classical data as explicit categorical structure one needs to go beyond dagger compactness. In \cite{AC04} Abramsky and Coecke used biproducts for this purpose. 
However, as argued in \cite{deLL}, the biproduct requirement prevents the passage from the vectorial to the projective realm, a step which is essential to eliminate redundant global phase data. This in particular meant that the approach did not capture the  essential \em decoherence \em component of quantum measurements, which transforms superposed states into classical  mixtures. This problem was solved in \cite{deLL} and \cite{Sel05} by introducing density operators.  

But, non of these \em additive \em structures admits elegant diagrammatic representation.
On the other hand, the even more abstract \em classical objects \em introduced by Coecke and Pavlovic in \cite{CP}, inspired on Carboni and Walter's axiomatisation of the category of relations in terms of their \em Frobenius law \em \cite{CarboniWalters},  are expressed entirely in terms of the \em multiplicative \em tensor structure.  In \cite{CPaquette} it was shown 
by two of the present authors that these classical objects did allow elegant diagrammatic representation: computation proceeds by the purely diagrammatic so-called \em spider theorem \em -- variants of this theorem also known in other contexts such as topological quantum field theories \cite{Kock},  abstract category theory \cite{Lack} and representation theory \cite[and references therein]{Morrison}. Classical objects moreover admit an operational interpretation in terms of copying and erasing \cite{CP}. More precisely, they exploit the fact that while classical data can be arbitrarily copied and erased, quantum data can't.  Formally, classical objects `refine' the dagger compact structures of \cite{AC04} in the sense that if an object comes with a classical object structure than it also admits a compact structure \cite{CPP}. Recently, Coecke, Pavlovic and Vicary showed that for finite dimensional Hilbert spaces these classical objects, or more precisely, \em special co-commutative dagger Frobenius comonoids\em, are in one-to-one correspondence with orthonormal bases \cite{CPV}. Also recently, Coecke and Duncan showed that they enable to axiomatise the key quantum mechanical notion of \em complementary observables \em \cite{BobRoss}. This enabled to reach a milestone in this research program, namely, abstract computation of the quantum Fourier transform, the quantum component of Shor's factoring algorithm.

  
    However, all wasn't that shiny and bright. 
    
    Firstly, these classical objects forced objects to be self-dual relative to the compact structure, that is, $A^*=A$. Concretely, a classical object consists of an object $A$, together with a \em copying operation \em $\delta:A\to A\otimes A$ and a \em deleting operation \em $\gamma:A\to I$. The induced compact structure is then given by $\delta\circ\gamma^\dagger:I\to A\otimes A$.
    But general compact structures are of type  $I\to A^*\otimes A$. Hence the ones induced by classical objects in addition satisfy $A^*=A$.  This is problematic for the graphical calculus in which a compact structure depicts as a `cup':\vskip 4mm
\centerline{
\ifx\JPicScale\undefined\def\JPicScale{1}\fi
\psset{unit=\JPicScale mm}
\psset{linewidth=0.3,dotsep=1,hatchwidth=0.3,hatchsep=1.5,shadowsize=1,dimen=middle}
\psset{dotsize=0.7 2.5,dotscale=1 1,fillcolor=black}
\psset{arrowsize=1 2,arrowlength=1,arrowinset=0.25,tbarsize=0.7 5,bracketlength=0.15,rbracketlength=0.15}
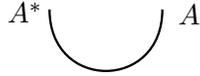
\begin{pspicture}(0,0)(24.5,9.1)
\psbezier(20.9,9.1)(20.9,-1.8)(5.9,-1.8)(5.9,9.1)
\rput(2.6,8.7){$A^*$}
\rput(24.5,8.4){$A$}
\end{pspicture}
} 
\vskip 0mm
\noindent A key property of these cups  is that `boxes' can be slided along these \cite{Kindergarten,Sel05}:
\vskip 3mm
 \centerline{
\ifx\JPicScale\undefined\def\JPicScale{1}\fi
\psset{unit=\JPicScale mm}
\psset{linewidth=0.3,dotsep=1,hatchwidth=0.3,hatchsep=1.5,shadowsize=1,dimen=middle}
\psset{dotsize=0.7 2.5,dotscale=1 1,fillcolor=black}
\psset{arrowsize=1 2,arrowlength=1,arrowinset=0.25,tbarsize=0.7 5,bracketlength=0.15,rbracketlength=0.15}
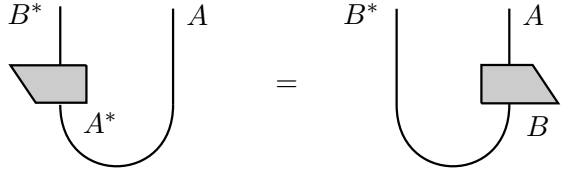
\begin{pspicture}(0,0)(72.1,22.2)
\psline(5.9,9.4)(5.9,22.2)
\psbezier(20.9,9.1)(20.9,-1.8)(5.9,-1.8)(5.9,9.1)
\rput(11.2,6.8){$A^*$}
\rput(24.1,21.1){$A$}
\newrgbcolor{userFillColour}{0.8 0.8 0.8}
\psline[fillcolor=userFillColour,fillstyle=solid](9.38,14.36)
(9.38,9.38)
(2.64,9.38)
(-0.72,14.36)(9.38,14.36)
\psline(20.9,9.1)(20.9,21.9)
\rput(1.3,21.2){$B^*$}
\rput(36,11.7){$=$}
\psline(50.62,9.08)(50.62,21.88)
\psbezier(65.6,9.1)(65.6,-1.8)(50.62,-1.8)(50.62,9.1)
\rput(68.8,21.1){$A$}
\psline(65.6,9.1)(65.6,21.9)
\rput(45.9,21.2){$B^*$}
\newrgbcolor{userFillColour}{0.8 0.8 0.8}
\psline[fillcolor=userFillColour,fillstyle=solid](61.89,9.3)
(61.89,14.4)
(68.7,14.4)
(72.1,9.3)(61.89,9.3)
\rput(69.38,6.25){$B$}
\end{pspicture}
} 
\vskip 0mm
\noindent But when considering `compound wires' this becomes:
\vskip 3mm
\centerline{
\ifx\JPicScale\undefined\def\JPicScale{1}\fi
\psset{unit=\JPicScale mm}
\psset{linewidth=0.3,dotsep=1,hatchwidth=0.3,hatchsep=1.5,shadowsize=1,dimen=middle}
\psset{dotsize=0.7 2.5,dotscale=1 1,fillcolor=black}
\psset{arrowsize=1 2,arrowlength=1,arrowinset=0.25,tbarsize=0.7 5,bracketlength=0.15,rbracketlength=0.15}
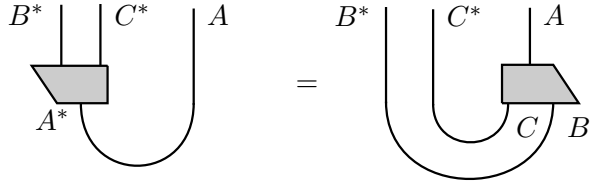
\begin{pspicture}(0,0)(76.25,23.82)
\psline(7.45,10.92)(7.45,23.72)
\psline(12.65,11.02)(12.65,23.82)
\psbezier(25.05,10.62)(25.05,-0.47)(10.05,-0.47)(10.05,10.62)
\rput(6.25,8.42){$A^*$}
\rput(28.25,22.42){$A$}
\newrgbcolor{userFillColour}{0.8 0.8 0.8}
\psline[fillcolor=userFillColour,fillstyle=solid](13.6,15.6)
(13.6,10.62)
(6.88,10.62)
(3.52,15.6)(13.6,15.6)
\psline(25.05,10.42)(25.05,23.22)
\rput(2.75,22.42){$B^*$}
\rput(40.15,13.02){$=$}
\psline(56.88,10.33)(56.88,23.12)
\psbezier(66.85,10.52)(66.85,3.72)(56.88,3.72)(56.88,10.52)
\rput(72.95,22.42){$A$}
\psline(69.75,11.62)(69.75,23.32)
\rput(46.15,22.22){$B^*$}
\rput(16.85,22.42){$C^*$}
\psbezier(72.85,10.82)(72.85,-2.88)(50.62,-2.88)(50.62,10.82)
\psline(50.62,10.62)(50.62,23.42)
\rput(61.05,22.22){$C^*$}
\newrgbcolor{userFillColour}{0.8 0.8 0.8}
\psline[fillcolor=userFillColour,fillstyle=solid](66.04,10.62)
(66.04,15.73)
(72.85,15.73)
(76.25,10.62)(66.04,10.62)
\rput(76.25,7.5){$B$}
\rput(69.38,7.5){$C$}
\end{pspicture}
} 
\vskip 0mm
\noindent and in the case of self-dual compact structure we have $B^*=B$, $C^*=C$, and in particular, for the compound cup, $C\otimes B=B^*\otimes C^*=B\otimes C$.
    
    Secondly, when considering several classical objects at once, the induced compact structures do not necessarily coincide, which has severe consequences for the diagrammatic calculus.  For example, while in \cite{BobRoss} the authors were able to axiomatise the complementary $X$- and $Z$-observables, by no means they could adjoin the $Y$-observable on the same footing, exactly because it induces a different compact structure than the $X$- and $Z$-observables do -- see Lemma \ref{non_equal_class_struc} below.  In the conclusion to this paper we provide explicit calculations which support this fact.
    
Thirdly, maybe less important to some, is that in $\bf FdHilb$, the category of finite dimensional Hilbert spaces and linear maps, basis-independence of compact structures requires them to be non-self-dual, something which directly related to the fact that there is no canonical isomorphism of type ${\cal H}\to{\cal H}^*$, where ${\cal H}^*$ is either the dual or conjugate Hilbert space. More precisely, there is a basis-independent compact structure of type $\mathbb{C}\to {\cal H}^*\otimes{\cal H}$, namely the counterpart to the identity via the canonical correspondence ${\cal H}^*\otimes{\cal H}\simeq[{\cal H}\multimap{\cal H}]$, while a canonical map of type $\mathbb{C}\to {\cal H}\otimes{\cal H}$ would induce a canonical one of type ${\cal H}\to{\cal H}^*$, which there isn't.

In this paper, firstly, we introduce an elaboration on the classical objects of \cite{CP}, to which we refer as \em basis structures\em,  which bypasses these problems.
While classical objects `factorise'  compact structures in terms of a comonoid multiplication and its unit, we introduce a third component, namely, a unitary comonoid homomorphism. This homomorphism is an \em explicit witness for the passage from a space to its dual\em.   A different but equivalent perspective is that in dagger symmetric monoidal categories, classical objects do not refine but are complementary to compact structures.  Together they then induce our basis structures. 
These bases structures allow for the `non-trivial duals',  required, for example,  to accommodate Selinger's diagrammatic representation of mixed states and completely positive maps.  
Whenever dealing with several observables, for example the $X$-, the $Z$- as well as the $Y$-observable, we can   model them relative to a unique compact structure. 
 
Next, we show that these basis structures enable a diagrammatic description, among
other protocols, of  Perdrix' state transfer protocol \cite{Per05a}. State transfer is
key to the unification of measurement-only and one-way models of quantum
computation \cite{JP05a}. Moreover, state transfer, as a substitute for teleportation,
is a key feature for optimising the resources of measurement-only quantum
computation \cite{Per07a}.  More generally, measurement-based quantum computational models have recently become very prominent within the landscape of quantum computing
due to the their great promise for actual implementation \cite{Jozsa}.
The diagrammatic analysis of both teleportation and state
transfer reveals important structural differences between these two measurement-only
quantum computational models.  

In this paper we chose to present classical data by indices.  We do this to stress that, for state transfer, contra teleportation, basis structures are already required even when representing classical data as indices. Since classical objects were specifically crafted to represent classical information flow as categorical structure, and since basis structures extend classical objects, we could have easily provided a fully comprehensive purely categorical description of classical data.   But then the main point we wanted to stress here wouldn't have come out as clear.


\section{Categorical semantics and graphical language}

\subsection{Dagger symmetric monoidal category}

A \emph{symmetric monoidal category}  
consists of a category $\mathcal C$, a bifunctor {$-\otimes-: \mathcal C \times  \mathcal C \to  \mathcal C$,} a unit object $I$ and natural isomorphisms $\lambda_A:A \simeq A\otimes I$, $\alpha_{A,B,C}:A\otimes (B\otimes C) \simeq (A\otimes B)\otimes C$ and $\sigma_{A,B}: A\otimes B \simeq B\otimes A$ satisfying the usual coherence conditions.

A \emph{$\dagger$-symmetric monoidal category} ($\dagger$-SMC) \cite{Sel05}  
is a symmetric monoidal category together with an involutive, identity-on-objects, contravariant endofunctor $\dagger:\mathcal C \to \mathcal C$, which preserves the monoidal structure, i.e.
\[
(g\circ f)^\dagger = f^\dagger \circ g^\dagger\qquad\qquad
f^{\dagger\dagger}= f\qquad\qquad
(f\otimes g)^\dagger = f^\dagger \otimes g^\dagger
\]
and also $\alpha_{A,B,C}^\dagger = \alpha_{A,B,C}^{-1}$, 
$\lambda_{A}^\dagger = \lambda_{A}^{-1}$ and
$\sigma_{A,B}^\dagger = \sigma_{A,B}^{-1}$, that is, the natural isomorphisms of the structures are `unitary'. Indeed, in a $\dagger$-SMC, a morphism $f:A \to B$ is \em unitary \em if it is an isomorphism such that $f^{-1} =f ^\dagger$.  In what follows, for convenience, we will take $\alpha$, $\lambda$ and  $\rho$ to be strict.

\begin{example}
The category $\bf FdHilb$, of finite dim.~Hilbert spaces, linear maps and tensor products, is a  $\dagger$-SMC, where $(-)^\dagger$ is the adjoint.
\end{example}

A rigorous graphical language for symmetric monoidal categories has been introduced by Joyal and Street \cite{JS} and extended to $\dagger$-SMCs by Selinger \cite{Sel05}. Such a graphical calculus is handy not only to get a representation of the information flow but is also a powerful proof technique. Indeed, in a $\dagger$-SMC (and richer structures which we introduce below), an equation holds if and only if there is an equality between their respective graphical representations~\cite{JS,Sel05}. Elementary components of this calculus are as follows:
\noindent - The identity $1_I:I\rightarrow I$ is represented by the empty picture.
\noindent - A morphism $f:A_1\otimes ...\otimes A_n\rightarrow B_1\otimes ...\otimes B_m$ and the identity $1_A:A\rightarrow A$ are depicted respectively as

\vspace{3mm}
\centerline{
\ifx\JPicScale\undefined\def\JPicScale{1}\fi
\psset{unit=\JPicScale mm}
\psset{linewidth=0.3,dotsep=1,hatchwidth=0.3,hatchsep=1.5,shadowsize=1,dimen=middle}
\psset{dotsize=0.7 2.5,dotscale=1 1,fillcolor=black}
\psset{arrowsize=1 2,arrowlength=1,arrowinset=0.25,tbarsize=0.7 5,bracketlength=0.15,rbracketlength=0.15}
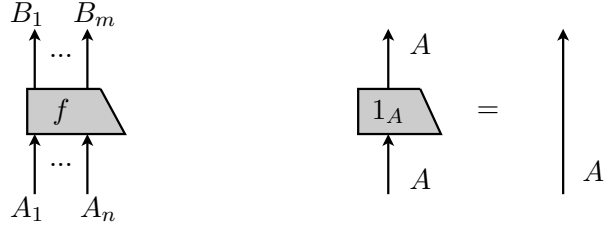
\begin{pspicture}(0,0)(79.25,28)
\newrgbcolor{userFillColour}{0.8 0.8 0.8}
\psline[fillcolor=userFillColour,fillstyle=solid](13.75,18)
(17,12)
(4,12)
(4,18)(13.75,18)
\psline{->}(5,18)(5,26)
\psline{->}(12,18)(12,26)
\psline{->}(5,4)(5,12)
\psline{->}(12,4)(12,12)
\rput(8.5,22.5){$...$}
\rput(8.5,8){$...$}
\rput(4,28){$B_1$}
\rput(13,28){$B_m$}
\rput(13.5,2){$A_n$}
\rput(4,2){$A_1$}
\rput(8.5,15){$f$}
\newrgbcolor{userFillColour}{0.8 0.8 0.8}
\psline[fillcolor=userFillColour,fillstyle=solid](56.25,18)
(59,12)
(48,12)
(48,18)(56.25,18)
\psline{->}(52,18)(52,26)
\psline{->}(52,4)(52,12)
\rput(56.25,24){$A$}
\rput(56.25,6){$A$}
\rput(52,15){$1_A$}
\rput(65.25,15){$=$}
\psline{->}(75.25,4)(75.25,26)
\rput(79.25,7){$A$}
\end{pspicture}
}

\noindent - The composite $g\circ f:A\rightarrow C$ for $f:A\rightarrow B$ and $g:B\rightarrow C$ 
and tensor $f\otimes g:A\otimes C\rightarrow B\otimes D$ for $f:A\rightarrow B$ and $g:C\rightarrow D$ are graphically represented as\vspace{3mm}

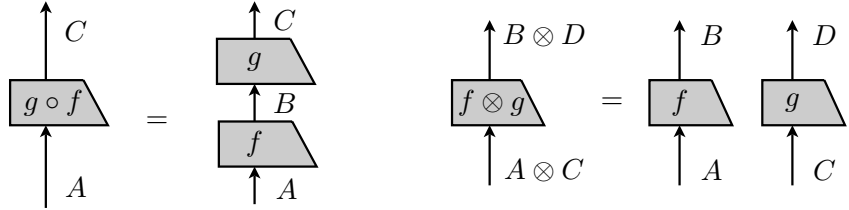
\centerline{
\ifx\JPicScale\undefined\def\JPicScale{1}\fi
\psset{unit=\JPicScale mm}
\psset{linewidth=0.3,dotsep=1,hatchwidth=0.3,hatchsep=1.5,shadowsize=1,dimen=middle}
\psset{dotsize=0.7 2.5,dotscale=1 1,fillcolor=black}
\psset{arrowsize=1 2,arrowlength=1,arrowinset=0.25,tbarsize=0.7 5,bracketlength=0.15,rbracketlength=0.15}
\begin{pspicture}(0,0)(113,37.5)
\newrgbcolor{userFillColour}{0.8 0.8 0.8}
\psline[fillcolor=userFillColour,fillstyle=solid](11.75,27)
(15,21)
(2,21)
(2,27)(11.75,27)
\psline{->}(6.75,27)(6.75,37.5)
\psline{->}(6.75,10)(6.75,21)
\rput(10.75,33.56){$C$}
\rput(10.75,12.75){$A$}
\rput(7.75,24){$g\circ f$}
\rput(21.5,21.5){$=$}
\newrgbcolor{userFillColour}{0.8 0.8 0.8}
\psline[fillcolor=userFillColour,fillstyle=solid](39.5,21.5)
(42.75,15.5)
(29.75,15.5)
(29.75,21.5)(39.5,21.5)
\psline{->}(34.5,10.5)(34.5,15.5)
\rput(38.5,35){$C$}
\rput(38.75,12.5){$A$}
\newrgbcolor{userFillColour}{0.8 0.8 0.8}
\psline[fillcolor=userFillColour,fillstyle=solid](39.25,32.5)
(42.5,26.5)
(29.5,26.5)
(29.5,32.5)(39.25,32.5)
\psline{->}(34.5,21.5)(34.5,26.5)
\rput(38.5,24){$B$}
\rput(34.5,29.5){$g$}
\rput(34.5,18.5){$f$}
\psline{->}(34.5,32.5)(34.5,37.5)
\newrgbcolor{userFillColour}{0.8 0.8 0.8}
\psline[fillcolor=userFillColour,fillstyle=solid](69.94,27)
(73,21)
(60.75,21)
(60.75,27)(69.94,27)
\psline{->}(65.75,27)(65.75,35)
\psline{->}(65.75,13)(65.75,21)
\rput(73,33){$B\otimes D$}
\rput(73,15){$A\otimes C$}
\rput(66,24){$f\otimes g$}
\rput(82,24){$=$}
\newrgbcolor{userFillColour}{0.8 0.8 0.8}
\psline[fillcolor=userFillColour,fillstyle=solid](95.25,27)
(98,21)
(87,21)
(87,27)(95.25,27)
\psline{->}(91,27)(91,35)
\psline{->}(91,13)(91,21)
\rput(95.25,33){$B$}
\rput(95.25,15){$A$}
\rput(91,24){$f$}
\newrgbcolor{userFillColour}{0.8 0.8 0.8}
\psline[fillcolor=userFillColour,fillstyle=solid](110.25,27)
(113,21)
(102,21)
(102,27)(110.25,27)
\psline{->}(106,27)(106,35)
\psline{->}(106,13)(106,21)
\rput(110.25,33){$D$}
\rput(110.25,15){$C$}
\rput(106,24){$g$}
\end{pspicture}
}\vspace{-9mm}

\noindent - Given $A$ and $B$, a component of the symmetry natural isomorphism $\sigma_{A,B}:A\otimes B\rightarrow B\otimes A$ and the dagger of a morphism $f:A\rightarrow B$ are depicted as\vspace{2mm}
\centerline{
\ifx\JPicScale\undefined\def\JPicScale{1}\fi
\psset{unit=\JPicScale mm}
\psset{linewidth=0.3,dotsep=1,hatchwidth=0.3,hatchsep=1.5,shadowsize=1,dimen=middle}
\psset{dotsize=0.7 2.5,dotscale=1 1,fillcolor=black}
\psset{arrowsize=1 2,arrowlength=1,arrowinset=0.25,tbarsize=0.7 5,bracketlength=0.15,rbracketlength=0.15}
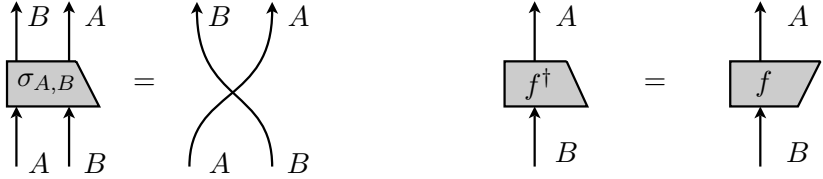
\begin{pspicture}(0,0)(40.5,26)
\newrgbcolor{userFillColour}{0.8 0.8 0.8}
\psline[fillcolor=userFillColour,fillstyle=solid](10.94,18)
(14,12)
(1.75,12)
(1.75,18)(10.94,18)
\psline{->}(10,18)(10,26)
\psline{->}(3,4)(3,12)
\rput(7,15){$\sigma_{A,B}$}
\rput(20,15){$=$}
\rput(6,24){$B$}
\psline{->}(3,18)(3,26)
\psline{->}(10,4)(10,12)
\rput(13.5,4.5){$B$}
\rput(13.5,24){$A$}
\rput(6,4.5){$A$}
\psbezier{->}(26,4)(26,14.43)(37,12.35)(37,25.91)
\psbezier{->}(37,4)(37,14.43)(27,12.35)(27,25.91)
\rput(30,24){$B$}
\rput(40.5,24){$A$}
\rput(40.5,4.5){$B$}
\rput(30,4.5){$A$}
\end{pspicture}
~~~~~~~~~~~~~~~~~~
\ifx\JPicScale\undefined\def\JPicScale{1}\fi
\psset{unit=\JPicScale mm}
\psset{linewidth=0.3,dotsep=1,hatchwidth=0.3,hatchsep=1.5,shadowsize=1,dimen=middle}
\psset{dotsize=0.7 2.5,dotscale=1 1,fillcolor=black}
\psset{arrowsize=1 2,arrowlength=1,arrowinset=0.25,tbarsize=0.7 5,bracketlength=0.15,rbracketlength=0.15}
\begin{pspicture}(0,0)(45,26)
\rput(23,15){$=$}
\newrgbcolor{userFillColour}{0.8 0.8 0.8}
\psline[fillcolor=userFillColour,fillstyle=solid](11.25,18)
(14,12)
(3,12)
(3,18)(11.25,18)
\psline{->}(7,18)(7,26)
\psline{->}(7,4)(7,12)
\rput(11.25,24){$A$}
\rput(11.25,6){$B$}
\newrgbcolor{userFillColour}{0.8 0.8 0.8}
\psline[fillcolor=userFillColour,fillstyle=solid](45,18)
(42,12)
(33,12)
(33,18)(45,18)
\psline{->}(37,18)(37,26)
\psline{->}(37,4)(37,12)
\rput(39,6){}
\rput(42,6){}
\rput(42,6){}
\rput(42,6){}
\rput(42,6){$B$}
\rput(42,24){$A$}
\rput(37.5,15){$f$}
\rput(7.5,15){$f^\dagger$}
\end{pspicture}
}\vspace{-2mm}
\noindent
that is, $(-)^\dagger$ is graphically represented by `vertical reflection'.

\noindent - Finally, for $g\circ f$ and $f\otimes g$ as above, $(g\circ f)^\dagger$ and $(f\otimes g)^\dagger$ are depicted as\vspace{2mm}
 
\centerline{
\ifx\JPicScale\undefined\def\JPicScale{1}\fi
\psset{unit=\JPicScale mm}
\psset{linewidth=0.3,dotsep=1,hatchwidth=0.3,hatchsep=1.5,shadowsize=1,dimen=middle}
\psset{dotsize=0.7 2.5,dotscale=1 1,fillcolor=black}
\psset{arrowsize=1 2,arrowlength=1,arrowinset=0.25,tbarsize=0.7 5,bracketlength=0.15,rbracketlength=0.15}
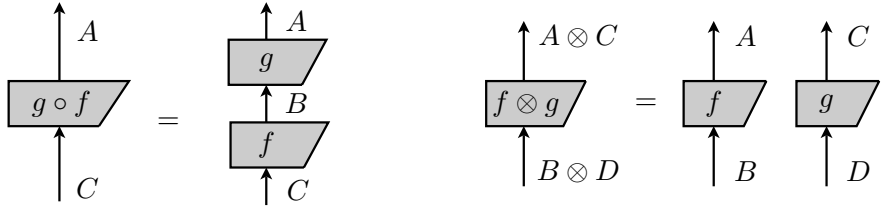
\begin{pspicture}(0,0)(43.25,28.5)
\newrgbcolor{userFillColour}{0.8 0.8 0.8}
\psline[fillcolor=userFillColour,fillstyle=solid](12.75,12)
(16.75,18)
(0.75,18)
(0.75,12)(12.75,12)
\psline{->}(7.5,18)(7.5,28.5)
\psline{->}(7.5,2)(7.5,12)
\rput(11.25,24.56){$A$}
\rput(11.25,3.75){$C$}
\rput(7.75,15){$g\circ f$}
\rput(22,12.5){$=$}
\newrgbcolor{userFillColour}{0.8 0.8 0.8}
\psline[fillcolor=userFillColour,fillstyle=solid](40,6.5)
(43.25,12.5)
(30.25,12.5)
(30.25,6.5)(40,6.5)
\psline{->}(35,1.5)(35,6.5)
\rput(39,26){$A$}
\rput(39.25,3.5){$C$}
\newrgbcolor{userFillColour}{0.8 0.8 0.8}
\psline[fillcolor=userFillColour,fillstyle=solid](39.75,17.5)
(43,23.5)
(30,23.5)
(30,17.5)(39.75,17.5)
\psline{->}(35,12.5)(35,17.5)
\rput(39,15){$B$}
\rput(35,20.5){$g$}
\rput(35,9.5){$f$}
\psline{->}(35,23.5)(35,28.5)
\end{pspicture}
~~~~~~~~~~~~~~
\ifx\JPicScale\undefined\def\JPicScale{1}\fi
\psset{unit=\JPicScale mm}
\psset{linewidth=0.3,dotsep=1,hatchwidth=0.3,hatchsep=1.5,shadowsize=1,dimen=middle}
\psset{dotsize=0.7 2.5,dotscale=1 1,fillcolor=black}
\psset{arrowsize=1 2,arrowlength=1,arrowinset=0.25,tbarsize=0.7 5,bracketlength=0.15,rbracketlength=0.15}
\begin{pspicture}(0,0)(54,26)
\newrgbcolor{userFillColour}{0.8 0.8 0.8}
\psline[fillcolor=userFillColour,fillstyle=solid](15,18)
(12,12)
(1.75,12)
(1.75,18)(15,18)
\psline{->}(6.75,18)(6.75,26)
\psline{->}(6.75,4)(6.75,12)
\rput(14,24){$A\otimes C$}
\rput(14,6){$B\otimes D$}
\rput(7,15){$f\otimes g$}
\rput(23,15){$=$}
\newrgbcolor{userFillColour}{0.8 0.8 0.8}
\psline[fillcolor=userFillColour,fillstyle=solid](39,18)
(36,12)
(28,12)
(28,18)(39,18)
\psline{->}(32,18)(32,26)
\psline{->}(32,4)(32,12)
\rput(36.25,24){$A$}
\rput(36.25,6){$B$}
\rput(32,15){$f$}
\newrgbcolor{userFillColour}{0.8 0.8 0.8}
\psline[fillcolor=userFillColour,fillstyle=solid](54,18)
(51,12)
(43,12)
(43,18)(54,18)
\psline{->}(47,18)(47,26)
\psline{->}(47,4)(47,12)
\rput(51.25,24){$C$}
\rput(51.25,6){$D$}
\rput(47,15){$g$}
\end{pspicture}
}\vspace{-4mm}

\subsection{Dagger compact structure}

The following definition of $\dagger$-compact structure `localises' the $\dagger$-compact categories that are key to Abramsky and Coecke's derivation of quantum teleportation in  \cite{AC04}.

\begin{definition}
A \emph{$\dagger$-compact structure} in a $\dagger$-SMC is a pair $(A,\epsilon_A:A\otimes A^* \to I)$ such that 
\[
( \epsilon_A \otimes 1_A)\circ (1_A \otimes \sigma_{A,A^*})\circ (1_A \otimes\epsilon_A^\dagger)=1_A
\]
or, graphically,

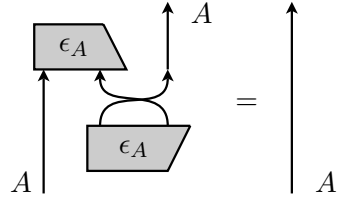
\centerline{
\ifx\JPicScale\undefined\def\JPicScale{1}\fi
\psset{unit=\JPicScale mm}
\psset{linewidth=0.3,dotsep=1,hatchwidth=0.3,hatchsep=1.5,shadowsize=1,dimen=middle}
\psset{dotsize=0.7 2.5,dotscale=1 1,fillcolor=black}
\psset{arrowsize=1 2,arrowlength=1,arrowinset=0.25,tbarsize=0.7 5,bracketlength=0.15,rbracketlength=0.15}
\begin{pspicture}(0,0)(43.5,25.5)
\newrgbcolor{userFillColour}{0.8 0.8 0.8}
\psline[fillcolor=userFillColour,fillstyle=solid](13.94,22.5)
(17,16.5)
(4.75,16.5)
(4.75,22.5)(13.94,22.5)
\psline{->}(6,0)(6,16.5)
\rput(10,19.5){$\epsilon_A$}
\rput(33,12){$=$}
\rput(27,24){$A$}
\rput(3,1.5){$A$}
\rput(43.5,1.5){$A$}
\psbezier{->}(22.5,9)(22.5,15)(13.5,10.5)(13.5,16.5)
\psbezier{->}(13.5,9)(13.5,15)(22.5,10)(22.5,16.5)
\newrgbcolor{userFillColour}{0.8 0.8 0.8}
\psline[fillcolor=userFillColour,fillstyle=solid](25.5,9)
(22.5,3)
(11.81,3)
(11.81,9)(25.5,9)
\rput(18,6){$\epsilon_A$}
\psline{->}(22.5,16.5)(22.5,25.5)
\psline{->}(39,0)(39,25.5)
\end{pspicture}
}
\end{definition}

\begin{example}
The \em conjugate Hilbert space \em $\HH^*$ of a Hilbert space $\HH$ is the Hilbert space with the same vectors as $\HH$ but with scalar multiplication and inner-product conjugated, that is, explicitly,
\[
c\bullet_{\HH^*}\psi=\bar{c}\bullet_{\HH}\psi\qquad\text{and}\qquad
\langle\psi\mid\phi\rangle_{\HH^*}=\langle\phi\mid\psi\rangle_{\HH}\,.
\]
In $\bf FdHilb$, for any $\HH\in |{\bf FdHilb}|$, $(\HH, \epsilon_\HH : \HH\otimes \HH^* \to \mathbb C)$ is a $\dagger$-compact structure where $\HH^*$ is the conjugate space of $\HH$, and \[
\epsilon_\HH : \HH \otimes \HH^* \to \mathbb C :: e_i \otimes \bar e_j \mapsto \ip{e_i}{e_j}\,.
\]
\end{example}

\begin{definition}
A \emph{$\dagger$-compact category} is a $\dagger$-SMC  
where each object $A$ comes with a $\dagger$-compact structure, and where the $\dagger$-compact structures on an object $A$ and its \em dual \em $A^*$ are connected by  $\epsilon_{A^*} = \epsilon_A\circ \sigma_{A^*,A}$,  which depicts as \vskip .4em

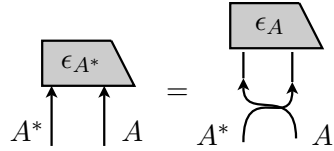
\centerline{
\ifx\JPicScale\undefined\def\JPicScale{1}\fi
\psset{unit=\JPicScale mm}
\psset{linewidth=0.3,dotsep=1,hatchwidth=0.3,hatchsep=1.5,shadowsize=1,dimen=middle}
\psset{dotsize=0.7 2.5,dotscale=1 1,fillcolor=black}
\psset{arrowsize=1 2,arrowlength=1,arrowinset=0.25,tbarsize=0.7 5,bracketlength=0.15,rbracketlength=0.15}
\begin{pspicture}(0,0)(42,20)
\newrgbcolor{userFillColour}{0.8 0.8 0.8}
\psline[fillcolor=userFillColour,fillstyle=solid](13.94,15)
(17,9)
(4.75,9)
(4.75,15)(13.94,15)
\psline{->}(6,1)(6,9)
\rput(10,12){$\epsilon_{A^*}$}
\rput(22.75,8){$=$}
\psline{->}(13,1)(13,9)
\rput(16.75,3){$A$}
\rput(2.75,3){$A^*$}
\newrgbcolor{userFillColour}{0.8 0.8 0.8}
\psline[fillcolor=userFillColour,fillstyle=solid](38.94,20)
(42,14)
(29.75,14)
(29.75,20)(38.94,20)
\psline(31.5,10)(31.5,13.5)
\rput(35,17){$\epsilon_{A}$}
\psline(38,9.5)(38,13.5)
\rput(27.5,2.5){$A^*$}
\rput(42,2.5){$A$}
\psbezier{->}(38.5,2)(38.5,9.5)(31.5,4)(31.5,10)
\psbezier{->}(31.5,1.5)(31.5,10)(38,3)(38,9.5)
\end{pspicture}
~\vspace{-1.2em}}
\end{definition}

The graphical language of $\dagger$-SMCs can be extended to $\dagger$-compact categories as follows \cite{Sel05}. The identity $1_{A^*}$ is represented as an arrow with opposite orientation and labeled by $A$:\vspace{-5mm}

~~~~~~~~\centerline{
%
%
\ifx\JPicScale\undefined\def\JPicScale{1}\fi
\psset{unit=\JPicScale mm}
\psset{linewidth=0.3,dotsep=1,hatchwidth=0.3,hatchsep=1.5,shadowsize=1,dimen=middle}
\psset{dotsize=0.7 2.5,dotscale=1 1,fillcolor=black}
\psset{arrowsize=1 2,arrowlength=1,arrowinset=0.25,tbarsize=0.7 5,bracketlength=0.15,rbracketlength=0.15}
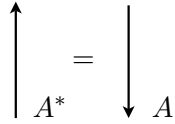
\begin{pspicture}(0,0)(45,22.5)
\psline{->}(6,1.5)(6,17)
\rput(15,9){$=$}
\psline{<-}(21,1.5)(21,16.5)
\rput(25.5,3){$A$}
\rput(10.5,3){$A^*$}
\rput(45,22.5){}
\rput(12,9){}
\end{pspicture}
}\vspace{-2mm}

\noindent
and $\epsilon_X:X\otimes X^*\rightarrow I$ with $X\in\{A,A^*\}$ and their adjoints depict as:\vspace{2mm} 

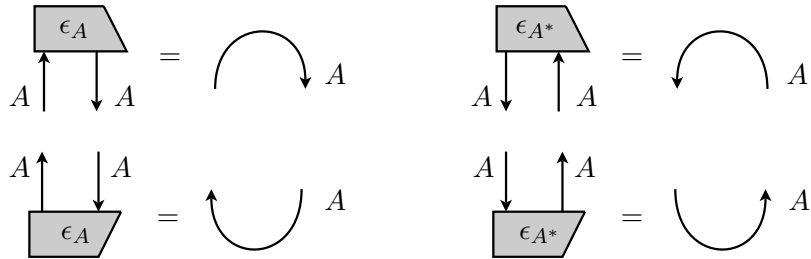
\centerline{
\ifx\JPicScale\undefined\def\JPicScale{1}\fi
\psset{unit=\JPicScale mm}
\psset{linewidth=0.3,dotsep=1,hatchwidth=0.3,hatchsep=1.5,shadowsize=1,dimen=middle}
\psset{dotsize=0.7 2.5,dotscale=1 1,fillcolor=black}
\psset{arrowsize=1 2,arrowlength=1,arrowinset=0.25,tbarsize=0.7 5,bracketlength=0.15,rbracketlength=0.15}
\begin{pspicture}(0,0)(44.75,15)
\newrgbcolor{userFillColour}{0.8 0.8 0.8}
\psline[fillcolor=userFillColour,fillstyle=solid](13.94,15)
(17,9)
(4.75,9)
(4.75,15)(13.94,15)
\psline{->}(6,1)(6,9)
\rput(10,12){$\epsilon_A$}
\rput(22.75,8){$=$}
\psline{<-}(13,1)(13,9)
\rput(16.75,3){$A$}
\rput(2.75,3){$A$}
\psbezier{->}(28.75,4)(28.75,14)(40.75,14)(40.75,4)
\rput(44.75,6){$A$}
\end{pspicture}
~~~~~~~~~~~~
\ifx\JPicScale\undefined\def\JPicScale{1}\fi
\psset{unit=\JPicScale mm}
\psset{linewidth=0.3,dotsep=1,hatchwidth=0.3,hatchsep=1.5,shadowsize=1,dimen=middle}
\psset{dotsize=0.7 2.5,dotscale=1 1,fillcolor=black}
\psset{arrowsize=1 2,arrowlength=1,arrowinset=0.25,tbarsize=0.7 5,bracketlength=0.15,rbracketlength=0.15}
\begin{pspicture}(0,0)(44.75,15)
\newrgbcolor{userFillColour}{0.8 0.8 0.8}
\psline[fillcolor=userFillColour,fillstyle=solid](13.94,15)
(17,9)
(4.75,9)
(4.75,15)(13.94,15)
\psline{<-}(6,1)(6,9)
\rput(10,12){$\epsilon_{A^*}$}
\rput(22.75,8){$=$}
\psline{->}(13,1)(13,9)
\rput(16.75,3){$A$}
\rput(2.75,3){$A$}
\psbezier{<-}(28.75,4)(28.75,14)(40.75,14)(40.75,4)
\rput(44.75,6){$A$}
\end{pspicture}
}\vspace{4mm}
\centerline{
\ifx\JPicScale\undefined\def\JPicScale{1}\fi
\psset{unit=\JPicScale mm}
\psset{linewidth=0.3,dotsep=1,hatchwidth=0.3,hatchsep=1.5,shadowsize=1,dimen=middle}
\psset{dotsize=0.7 2.5,dotscale=1 1,fillcolor=black}
\psset{arrowsize=1 2,arrowlength=1,arrowinset=0.25,tbarsize=0.7 5,bracketlength=0.15,rbracketlength=0.15}
\begin{pspicture}(0,0)(45,17)
\newrgbcolor{userFillColour}{0.8 0.8 0.8}
\psline[fillcolor=userFillColour,fillstyle=solid](16.5,9)
(13.5,3)
(4.31,3)
(4.31,9)(16.5,9)
\psline{->}(6,9)(6,17)
\rput(10.5,6){$\epsilon_{A}$}
\rput(22.75,8){$=$}
\psline{<-}(13.5,9)(13.5,17)
\rput(16.5,15){$A$}
\rput(3,15){$A$}
\psbezier{<-}(28.5,12)(28.5,1.5)(40.5,1.5)(40.5,12)
\rput(45,10.5){$A$}
\end{pspicture}
~~~~~~~~~~~~
\ifx\JPicScale\undefined\def\JPicScale{1}\fi
\psset{unit=\JPicScale mm}
\psset{linewidth=0.3,dotsep=1,hatchwidth=0.3,hatchsep=1.5,shadowsize=1,dimen=middle}
\psset{dotsize=0.7 2.5,dotscale=1 1,fillcolor=black}
\psset{arrowsize=1 2,arrowlength=1,arrowinset=0.25,tbarsize=0.7 5,bracketlength=0.15,rbracketlength=0.15}
\begin{pspicture}(0,0)(45,17)
\newrgbcolor{userFillColour}{0.8 0.8 0.8}
\psline[fillcolor=userFillColour,fillstyle=solid](16.5,9)
(13.5,3)
(4.31,3)
(4.31,9)(16.5,9)
\psline{<-}(6,9)(6,17)
\rput(10.5,6){$\epsilon_{A^*}$}
\rput(22.75,8){$=$}
\psline{->}(13.5,9)(13.5,17)
\rput(16.5,15){$A$}
\rput(3,15){$A$}
\psbezier{->}(28.5,12)(28.5,1.5)(40.5,1.5)(40.5,12)
\rput(45,10.5){$A$}
\end{pspicture}
}
The  axiomatic requirements for $\dagger$-compact categories depict as:\vspace{2mm}\\ 
\centerline{
\ifx\JPicScale\undefined\def\JPicScale{1}\fi
\psset{unit=\JPicScale mm}
\psset{linewidth=0.3,dotsep=1,hatchwidth=0.3,hatchsep=1.5,shadowsize=1,dimen=middle}
\psset{dotsize=0.7 2.5,dotscale=1 1,fillcolor=black}
\psset{arrowsize=1 2,arrowlength=1,arrowinset=0.25,tbarsize=0.7 5,bracketlength=0.15,rbracketlength=0.15}
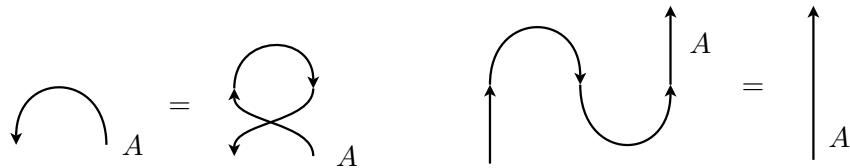
\begin{pspicture}(0,0)(45,18)
\rput(22.75,8){$=$}
\psbezier{<-}(1,3)(1,13)(13,13)(13,3)
\rput(16.5,3){$A$}
\psbezier{->}(30,10.5)(30,18)(40.5,18)(40.5,10.5)
\rput(45,1.5){$A$}
\psbezier{->}(40.5,1.5)(40.5,6)(30,6)(30,10.5)
\psbezier{<-}(30,1.5)(30,6)(40.5,6)(40.5,10.5)
\end{pspicture}
 ~~~~~~~~~~~ 
\ifx\JPicScale\undefined\def\JPicScale{1}\fi
\psset{unit=\JPicScale mm}
\psset{linewidth=0.3,dotsep=1,hatchwidth=0.3,hatchsep=1.5,shadowsize=1,dimen=middle}
\psset{dotsize=0.7 2.5,dotscale=1 1,fillcolor=black}
\psset{arrowsize=1 2,arrowlength=1,arrowinset=0.25,tbarsize=0.7 5,bracketlength=0.15,rbracketlength=0.15}
\begin{pspicture}(0,0)(47.5,21.5)
\rput(36,10.5){$=$}
\psbezier{->}(1,11)(1,21)(13,21)(13,11)
\rput(29,16.5){$A$}
\psbezier{->}(13,11)(13,0.5)(25,0.5)(25,11)
\psline{->}(25,11)(25,21.5)
\psline{->}(1,0.5)(1,11)
\rput(47.5,4){$A$}
\psline{->}(44,1)(44,21.5)
\end{pspicture}
}\vspace{-3mm} 

\begin{definition}\label{def:strict}
A $\dagger$-compact category is \emph{strict} if $(A\otimes B)^* = B^* \otimes A^*$ and 
\[
\epsilon_{A\otimes B} =   \epsilon_A \circ(1_A\otimes \epsilon_B\otimes 1_{A^*} )\,,
\]
diagrammatically this is \vspace{-4mm}

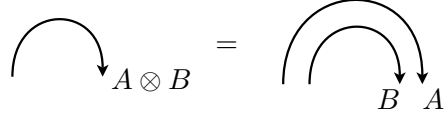
\centerline{
\ifx\JPicScale\undefined\def\JPicScale{1}\fi
\psset{unit=\JPicScale mm}
\psset{linewidth=0.3,dotsep=1,hatchwidth=0.3,hatchsep=1.5,shadowsize=1,dimen=middle}
\psset{dotsize=0.7 2.5,dotscale=1 1,fillcolor=black}
\psset{arrowsize=1 2,arrowlength=1,arrowinset=0.25,tbarsize=0.7 5,bracketlength=0.15,rbracketlength=0.15}
\begin{pspicture}(0,0)(57,19.5)
\rput(29.5,9.5){$=$}
\psbezier{->}(1,5.5)(1,15.5)(13,15.5)(13,5.5)
\rput(19.5,5){$A\otimes B$}
\psbezier{->}(37,4.5)(37,19)(55.5,19.5)(55.5,4.5)
\psbezier{->}(40.5,4.5)(40.5,14.5)(52.5,14.5)(52.5,4.5)
\rput(51,2.5){$B$}
\rput(57,2.5){$A$}
\end{pspicture}
}\vspace{-4mm} 
\end{definition}

In any compact category, the assignment $(-)^*$ on objects can be extended to a contravariant functor whose assignment on morphisms maps $f:A\to B$ to 
\[
f^*:B^* \to A^* := (1_{A^*}\otimes \epsilon_B)\circ (1_{A^*}\otimes f\otimes 1_{B^*})\circ (\eta_A \otimes 1_{B^*})\,.
 \]
 Such a mapping is depicted as:\vspace{1mm} 
 
\centerline{
\ifx\JPicScale\undefined\def\JPicScale{1}\fi
\psset{unit=\JPicScale mm}
\psset{linewidth=0.3,dotsep=1,hatchwidth=0.3,hatchsep=1.5,shadowsize=1,dimen=middle}
\psset{dotsize=0.7 2.5,dotscale=1 1,fillcolor=black}
\psset{arrowsize=1 2,arrowlength=1,arrowinset=0.25,tbarsize=0.7 5,bracketlength=0.15,rbracketlength=0.15}
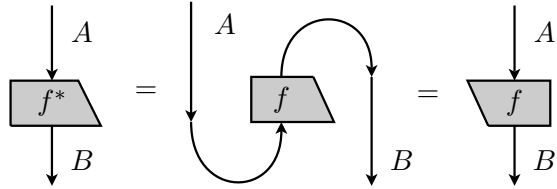
\begin{pspicture}(0,0)(74.75,25.5)
\rput(74,4.5){$B$}
\psbezier{<-}(51,15.5)(51,25.5)(39,25.5)(39,15.5)
\rput(31.5,22.5){$A$}
\psbezier{<-}(39,9.5)(39,-1)(27,-1)(27,9.5)
\psline{<-}(27,9.5)(27,25.5)
\psline{<-}(51,1)(51,15.5)
\newrgbcolor{userFillColour}{0.8 0.8 0.8}
\psline[fillcolor=userFillColour,fillstyle=solid](43.25,15.5)
(46,9.5)
(35,9.5)
(35,15.5)(43.25,15.5)
\rput(39,12.5){$f$}
\rput(55,4.5){$B$}
\psline{<-}(70,1)(70,12)
\newrgbcolor{userFillColour}{0.8 0.8 0.8}
\psline[fillcolor=userFillColour,fillstyle=solid](66.5,9)
(63.75,15)
(74.75,15)
(74.75,9)(66.5,9)
\psline{<-}(70,15)(70,25)
\rput(74,21.5){$A$}
\rput(70,12){$f$}
\rput(58.5,13){$=$}
\rput(12.5,4.5){$B$}
\psline{<-}(8.5,1)(8.5,12)
\newrgbcolor{userFillColour}{0.8 0.8 0.8}
\psline[fillcolor=userFillColour,fillstyle=solid](3,9)
(3,15)
(12,15)
(15,9)(3,9)
\psline{<-}(8.5,15)(8.5,25)
\rput(12.5,21.5){$A$}
\rput(8.5,12){$f^*$}
\rput(21,13.5){$=$}
\end{pspicture}
}

Moreover, we can define a covariant functor $(-)_*:=(-)^{\dagger *}=(-)^{*\dagger}$ whose assignment on morphisms is given by\vspace{2mm}

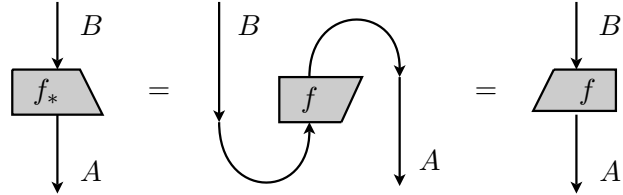
\centerline{
\ifx\JPicScale\undefined\def\JPicScale{1}\fi
\psset{unit=\JPicScale mm}
\psset{linewidth=0.3,dotsep=1,hatchwidth=0.3,hatchsep=1.5,shadowsize=1,dimen=middle}
\psset{dotsize=0.7 2.5,dotscale=1 1,fillcolor=black}
\psset{arrowsize=1 2,arrowlength=1,arrowinset=0.25,tbarsize=0.7 5,bracketlength=0.15,rbracketlength=0.15}
\begin{pspicture}(0,0)(84.75,25.5)
\rput(84,3){$A$}
\psbezier{<-}(56,15.5)(56,25.5)(44,25.5)(44,15.5)
\rput(36,22.5){$B$}
\psbezier{<-}(44,9.5)(44,-1)(32,-1)(32,9.5)
\psline{<-}(32,9.5)(32,25.5)
\psline{<-}(56,1)(56,15.5)
\newrgbcolor{userFillColour}{0.8 0.8 0.8}
\psline[fillcolor=userFillColour,fillstyle=solid](48.25,9.5)
(51,15.5)
(40,15.5)
(40,9.5)(48.25,9.5)
\rput(44,12.5){$f$}
\rput(60,4.5){$A$}
\psline{<-}(79.5,0)(79.5,10.5)
\newrgbcolor{userFillColour}{0.8 0.8 0.8}
\psline[fillcolor=userFillColour,fillstyle=solid](76.5,16.5)
(73.75,11)
(84.75,11)
(84.75,16.5)(76.5,16.5)
\psline{<-}(79.5,16.5)(79.5,25.5)
\rput(84,22.5){$B$}
\rput(81,13.5){$f$}
\rput(67.5,13.5){$=$}
\rput(15,3){$A$}
\psline{<-}(10.5,0)(10.5,10.5)
\newrgbcolor{userFillColour}{0.8 0.8 0.8}
\psline[fillcolor=userFillColour,fillstyle=solid](4.5,16.5)
(4.5,10.5)
(16.5,10.5)
(13.5,16.5)(4.5,16.5)
\psline{<-}(10.5,16.5)(10.5,25.5)
\rput(15,22.5){$B$}
\rput(9,13.5){$f_*$}
\rput(24,13.5){$=$}
\end{pspicture}
}
Thus, given an $f:A \to B$, we get the following graphical notation \cite{Sel05}:

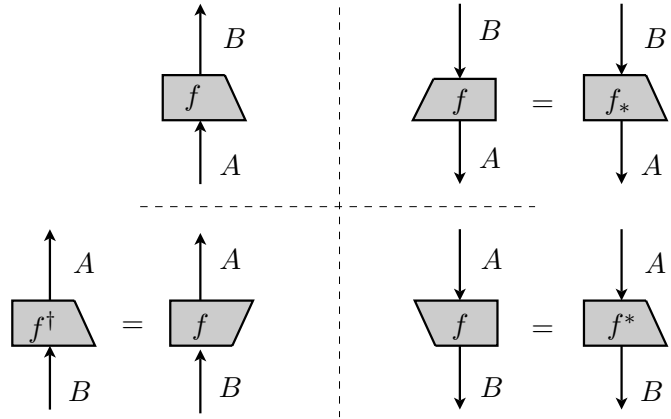
\centerline{
\ifx\JPicScale\undefined\def\JPicScale{1}\fi
\psset{unit=\JPicScale mm}
\psset{linewidth=0.3,dotsep=1,hatchwidth=0.3,hatchsep=1.5,shadowsize=1,dimen=middle}
\psset{dotsize=0.7 2.5,dotscale=1 1,fillcolor=black}
\psset{arrowsize=1 2,arrowlength=1,arrowinset=0.25,tbarsize=0.7 5,bracketlength=0.15,rbracketlength=0.15}
\begin{pspicture}(0,0)(81,57)
\psline{->}(19,1)(19,9.5)
\psline{->}(19,15.5)(19,25)
\rput(57.5,35){$A$}
\rput(23,4.5){$B$}
\newrgbcolor{userFillColour}{0.8 0.8 0.8}
\psline[fillcolor=userFillColour,fillstyle=solid](23.25,10)
(26,16)
(15,16)
(15,10)(23.25,10)
\rput(19,12.5){$f$}
\rput(23,21.5){$A$}
\psline{<-}(53.5,31.5)(53.5,42.5)
\newrgbcolor{userFillColour}{0.8 0.8 0.8}
\psline[fillcolor=userFillColour,fillstyle=solid](50,45.5)
(47.25,40)
(58.25,40)
(58.25,45.5)(50,45.5)
\psline{<-}(53.5,45.5)(53.5,55.5)
\rput(57.5,52){$B$}
\rput(53.5,42.5){$f$}
\rput(57.75,4.5){$B$}
\psline{<-}(53.5,1.5)(53.5,10)
\newrgbcolor{userFillColour}{0.8 0.8 0.8}
\psline[fillcolor=userFillColour,fillstyle=solid](50.25,10)
(47.5,16)
(58.5,16)
(58.5,10)(50.25,10)
\psline{<-}(53.5,16)(53.5,25.5)
\rput(57.75,21.5){$A$}
\rput(53.5,12.5){$f$}
\newrgbcolor{userFillColour}{0.8 0.8 0.8}
\psline[fillcolor=userFillColour,fillstyle=solid](22.25,46)
(25,40)
(14,40)
(14,46)(22.25,46)
\rput(18,43){$f$}
\psline{->}(19,31.5)(19,40)
\psline{->}(19,46)(19,55.5)
\rput(23.5,51){$B$}
\rput(23,34){$A$}
\psline[linewidth=0.1,linestyle=dashed,dash=1 1](11,28.5)(63.5,28.5)
\psline[linewidth=0.1,linestyle=dashed,dash=1 1](37.5,57)(37.5,0.5)
\newrgbcolor{userFillColour}{0.8 0.8 0.8}
\psline[fillcolor=userFillColour,fillstyle=solid](2.25,16)
(5,10)
(-6,10)
(-6,16)(2.25,16)
\rput(-2,12.5){$f^\dagger$}
\psline{->}(-1,1.5)(-1,10)
\psline{->}(-1,16)(-1,25.5)
\rput(3.5,21){$A$}
\rput(3,4){$B$}
\newrgbcolor{userFillColour}{0.8 0.8 0.8}
\psline[fillcolor=userFillColour,fillstyle=solid](78.25,46)
(81,40)
(70,40)
(70,46)(78.25,46)
\rput(74.5,42.5){$f_*$}
\psline{<-}(75,31.5)(75,40)
\psline{<-}(75,46)(75,55.5)
\rput(79.5,51){$B$}
\rput(79,34){$A$}
\newrgbcolor{userFillColour}{0.8 0.8 0.8}
\psline[fillcolor=userFillColour,fillstyle=solid](78.25,16)
(81,10)
(70,10)
(70,16)(78.25,16)
\rput(75,12.5){$f^*$}
\psline{<-}(75,1.5)(75,10)
\psline{<-}(75,16)(75,25.5)
\rput(79.5,21){$A$}
\rput(79,4){$B$}
\rput(64.5,42){$=$}
\rput(64.5,12){$=$}
\rput(10,12.5){$=$}
\end{pspicture}
}

\noindent
that is, $(-)_*$ is graphically represented by horizontal reflection and $(-)^*$  by $180^\circ$ rotation. This captures $(-)^\dagger=((-)^*)_*$ and similar equations.

\section{Bases axiomatisation}
   
\subsection{Cloning vs copying}

The no-cloning theorem \cite{WZ}, a well-known result of quantum information theory, states that for any Hilbert space of finite dimension greater than two,  there is no quantum evolution $f : \HH \to \HH\otimes \HH$ such that for any $\ket \phi\in \HH$, $f(\ket \phi) = \ket \phi \otimes \ket \phi$. Despite of the no-cloning theorem, \emph{copying} is allowed by quantum mechanics: for a given orthonormal basis $\{e_k, k\in K\}$, the linear operator $\delta: \HH \to \HH\otimes \HH :: e_k \mapsto e_k\otimes e_k$ is an isometry (i.e. $\delta^\dagger \circ \delta = 1_\HH$), thus a valid quantum evolution.  Here, the only vectors that are truly copied are the basis vectors.
Coecke and Pavlovic \cite{CP} relied  on this fact when axiomatising bases as $\dagger$-Frobenius structures.

\begin{definition} 
A \emph{$\dagger$-Frobenius structure} in a $\dagger$-SMC is an internal co-commutative comonoid 
$(A,\delta_A:A\to A\otimes A, \gamma_A:A\to I)$
such that
\[
\delta_A^\dagger \circ \delta_A = 1_A\qquad\mbox{\rm and}\qquad
\delta_A\circ \delta_A^\dagger= (\delta^\dagger_A \otimes 1_A)\circ (1_A \otimes \delta_A)\,. 
\]
\end{definition} 

\noindent
The structural morphisms therein are diagrammatically represented as \cite{CPaquette}: \\~\\ \vspace{-5mm}

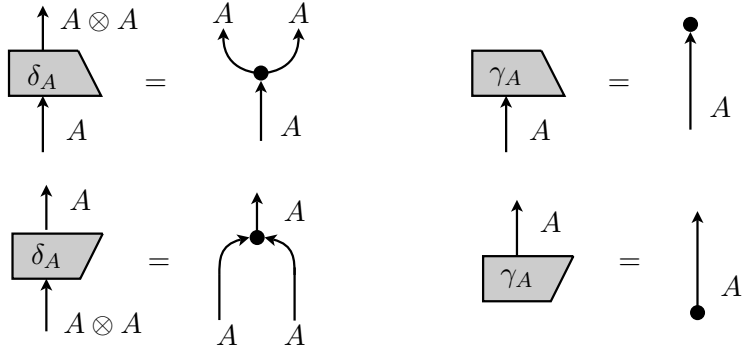
\centerline{
\ifx\JPicScale\undefined\def\JPicScale{1}\fi
\psset{unit=\JPicScale mm}
\psset{linewidth=0.3,dotsep=1,hatchwidth=0.3,hatchsep=1.5,shadowsize=1,dimen=middle}
\psset{dotsize=0.7 2.5,dotscale=1 1,fillcolor=black}
\psset{arrowsize=1 2,arrowlength=1,arrowinset=0.25,tbarsize=0.7 5,bracketlength=0.15,rbracketlength=0.15}
\begin{pspicture}(0,0)(44.5,21.5)
\psline{->}(39.5,3.5)(39.5,11.5)
\rput{0}(39.5,12.5){\psellipse[linestyle=none,fillstyle=solid](0,0)(1,-1)}
\psbezier{<-}(44.5,18.5)(44.5,12.5)(39.5,12.5)(39.5,12.5)
\psbezier{<-}(34.5,18.5)(34.5,12.5)(39.5,12.5)(39.5,12.5)
\rput(34.5,20.5){$A$}
\rput(44.5,20.5){$A$}
\rput(43.5,5.5){$A$}
\psline{->}(10.5,15.5)(10.5,21.5)
\psline{->}(10.5,2)(10.5,9.5)
\rput(15,5){$A$}
\newrgbcolor{userFillColour}{0.8 0.8 0.8}
\psline[fillcolor=userFillColour,fillstyle=solid](15.19,15.5)
(18.25,9.5)
(6,9.5)
(6,15.5)(15.19,15.5)
\rput(25.5,11){$=$}
\rput(10.5,12){$\delta_A$}
\rput(18,19.5){$A\otimes A$}
\end{pspicture}
 ~~~~~~~~~
\ifx\JPicScale\undefined\def\JPicScale{1}\fi
\psset{unit=\JPicScale mm}
\psset{linewidth=0.3,dotsep=1,hatchwidth=0.3,hatchsep=1.5,shadowsize=1,dimen=middle}
\psset{dotsize=0.7 2.5,dotscale=1 1,fillcolor=black}
\psset{arrowsize=1 2,arrowlength=1,arrowinset=0.25,tbarsize=0.7 5,bracketlength=0.15,rbracketlength=0.15}
\begin{pspicture}(0,0)(42,20)
\psline{->}(38,5)(38,18)
\rput{0}(38,19){\psellipse[linestyle=none,fillstyle=solid](0,0)(1,-1)}
\rput(42,8){$A$}
\psline{->}(13.5,2)(13.5,9.5)
\rput(18,5){$A$}
\newrgbcolor{userFillColour}{0.8 0.8 0.8}
\psline[fillcolor=userFillColour,fillstyle=solid](18.19,15.5)
(21.25,9.5)
(9,9.5)
(9,15.5)(18.19,15.5)
\rput(28.5,11){$=$}
\rput(13.5,12){$\gamma_A$}
\end{pspicture}
}\vspace{2mm}

\centerline{\, 
\ifx\JPicScale\undefined\def\JPicScale{1}\fi
\psset{unit=\JPicScale mm}
\psset{linewidth=0.3,dotsep=1,hatchwidth=0.3,hatchsep=1.5,shadowsize=1,dimen=middle}
\psset{dotsize=0.7 2.5,dotscale=1 1,fillcolor=black}
\psset{arrowsize=1 2,arrowlength=1,arrowinset=0.25,tbarsize=0.7 5,bracketlength=0.15,rbracketlength=0.15}
\begin{pspicture}(0,0)(43.5,21.5)
\rput(34.5,1.5){$A$}
\psline{->}(10.5,15.5)(10.5,21.5)
\psline{->}(10.5,2)(10.5,9)
\rput(15,19.5){$A$}
\newrgbcolor{userFillColour}{0.8 0.8 0.8}
\psline[fillcolor=userFillColour,fillstyle=solid](18,15)
(15,9)
(6,9)
(6,15)(18,15)
\rput(25.5,11){$=$}
\rput(10.5,12){$\delta_A$}
\rput(18,3){$A\otimes A$}
\psline{->}(38.5,14.93)(38.5,20.5)
\rput{0}(38.5,14.5){\psellipse[linestyle=none,fillstyle=solid](0,0)(1,-1)}
\psbezier{<-}(39.5,14.5)(43.5,13.5)(43.5,11.5)(43.5,9.5)
\psbezier{<-}(37.5,14.5)(33.5,13.5)(33.5,11.5)(33.5,9.5)
\psline(43.5,3.43)(43.5,10.5)
\psline(33.5,3.5)(33.5,9.5)
\rput(43.5,1.5){$A$}
\rput(43.5,18){$A$}
\end{pspicture}
 ~~~~~~~~~\, 
\ifx\JPicScale\undefined\def\JPicScale{1}\fi
\psset{unit=\JPicScale mm}
\psset{linewidth=0.3,dotsep=1,hatchwidth=0.3,hatchsep=1.5,shadowsize=1,dimen=middle}
\psset{dotsize=0.7 2.5,dotscale=1 1,fillcolor=black}
\psset{arrowsize=1 2,arrowlength=1,arrowinset=0.25,tbarsize=0.7 5,bracketlength=0.15,rbracketlength=0.15}
\begin{pspicture}(0,0)(42,19.5)
\psline{->}(37.5,5)(37.5,18)
\rput{0}(37.5,4.5){\psellipse[linestyle=none,fillstyle=solid](0,0)(1,-1)}
\rput(42,8){$A$}
\psline{->}(13.5,12)(13.5,19.5)
\rput(18,16.5){$A$}
\newrgbcolor{userFillColour}{0.8 0.8 0.8}
\psline[fillcolor=userFillColour,fillstyle=solid](21,12)
(18,6)
(9,6)
(9,12)(21,12)
\rput(28.5,11){$=$}
\rput(13.5,9){$\gamma_A$}
\end{pspicture}
}
\vspace{1mm}

\noindent
and their axiomatic conditions depict as:

~\\
\centerline{
\ifx\JPicScale\undefined\def\JPicScale{1}\fi
\psset{unit=\JPicScale mm}
\psset{linewidth=0.3,dotsep=1,hatchwidth=0.3,hatchsep=1.5,shadowsize=1,dimen=middle}
\psset{dotsize=0.7 2.5,dotscale=1 1,fillcolor=black}
\psset{arrowsize=1 2,arrowlength=1,arrowinset=0.25,tbarsize=0.7 5,bracketlength=0.15,rbracketlength=0.15}
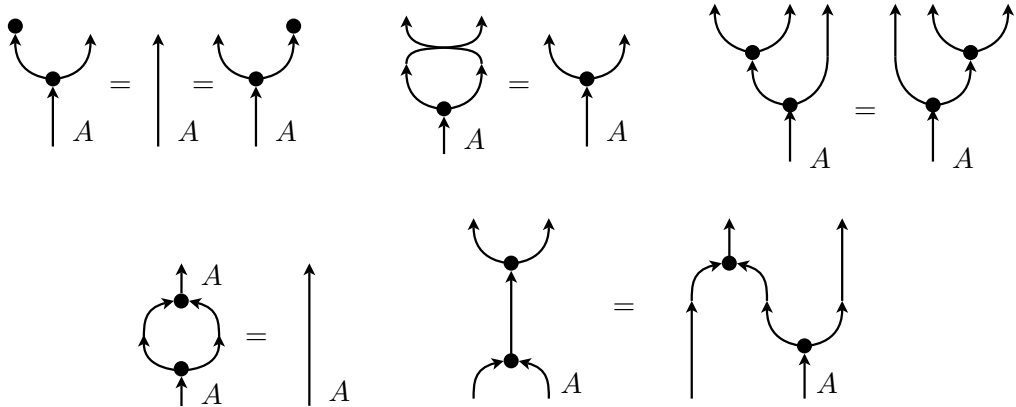
\begin{pspicture}(0,0)(138,23)
\psline{->}(11,4)(11,12)
\rput{0}(11,13){\psellipse[linestyle=none,fillstyle=solid](0,0)(1,-1)}
\psbezier{<-}(16,19)(16,13)(11,13)(11,13)
\psbezier{<-}(6,19)(6,13)(11,13)(11,13)
\rput(15,6){$A$}
\rput{0}(6,20){\psellipse[linestyle=none,fillstyle=solid](0,0)(1,-1)}
\psline{->}(38,4)(38,12)
\rput{0}(38,13){\psellipse[linestyle=none,fillstyle=solid](0,0)(1,-1)}
\psbezier{<-}(43,19)(43,13)(38,13)(38,13)
\psbezier{<-}(33,19)(33,13)(38,13)(38,13)
\rput(42,6){$A$}
\rput{0}(43,20){\psellipse[linestyle=none,fillstyle=solid](0,0)(1,-1)}
\psline{->}(25,4)(25,19)
\rput(29,6){$A$}
\rput(31,12){$=$}
\rput(20,12){$=$}
\psline{->}(63,3)(63,8)
\rput{0}(63,9){\psellipse[linestyle=none,fillstyle=solid](0,0)(1,-1)}
\psbezier{<-}(68,15)(68,9)(63,9)(63,9)
\psbezier{<-}(58,15)(58,9)(63,9)(63,9)
\rput(67,5){$A$}
\psbezier{->}(58,15)(58,19)(68,15)(68,21.43)
\psbezier{->}(68,15)(68,19)(58,15)(58,21.43)
\rput(73,12){$=$}
\psline{->}(82,4)(82,12)
\rput{0}(82,13){\psellipse[linestyle=none,fillstyle=solid](0,0)(1,-1)}
\psbezier{<-}(87,19)(87,13)(82,13)(82,13)
\psbezier{<-}(77,19)(77,13)(82,13)(82,13)
\rput(86,6){$A$}
\psline{->}(123,16)(123,23)
\rput(119,8.5){$=$}
\psline{->}(128,2)(128,9)
\rput{0}(128,9.5){\psellipse[linestyle=none,fillstyle=solid](0,0)(1,-1)}
\psbezier{<-}(133,15.5)(133,9.5)(128,9.5)(128,9.5)
\psbezier(123,16)(123,10)(128,9.5)(128,9.5)
\rput(132,2.5){$A$}
\rput{0}(133,16.5){\psellipse[linestyle=none,fillstyle=solid](0,0)(1,-1)}
\psbezier{<-}(138,23)(138,17)(133,16.5)(133,16.5)
\psbezier{<-}(128,23)(128,17)(133,16.5)(133,16.5)
\psline{->}(114,16)(114,23)
\psline{->}(109,2)(109,9)
\rput{0}(109,9.5){\psellipse[linestyle=none,fillstyle=solid](0,0)(1,-1)}
\psbezier(114,16)(114,10)(109,9.5)(109,9.5)
\psbezier{<-}(104,15.5)(104,9.5)(109,9.5)(109,9.5)
\rput(113,2.5){$A$}
\rput{0}(104,16.5){\psellipse[linestyle=none,fillstyle=solid](0,0)(1,-1)}
\psbezier{<-}(109,23)(109,17)(104,16.5)(104,16.5)
\psbezier{<-}(99,23)(99,17)(104,16.5)(104,16.5)
\end{pspicture}
}

~\\
\centerline{
\ifx\JPicScale\undefined\def\JPicScale{1}\fi
\psset{unit=\JPicScale mm}
\psset{linewidth=0.3,dotsep=1,hatchwidth=0.3,hatchsep=1.5,shadowsize=1,dimen=middle}
\psset{dotsize=0.7 2.5,dotscale=1 1,fillcolor=black}
\psset{arrowsize=1 2,arrowlength=1,arrowinset=0.25,tbarsize=0.7 5,bracketlength=0.15,rbracketlength=0.15}
\begin{pspicture}(0,0)(27,20)
\psline{->}(6,1)(6,4.96)
\rput{90}(6,5.98){\psellipse[linestyle=none,fillstyle=solid](0,0)(1.02,-1)}
\psbezier{<-}(11,10.5)(11,5.75)(6,5.75)(6,5.75)
\psbezier{<-}(1,10.5)(1,5.75)(6,5.75)(6,5.75)
\rput(10,18.42){$A$}
\rput(10,2.58){$A$}
\psline{<-}(6,20)(6,16.04)
\rput{0}(6,15){\psellipse[linestyle=none,fillstyle=solid](0,0)(1,1)}
\rput(16,10){$=$}
\psline{->}(23,1.04)(23,20)
\rput(27,3){$A$}
\psbezier{<-}(7,15)(11,14)(11,12)(11,10)
\psbezier{<-}(5,15)(1,14)(1,12)(1,10)
\end{pspicture}
~~~~~~~~~
\ifx\JPicScale\undefined\def\JPicScale{1}\fi
\psset{unit=\JPicScale mm}
\psset{linewidth=0.3,dotsep=1,hatchwidth=0.3,hatchsep=1.5,shadowsize=1,dimen=middle}
\psset{dotsize=0.7 2.5,dotscale=1 1,fillcolor=black}
\psset{arrowsize=1 2,arrowlength=1,arrowinset=0.25,tbarsize=0.7 5,bracketlength=0.15,rbracketlength=0.15}
\begin{pspicture}(0,0)(54,26)
\psline{->}(10,7)(10,19)
\rput{0}(10,20){\psellipse[linestyle=none,fillstyle=solid](0,0)(1,-1)}
\psbezier{<-}(15,26)(15,20)(10,20)(10,20)
\psbezier{<-}(5,26)(5,20)(10,20)(10,20)
\rput{0}(10,7){\psellipse[linestyle=none,fillstyle=solid](0,0)(1,-1)}
\psbezier{<-}(11,7)(15,6)(15,4)(15,2)
\psbezier{<-}(9,7)(5,6)(5,4)(5,2)
\psline{->}(49,2)(49,8)
\rput{0}(49,9){\psellipse[linestyle=none,fillstyle=solid](0,0)(1,1)}
\psbezier{<-}(44,15)(44,9)(49,9)(49,9)
\psbezier{<-}(54,15)(54,9)(49,9)(49,9)
\psline{->}(39,20.43)(39,26)
\rput{0}(39,20){\psellipse[linestyle=none,fillstyle=solid](0,0)(1,1)}
\psbezier{<-}(38,20)(34,19)(34,17)(34,15)
\psbezier{<-}(40,20)(44,19)(44,17)(44,15)
\psline{->}(54,15)(54,26)
\psline{->}(34,2)(34,15)
\rput(25,14){$=$}
\rput(18,4){$A$}
\rput(52,4){$A$}
\end{pspicture}
}

\begin{example}
Let $\{\ket 0, \ket 1\}$ be the so-called standard  basis of $\HH_2$, the Hilbert space of dimension $2$. Let 
\begin{equation}\label{eq:one_to_one}
\delta_{std} :\HH_2 \to \HH_2\otimes \HH_2 :: \ket j \mapsto \ket{jj}
\qquad\text{and}\qquad\gamma_{std}: \HH_2 \to \mathbb C :: \ket j \mapsto 1\,.
\end{equation}
Then $(\HH_2,\delta_{std}, \gamma_{std})$ is  a $\dagger$-Frobenius structure in $\bf FdHilb$.
\end{example}

\begin{theorem}{\rm\cite{CPV}}\label{thm:CPV}
There is a one-to-one correspondence between $\dagger$-Frobenius structure and orthonormal bases in $\bf FdHilb$; this correspondence is established by eqs.{\rm(\ref{eq:one_to_one})}.
\end{theorem}

Thus, $\dagger$-Frobenius structures truly axiomatise orthonormal bases. 

\begin{definition}Let $(A,\delta_A, \gamma_A)$ and $(B,\delta_B,\gamma_B)$ be two $\dagger$-Frobenius structures, then 
$f:A\to B$ is a \emph{partial map} if $\delta_B \circ f = (f\otimes f)\circ \delta_A$, diagrammatically \vspace{2mm}

\centerline{
\ifx\JPicScale\undefined\def\JPicScale{1}\fi
\psset{unit=\JPicScale mm}
\psset{linewidth=0.3,dotsep=1,hatchwidth=0.3,hatchsep=1.5,shadowsize=1,dimen=middle}
\psset{dotsize=0.7 2.5,dotscale=1 1,fillcolor=black}
\psset{arrowsize=1 2,arrowlength=1,arrowinset=0.25,tbarsize=0.7 5,bracketlength=0.15,rbracketlength=0.15}
\begin{pspicture}(0,0)(55,22)
\newrgbcolor{userFillColour}{0.8 0.8 0.8}
\psline[fillcolor=userFillColour,fillstyle=solid](15.75,11)
(18,5)
(9,5)
(9,11)(15.75,11)
\rput(12.27,8){$f$}
\psline{->}(13,1)(13,5)
\psline{->}(13,11)(13,15)
\rput(22,20){$B$}
\rput(16.75,3){$A$}
\rput{0}(13,16){\psellipse[linestyle=none,fillstyle=solid](0,0)(1,-1)}
\psbezier{<-}(18,22)(18,16)(13,16)(13,16)
\psbezier{<-}(8,22)(8,16)(13,16)(13,16)
\rput(4,20){$B$}
\newrgbcolor{userFillColour}{0.8 0.8 0.8}
\psline[fillcolor=userFillColour,fillstyle=solid](42.75,18)
(45,12)
(36,12)
(36,18)(42.75,18)
\rput(39.27,15){$f$}
\psline{->}(45.25,1)(45.25,5)
\rput(49,3){$A$}
\rput{0}(45.25,6){\psellipse[linestyle=none,fillstyle=solid](0,0)(1,-1)}
\psbezier{<-}(50.25,12)(50.25,6)(45.25,6)(45.25,6)
\psbezier{<-}(40.25,12)(40.25,6)(45.25,6)(45.25,6)
\newrgbcolor{userFillColour}{0.8 0.8 0.8}
\psline[fillcolor=userFillColour,fillstyle=solid](52.75,18)
(55,12)
(46,12)
(46,18)(52.75,18)
\rput(49.27,15){$f$}
\psline{->}(40,18)(40,22)
\psline{->}(50,18)(50,22)
\rput(36,21){$B$}
\rput(54,21){$B$}
\rput(28,12){$=$}
\end{pspicture}
} 

\noindent Moreover, it is a \emph{total map} if also $\gamma_B \circ f = \gamma_A$, which depicts as \vspace{2mm}

\centerline{
\ifx\JPicScale\undefined\def\JPicScale{1}\fi
\psset{unit=\JPicScale mm}
\psset{linewidth=0.3,dotsep=1,hatchwidth=0.3,hatchsep=1.5,shadowsize=1,dimen=middle}
\psset{dotsize=0.7 2.5,dotscale=1 1,fillcolor=black}
\psset{arrowsize=1 2,arrowlength=1,arrowinset=0.25,tbarsize=0.7 5,bracketlength=0.15,rbracketlength=0.15}
\begin{pspicture}(0,0)(32,22)
\newrgbcolor{userFillColour}{0.8 0.8 0.8}
\psline[fillcolor=userFillColour,fillstyle=solid](7.48,13)
(9.73,7)
(0.73,7)
(0.73,13)(7.48,13)
\rput(4,10){$f$}
\psline{->}(4.73,1)(4.73,7)
\psline{->}(4.73,13)(4.73,20)
\rput(8.73,4){$A$}
\rput{0}(4.73,21){\psellipse[linestyle=none,fillstyle=solid](0,0)(1,-1)}
\rput(8.73,17){$B$}
\rput(16.73,12){$=$}
\psline{->}(27.73,3)(27.73,17)
\rput{0}(27.73,18){\psellipse[linestyle=none,fillstyle=solid](0,0)(1,-1)}
\rput(32,5){$A$}
\end{pspicture}
}

\noindent Finally, it is a \emph{permutation} if, in addition to the previous, $f$ is unitary.
\end{definition}

Carboni and Walters showed  in \cite{CarboniWalters} that in the category of finite sets, relations and the cartesian product, for a `suitably restricted' notion of $\dagger$-Frobenius structures, these partial and total maps, and permutations, correspond to the usual notion.  As a consequence of Theorem \ref{thm:CPV}, for arbitrary $\dagger$-Frobenius structures, in ${\bf FdHilb}$ these partial and total maps, and permutations, correspond to the usual notion -- a simple computation easily demonstrates this. They map the basis vectors of one classical structure on the basis vectors of the other classical structure.

\begin{definition} 
Let  $(A,\delta_A, \gamma_A)$ be a $\dagger$-Frobenius structure, then
a unitary morphism $f: A\to A$ is a \emph{phase map} if
\[
(f\otimes 1_A)\circ \delta_A = \delta_A \circ f = (1_A\otimes f)\circ \delta_A\,, 
\]
this is \vspace{2mm}

\centerline{
\ifx\JPicScale\undefined\def\JPicScale{1}\fi
\psset{unit=\JPicScale mm}
\psset{linewidth=0.3,dotsep=1,hatchwidth=0.3,hatchsep=1.5,shadowsize=1,dimen=middle}
\psset{dotsize=0.7 2.5,dotscale=1 1,fillcolor=black}
\psset{arrowsize=1 2,arrowlength=1,arrowinset=0.25,tbarsize=0.7 5,bracketlength=0.15,rbracketlength=0.15}
\begin{pspicture}(0,0)(89.75,22)
\newrgbcolor{userFillColour}{0.8 0.8 0.8}
\psline[fillcolor=userFillColour,fillstyle=solid](15.75,11)
(18,5)
(9,5)
(9,11)(15.75,11)
\rput(12.27,8){$f$}
\psline{->}(13,1)(13,5)
\psline{->}(13,11)(13,15)
\rput(22,20){$A$}
\rput(16.75,3){$A$}
\rput{0}(13,16){\psellipse[linestyle=none,fillstyle=solid](0,0)(1,-1)}
\psbezier{<-}(18,22)(18,16)(13,16)(13,16)
\psbezier{<-}(8,22)(8,16)(13,16)(13,16)
\rput(4,20){$A$}
\psline{->}(45.25,1)(45.25,5)
\rput(49,3){$A$}
\rput{0}(45.25,6){\psellipse[linestyle=none,fillstyle=solid](0,0)(1,-1)}
\psbezier{<-}(50.25,12)(50.25,6)(45.25,6)(45.25,6)
\psbezier{<-}(40,12)(40,6)(45.25,6)(45.25,6)
\newrgbcolor{userFillColour}{0.8 0.8 0.8}
\psline[fillcolor=userFillColour,fillstyle=solid](52.75,18)
(55,12)
(46,12)
(46,18)(52.75,18)
\rput(49.27,15){$f$}
\psline{->}(40,12)(40,22)
\psline{->}(50,18)(50,22)
\rput(36,21){$A$}
\rput(54,21){$A$}
\rput(28,12){$=$}
\newrgbcolor{userFillColour}{0.8 0.8 0.8}
\psline[fillcolor=userFillColour,fillstyle=solid](78.5,18)
(80.75,12)
(71.75,12)
(71.75,18)(78.5,18)
\rput(75.02,15){$f$}
\psline{->}(81,1)(81,5)
\rput(84.75,3){$A$}
\rput{0}(81,6){\psellipse[linestyle=none,fillstyle=solid](0,0)(1,-1)}
\psbezier{<-}(86,12)(86,6)(81,6)(81,6)
\psbezier{<-}(76,12)(76,6)(81,6)(81,6)
\psline{->}(75.75,18)(75.75,22)
\psline{->}(86.02,12)(86.02,22)
\rput(71.75,21){$A$}
\rput(89.75,21){$A$}
\rput(63,12){$=$}
\end{pspicture}
}
\vspace{-2mm}

\end{definition}

In ${\bf FdHilb}$ the equality $(f\otimes 1_\HH)\circ \delta_\HH = \delta_\HH \circ f$ implies 
for $f=\sum_{ij} f_{ij}|i\rangle\langle j|$ that  
$\sum_i f_{ij} |ik\rangle=\sum_i f_{ij} |ii\rangle$ for all $k$, hence $((f_{ij})_{ij})$ must be diagonal.
Unitarity assures that all these diagonal elements are of the form $e^{i\theta}$, hence the name `phase map'. 

\begin{lemma}{\rm\cite{CPP}}\label{lem:dagFSdagCS}
In a  $\dagger$-SMC, whenever $(A, \delta_A, \gamma_A)$ is a  $\dagger$-Frobenius structure, then $(A,\epsilon_A:=  \gamma_A \circ \delta^\dagger_A)$ is a $\dagger$-compact structure, with  $A^*=A$.
\end{lemma}

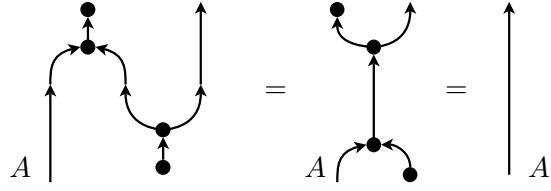
\begin{proof}

\centerline{
\ifx\JPicScale\undefined\def\JPicScale{1}\fi
\psset{unit=\JPicScale mm}
\psset{linewidth=0.3,dotsep=1,hatchwidth=0.3,hatchsep=1.5,shadowsize=1,dimen=middle}
\psset{dotsize=0.7 2.5,dotscale=1 1,fillcolor=black}
\psset{arrowsize=1 2,arrowlength=1,arrowinset=0.25,tbarsize=0.7 5,bracketlength=0.15,rbracketlength=0.15}
\begin{pspicture}(0,0)(67,26)
\psline{->}(17,5)(17,8)
\rput{0}(17,9){\psellipse[linestyle=none,fillstyle=solid](0,0)(1,1)}
\psbezier{<-}(12,15)(12,9)(17,9)(17,9)
\psbezier{<-}(22,15)(22,9)(17,9)(17,9)
\psline{->}(7,20.43)(7,24)
\rput{0}(7,20){\psellipse[linestyle=none,fillstyle=solid](0,0)(1,1)}
\psbezier{<-}(6,20)(2,19)(2,17)(2,15)
\psbezier{<-}(8,20)(12,19)(12,17)(12,15)
\psline{->}(22,15)(22,26)
\psline{->}(2,2)(2,15)
\rput(32,14){$=$}
\rput(-2,4){$A$}
\rput{0}(7,25){\psellipse[linestyle=none,fillstyle=solid](0,0)(1,1)}
\rput{0}(17,4){\psellipse[linestyle=none,fillstyle=solid](0,0)(1,1)}
\rput(56,14){$=$}
\psline{->}(45,7)(45,19)
\rput{90}(45,20){\psellipse[linestyle=none,fillstyle=solid](0,0)(1,0.97)}
\psbezier{<-}(40.15,24)(40.15,20)(45,20)(45,20)
\psbezier{<-}(49.85,26)(49.85,20)(45,20)(45,20)
\rput{90}(45,7){\psellipse[linestyle=none,fillstyle=solid](0,0)(1,0.97)}
\psbezier{<-}(44.03,7)(40.15,6)(40.15,4)(40.15,2)
\psbezier{<-}(45.97,7)(47.91,7)(49.85,6)(49.85,4)
\rput(37.24,4){$A$}
\rput{90}(49.85,3){\psellipse[linestyle=none,fillstyle=solid](0,0)(1,0.97)}
\rput{90}(40.15,25){\psellipse[linestyle=none,fillstyle=solid](0,0)(1,0.97)}
\psline{->}(63,3)(63,26)
\rput(67,4){$A$}
\end{pspicture}
}  

\end{proof}

So  $\dagger$-Frobenius structure `factorises' $\dagger$-compact structure. However, this forces $A^*=A$.\footnote{Note that this does not obstruct modelling bases in ${\bf FdHilb}$.  The reason of this is that there is no unique dagger compact structure on ${\bf FdHilb}$ but that many different ones can be chosen.  Those include the ones where we `pick' ${\cal H}^*$ to be the conjugate space as well as the ones where we `pick' ${\cal H}^*$ to be ${\cal H}$ itself.}  As  a consequence, a $\dagger$-compact category in which the $\dagger$-compact structure factorises as $\dagger$-Frobenius structure cannot be strict! (in the sense of Definition \ref{def:strict})

\begin{lemma}\label{non_equal_class_struc}
In a  $\dagger$-SMC, if $(A, \delta_A, \gamma_A)$ is a  $\dagger$-Frobenius structure, and $U:A\to A$ is unitary, then $(A, (U\otimes U)\circ\delta_A\circ U^\dagger, \gamma_A\circ U^\dagger)$ is also a $\dagger$-Frobenius structure.  These two $\dagger$-Frobenius structures induce the same $\dagger$-compact structure if and only if $U_*=U$.
\end{lemma}
\begin{proof}
The $\dagger$-compact structure  induced by $(A, (U\otimes U)\circ\delta_A\circ U^\dagger, \gamma_A\circ U^\dagger)$ is
\vskip 2mm
\centerline{
\ifx\JPicScale\undefined\def\JPicScale{1}\fi
\psset{unit=\JPicScale mm}
\psset{linewidth=0.3,dotsep=1,hatchwidth=0.3,hatchsep=1.5,shadowsize=1,dimen=middle}
\psset{dotsize=0.7 2.5,dotscale=1 1,fillcolor=black}
\psset{arrowsize=1 2,arrowlength=1,arrowinset=0.25,tbarsize=0.7 5,bracketlength=0.15,rbracketlength=0.15}
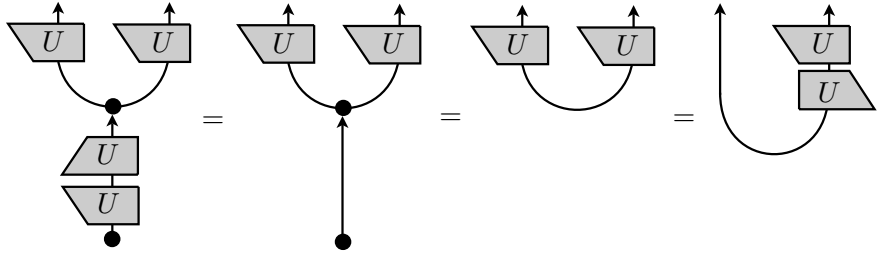
\begin{pspicture}(0,0)(115.54,33)
\psline{->}(109.5,20.5)(109.5,32.7)
\psline{->}(7,27.24)(7,33)
\psbezier(21.8,27.2)(21.8,16.3)(6.8,16.3)(6.8,27.2)
\newrgbcolor{userFillColour}{0.8 0.8 0.8}
\psline[fillcolor=userFillColour,fillstyle=solid](10.43,30.08)
(10.43,25.1)
(3.7,25.1)
(0.34,30.08)(10.43,30.08)
\psline{->}(21.8,27.2)(21.8,32.96)
\rput(27.6,17.1){$=$}
\rput(6.4,27.8){$U$}
\newrgbcolor{userFillColour}{0.8 0.8 0.8}
\psline[fillcolor=userFillColour,fillstyle=solid](24.54,29.98)
(24.54,25)
(17.81,25)
(14.45,29.98)(24.54,29.98)
\rput{0}(14.3,19.1){\psellipse[fillstyle=solid](0,0)(0.95,-0.94)}
\psline{<-}(14.2,18)(14.2,1.5)
\newrgbcolor{userFillColour}{0.8 0.8 0.8}
\psline[fillcolor=userFillColour,fillstyle=solid](17.6,10)
(17.6,15)
(10.87,15)
(7.51,10)(17.6,10)
\newrgbcolor{userFillColour}{0.8 0.8 0.8}
\psline[fillcolor=userFillColour,fillstyle=solid](17.69,8.38)
(17.69,3.4)
(10.96,3.4)
(7.6,8.38)(17.69,8.38)
\rput(13.7,6.2){$U$}
\rput(13.6,12.7){$U$}
\rput(21,27.7){$U$}
\psline{->}(37.6,27.04)(37.6,32.8)
\psbezier(52.4,27)(52.4,16.1)(37.4,16.1)(37.4,27)
\newrgbcolor{userFillColour}{0.8 0.8 0.8}
\psline[fillcolor=userFillColour,fillstyle=solid](41.03,29.88)
(41.03,24.9)
(34.3,24.9)
(30.94,29.88)(41.03,29.88)
\psline{->}(52.4,27)(52.4,32.76)
\rput(37,27.6){$U$}
\newrgbcolor{userFillColour}{0.8 0.8 0.8}
\psline[fillcolor=userFillColour,fillstyle=solid](55.14,29.78)
(55.14,24.8)
(48.41,24.8)
(45.05,29.78)(55.14,29.78)
\rput{0}(44.9,18.9){\psellipse[fillstyle=solid](0,0)(0.95,-0.93)}
\psline{<-}(44.8,17.8)(44.8,1.3)
\rput(51.6,27.5){$U$}
\rput{0}(14.15,1.44){\psellipse[fillstyle=solid](0,0)(0.95,-0.94)}
\rput{0}(44.85,1.06){\psellipse[fillstyle=solid](0,0)(0.95,-0.94)}
\psline{->}(68.7,26.84)(68.7,32.6)
\psbezier(83.5,26.8)(83.5,15.9)(68.5,15.9)(68.5,26.8)
\newrgbcolor{userFillColour}{0.8 0.8 0.8}
\psline[fillcolor=userFillColour,fillstyle=solid](72.13,29.68)
(72.13,24.7)
(65.4,24.7)
(62.04,29.68)(72.13,29.68)
\psline{->}(83.5,26.8)(83.5,32.56)
\rput(68.1,27.4){$U$}
\newrgbcolor{userFillColour}{0.8 0.8 0.8}
\psline[fillcolor=userFillColour,fillstyle=solid](86.24,29.58)
(86.24,24.6)
(79.51,24.6)
(76.15,29.58)(86.24,29.58)
\rput(82.7,27.3){$U$}
\rput(59.2,17.4){$=$}
\rput(90.2,17.3){$=$}
\psline{->}(95,20.62)(95,32.82)
\psbezier(109.4,20.9)(109.4,10)(95,10)(95,20.9)
\newrgbcolor{userFillColour}{0.8 0.8 0.8}
\psline[fillcolor=userFillColour,fillstyle=solid](105.5,18.7)
(105.5,23.8)
(112.2,23.8)
(115.54,18.7)(105.5,18.7)
\rput(109.51,21.03){$U$}
\newrgbcolor{userFillColour}{0.8 0.8 0.8}
\psline[fillcolor=userFillColour,fillstyle=solid](112.24,29.78)
(112.24,24.8)
(105.51,24.8)
(102.15,29.78)(112.24,29.78)
\rput(108.7,27.5){$U$}
\end{pspicture}
} 
\vskip 2mm
\noindent and $U\circ U^*=1_A$ if and only if $U_*=U$.
\end{proof}

In ${\bf FdHilb}$ the equation $U_*=U$ implies that the matrix representation of $U$ only involves real numbers.  It then easily follows that the $X$-, $Y$- and $Z$-bases, i.e., 
\[
\{|0\rangle+|1\rangle,|0\rangle-|1\rangle \}\qquad \quad
\{|0\rangle+i |1\rangle,|0\rangle-i |1\rangle \}\qquad \quad
\{|0\rangle, |1\rangle\}
\]
cannot be cast as $\dagger$-Frobenius structures which share the same compact structure, since transforming them in each other requires complex matrix entries.

\subsection{Dagger dual Frobenius structure}

We now introduce a different `factorisation' of compact structures, as \emph{$\dagger$-dual Frobenius structure}.  This does not impose $A^*=A$. Intuitively, the axiomatisation of bases in a dagger compact category implies that every object $A$ has two duals: 
\begin{itemize}
\item[-] First, the object $A^*$ which comes from the $\dagger$-compact structure and
\item[-] The object $A$ itself as it is self-dual when equipped with a $\dagger$-Frobenius structure $(A, \delta_A,\gamma_A)$ (see lemma \ref{lem:dagFSdagCS}). 
\end{itemize}
These two duals of $A$ are isomorphic \cite{KL81}. Instead of requiring $A=A^*$ as was done in \cite{CP,CPP}, we make this isomorphism -- the \emph{dualiser} between $A$ and $A^*$ -- explicit.

\begin{definition}
A \emph{$\dagger$-dual Frobenius structure} in a $\dagger$-SMC is a quadruple
$$(A,\delta_A:A\to A\otimes A,\gamma_A : A\to I, d_A:A\to A^*)$$
such that
\begin{itemize}
\item $(A,\delta_A,\gamma_A)$ is a $\dagger$-Frobenius structure.
\item $d_A$ is unitary. 
\end{itemize} 
\end{definition}

\begin{theorem}\label{thm:dagDFdagCdagF}
For a given  $\dagger$-SMC, 
\begin{itemize}
\item[{\rm(i)}] If $(A,\delta_A, \gamma_A)$ is a $\dagger$-Frobenius structure and $(A, \epsilon_A)$ is a $\dagger$-compact structure, then $(A, \delta_A, \gamma_A, d_A)$ is a $\dagger$-dual Frobenius structure, where 
\[
d_A:= (\gamma_A^\dagger\otimes 1_{A^*})\circ (\delta_A^\dagger \otimes 1_{A^*})\circ (1_A\otimes \epsilon_A^\dagger)\,,
\]
that is,  graphically, \\~

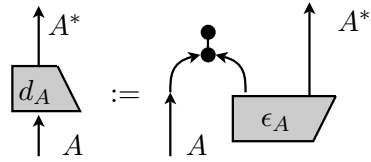
\centerline{
\ifx\JPicScale\undefined\def\JPicScale{1}\fi
\psset{unit=\JPicScale mm}
\psset{linewidth=0.3,dotsep=1,hatchwidth=0.3,hatchsep=1.5,shadowsize=1,dimen=middle}
\psset{dotsize=0.7 2.5,dotscale=1 1,fillcolor=black}
\psset{arrowsize=1 2,arrowlength=1,arrowinset=0.25,tbarsize=0.7 5,bracketlength=0.15,rbracketlength=0.15}
\begin{pspicture}(0,0)(50.5,27.5)
\newrgbcolor{userFillColour}{0.8 0.8 0.8}
\psline[fillcolor=userFillColour,fillstyle=solid](48,15)
(45,9)
(34.31,9)
(34.31,15)(48,15)
\rput(40,11.5){$\epsilon_A$}
\psline{->}(44.5,15)(44.5,27.5)
\psline(31,20.93)(31,23.5)
\rput{0}(31,20.5){\psellipse[linestyle=none,fillstyle=solid](0,0)(1,-1)}
\psbezier{<-}(32,20.5)(36,19.5)(36,17.5)(36,15.5)
\psbezier{<-}(30,20.5)(26,19.5)(26,17.5)(26,15.5)
\psline{->}(26,7)(26,15.5)
\rput{0}(31,23.5){\psellipse[linestyle=none,fillstyle=solid](0,0)(1,-1)}
\rput(29.5,8.5){$A$}
\psline{->}(8.5,7)(8.5,12.5)
\newrgbcolor{userFillColour}{0.8 0.8 0.8}
\psline[fillcolor=userFillColour,fillstyle=solid](11,19)
(14,13)
(5,13)
(5,19)(10.94,19)
\psline{->}(8.5,19)(8.5,27)
\rput(13,8.5){$A$}
\rput(12,24.5){$A^*$}
\rput(8,15.5){$d_A$}
\rput(20,15){$:=$}
\rput(50.5,25){$A^*$}
\end{pspicture}
}

\item[{\rm(ii)}] If $(A, \delta_A, \gamma, d_A)$ is a $\dagger$-dual Frobenius structure then $(A, \delta_A, \gamma_A)$ is a $\dagger$-Frobenius structure and $(A,\epsilon_A)$  is a $\dagger$-compact structure, where 
\[
\epsilon_A:= \gamma_A\circ \delta_A^\dagger\circ (1_A \otimes d^\dagger_A))
\]
which is
\end{itemize}
\end{theorem}
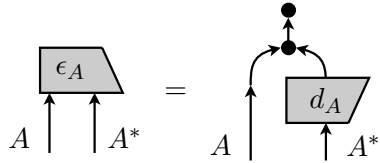
\centerline{
\ifx\JPicScale\undefined\def\JPicScale{1}\fi
\psset{unit=\JPicScale mm}
\psset{linewidth=0.3,dotsep=1,hatchwidth=0.3,hatchsep=1.5,shadowsize=1,dimen=middle}
\psset{dotsize=0.7 2.5,dotscale=1 1,fillcolor=black}
\psset{arrowsize=1 2,arrowlength=1,arrowinset=0.25,tbarsize=0.7 5,bracketlength=0.15,rbracketlength=0.15}
\begin{pspicture}(0,0)(49.75,22)
\newrgbcolor{userFillColour}{0.8 0.8 0.8}
\psline[fillcolor=userFillColour,fillstyle=solid](14.25,16)
(17,10)
(6,10)
(6,16)(14.25,16)
\psline{->}(7.25,2)(7.25,10)
\rput(17.25,4){$A^*$}
\rput(10,13){$\epsilon_A$}
\rput(24,10){$=$}
\psline{->}(13.25,2)(13.25,10)
\rput(3.25,4){$A$}
\psline{->}(39,16.43)(39,20)
\rput{0}(39,16){\psellipse[linestyle=none,fillstyle=solid](0,0)(1,1)}
\psbezier{<-}(38,16)(34,15)(34,13)(34,11)
\psbezier{<-}(40,16)(44,15)(44,13)(44,11)
\rput{0}(39,21){\psellipse[linestyle=none,fillstyle=solid](0,0)(1,1)}
\newrgbcolor{userFillColour}{0.8 0.8 0.8}
\psline[fillcolor=userFillColour,fillstyle=solid](47,6)
(49.75,12)
(38.75,12)
(38.75,6)(47,6)
\rput(44,9){$d_A$}
\psline{->}(34,1)(34,11)
\psline{->}(44,1)(44,6)
\rput(49,3){$A^*$}
\rput(30,3){$A$}
\end{pspicture}
}
\begin{proof}
\noindent \emph{{\rm(i)}} Unitarity of $d_A$ means $d^\dagger_A\circ d_A = 1_A$ and $d_A\circ d^\dagger_A =1_{A^*}$ which holds since\vspace{3mm}

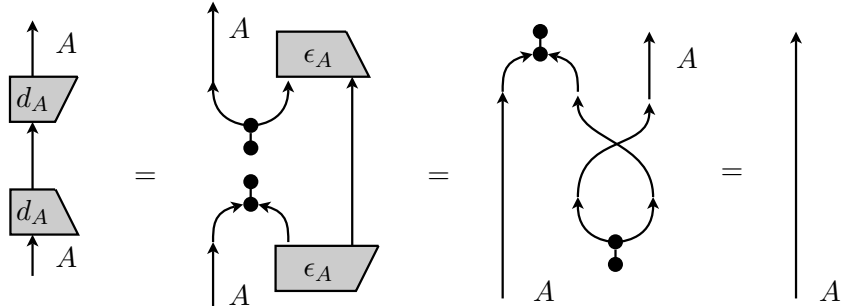
\centerline{
\ifx\JPicScale\undefined\def\JPicScale{1}\fi
\psset{unit=\JPicScale mm}
\psset{linewidth=0.3,dotsep=1,hatchwidth=0.3,hatchsep=1.5,shadowsize=1,dimen=middle}
\psset{dotsize=0.7 2.5,dotscale=1 1,fillcolor=black}
\psset{arrowsize=1 2,arrowlength=1,arrowinset=0.25,tbarsize=0.7 5,bracketlength=0.15,rbracketlength=0.15}
\begin{pspicture}(0,0)(118,47.5)
\newrgbcolor{userFillColour}{0.8 0.8 0.8}
\psline[fillcolor=userFillColour,fillstyle=solid](53.69,43.5)
(56.75,37.5)
(44.5,37.5)
(44.5,43.5)(53.69,43.5)
\rput(50,40){$\epsilon_A$}
\newrgbcolor{userFillColour}{0.8 0.8 0.8}
\psline[fillcolor=userFillColour,fillstyle=solid](58,15)
(55,9)
(44.31,9)
(44.31,15)(58,15)
\rput(50,11.5){$\epsilon_A$}
\psline{->}(54.5,15)(54.5,37.5)
\psline(41,20.93)(41,23.5)
\rput{0}(41,20.5){\psellipse[linestyle=none,fillstyle=solid](0,0)(1,-1)}
\psbezier{<-}(42,20.5)(46,19.5)(46,17.5)(46,15.5)
\psbezier{<-}(40,20.5)(36,19.5)(36,17.5)(36,15.5)
\psline{->}(36,7)(36,15.5)
\rput{0}(41,23.5){\psellipse[linestyle=none,fillstyle=solid](0,0)(1,-1)}
\psline(41,27.5)(41,30)
\rput{0}(41,31){\psellipse[linestyle=none,fillstyle=solid](0,0)(1,-1)}
\psbezier{<-}(46,37)(46,31)(41,31)(41,31)
\psbezier{<-}(36,37)(36,31)(41,31)(41,31)
\rput(39.5,8.5){$A$}
\rput(39.5,44){$A$}
\rput{0}(41,28){\psellipse[linestyle=none,fillstyle=solid](0,0)(1,-1)}
\psline{->}(36,36.5)(36,47.5)
\rput(66,24){$=$}
\psline(79.5,40.93)(79.5,43.5)
\rput{0}(79.5,40.5){\psellipse[linestyle=none,fillstyle=solid](0,0)(1,-1)}
\psbezier{<-}(80.5,40.5)(84.5,39.5)(84.5,37.5)(84.5,35.5)
\psbezier{<-}(78.5,40.5)(74.5,39.5)(74.5,37.5)(74.5,35.5)
\psline{->}(74.5,8)(74.5,35.5)
\rput{0}(79.5,43.5){\psellipse[linestyle=none,fillstyle=solid](0,0)(1,-1)}
\psline(89.5,12)(89.5,14.5)
\rput{0}(89.5,15.5){\psellipse[linestyle=none,fillstyle=solid](0,0)(1,-1)}
\psbezier{<-}(94.5,21.5)(94.5,15.5)(89.5,15.5)(89.5,15.5)
\psbezier{<-}(84.5,21.5)(84.5,15.5)(89.5,15.5)(89.5,15.5)
\rput(118,9){$A$}
\rput(99,40){$A$}
\rput{0}(89.5,12.5){\psellipse[linestyle=none,fillstyle=solid](0,0)(1,-1)}
\psline{->}(94,34.5)(94,43.5)
\rput(105,24.5){$=$}
\psbezier{->}(84.5,21.5)(84.5,30)(94,27.5)(94,34)
\psbezier{->}(94.5,21.5)(94.5,29)(84.5,29)(84.5,35)
\psline{->}(113.5,8)(113.5,43.5)
\rput(80,9){$A$}
\psline{->}(12,11)(12,16.5)
\newrgbcolor{userFillColour}{0.8 0.8 0.8}
\psline[fillcolor=userFillColour,fillstyle=solid](15.06,22.5)
(18.06,16.5)
(9.06,16.5)
(9.06,22.5)(15,22.5)
\psline{->}(12,22.5)(12,31.5)
\rput(16.5,13.5){$A$}
\rput(12,19.5){$d_A$}
\newrgbcolor{userFillColour}{0.8 0.8 0.8}
\psline[fillcolor=userFillColour,fillstyle=solid](18,37.5)
(15,31.5)
(9,31.5)
(9,37.5)(18,37.5)
\rput(12,34.5){$d_A$}
\psline{->}(12,37.5)(12,45)
\rput(27,24){$=$}
\rput(16.5,42){$A$}
\end{pspicture}
}\vspace{-2mm}

\centerline{
\ifx\JPicScale\undefined\def\JPicScale{1}\fi
\psset{unit=\JPicScale mm}
\psset{linewidth=0.3,dotsep=1,hatchwidth=0.3,hatchsep=1.5,shadowsize=1,dimen=middle}
\psset{dotsize=0.7 2.5,dotscale=1 1,fillcolor=black}
\psset{arrowsize=1 2,arrowlength=1,arrowinset=0.25,tbarsize=0.7 5,bracketlength=0.15,rbracketlength=0.15}
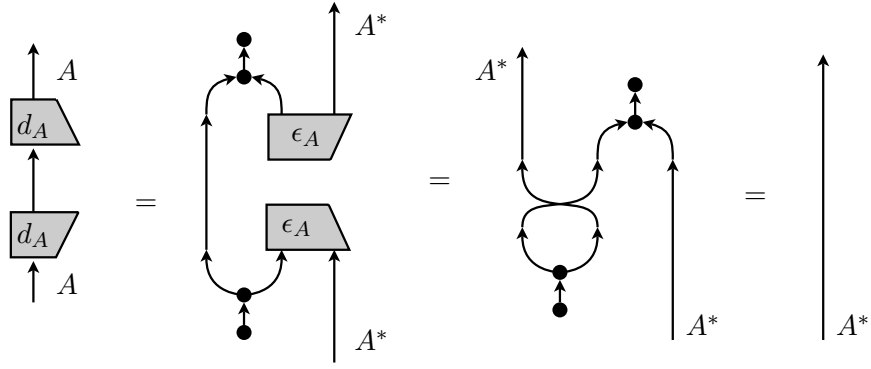
\begin{pspicture}(0,0)(121,49)
\newrgbcolor{userFillColour}{0.8 0.8 0.8}
\psline[fillcolor=userFillColour,fillstyle=solid](51.25,22)
(54,16)
(43,16)
(43,22)(51.25,22)
\psline{->}(52,34)(52,49)
\rput(47,19){$\epsilon_A$}
\rput(66,25){$=$}
\newrgbcolor{userFillColour}{0.8 0.8 0.8}
\psline[fillcolor=userFillColour,fillstyle=solid](51.5,28)
(54.25,34)
(43.25,34)
(43.25,28)(51.5,28)
\rput(48.25,31){$\epsilon_A$}
\psbezier{->}(77,19)(77,24.6)(87,19)(87,28)
\psbezier{->}(87,19)(87,24.6)(77,19)(77,28)
\psline{->}(35,16)(35,34)
\psline{->}(117,4)(117,42)
\rput(121,6){$A^*$}
\psline{->}(40,39.43)(40,43)
\rput{0}(40,39){\psellipse[linestyle=none,fillstyle=solid](0,0)(1,-1)}
\psbezier{<-}(41,39)(45,38)(45,36)(45,34)
\psbezier{<-}(39,39)(35,38)(35,36)(35,34)
\rput{0}(40,44){\psellipse[linestyle=none,fillstyle=solid](0,0)(1,-1)}
\psline{->}(52,1)(52,16)
\psline{->}(40,6)(40,9)
\rput{0}(40,10){\psellipse[linestyle=none,fillstyle=solid](0,0)(1,1)}
\psbezier{<-}(35,16)(35,10)(40,10)(40,10)
\psbezier{<-}(45,16)(45,10)(40,10)(40,10)
\rput{0}(40,5){\psellipse[linestyle=none,fillstyle=solid](0,0)(1,1)}
\rput(108,23){$=$}
\psline{->}(77,28)(77,43)
\psline{->}(97,4)(97,28)
\psline{->}(82,9)(82,12)
\rput{0}(82,13){\psellipse[linestyle=none,fillstyle=solid](0,0)(1,1)}
\psbezier{<-}(77,19)(77,13)(82,13)(82,13)
\psbezier{<-}(87,19)(87,13)(82,13)(82,13)
\rput{0}(82,8){\psellipse[linestyle=none,fillstyle=solid](0,0)(1,1)}
\psline{->}(92,33.43)(92,37)
\rput{0}(92,33){\psellipse[linestyle=none,fillstyle=solid](0,0)(1,-1)}
\psbezier{<-}(93,33)(97,32)(97,30)(97,28)
\psbezier{<-}(91,33)(87,32)(87,30)(87,28)
\rput{0}(92,38){\psellipse[linestyle=none,fillstyle=solid](0,0)(1,-1)}
\rput(101,6){$A^*$}
\rput(73,40){$A^*$}
\rput(57,46){$A^*$}
\rput(57,4){$A^*$}
\psline{->}(12,9)(12,14.5)
\newrgbcolor{userFillColour}{0.8 0.8 0.8}
\psline[fillcolor=userFillColour,fillstyle=solid](15.06,36)
(18.06,30)
(9.06,30)
(9.06,36)(15,36)
\psline{->}(12,20.5)(12,29.5)
\rput(16.5,11.5){$A$}
\newrgbcolor{userFillColour}{0.8 0.8 0.8}
\psline[fillcolor=userFillColour,fillstyle=solid](18,21)
(15,15)
(9,15)
(9,21)(18,21)
\rput(12,32.5){$d_A$}
\psline{->}(12,36)(12,43.5)
\rput(27,22){$=$}
\rput(16.5,40){$A$}
\rput(12,18){$d_A$}
\end{pspicture}
} 

\noindent \emph{{\rm(ii)}} We have $\dagger$-compactness for  $(A, \epsilon_A)$ since\vspace{2mm}

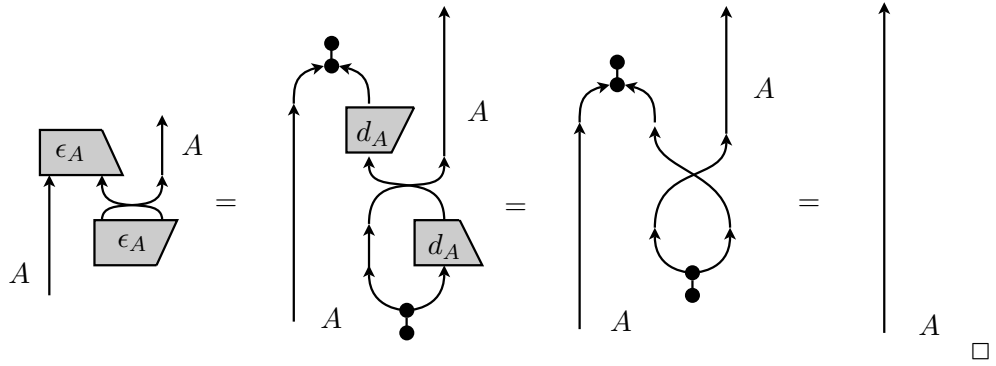
\centerline{
\ifx\JPicScale\undefined\def\JPicScale{1}\fi
\psset{unit=\JPicScale mm}
\psset{linewidth=0.3,dotsep=1,hatchwidth=0.3,hatchsep=1.5,shadowsize=1,dimen=middle}
\psset{dotsize=0.7 2.5,dotscale=1 1,fillcolor=black}
\psset{arrowsize=1 2,arrowlength=1,arrowinset=0.25,tbarsize=0.7 5,bracketlength=0.15,rbracketlength=0.15}
\begin{pspicture}(0,0)(124,54.5)
\newrgbcolor{userFillColour}{0.8 0.8 0.8}
\psline[fillcolor=userFillColour,fillstyle=solid](14,37.5)
(16.75,31.5)
(5.75,31.5)
(5.75,37.5)(14,37.5)
\psline{->}(22,31.5)(22,39.5)
\rput(9.75,34.5){$\epsilon_A$}
\newrgbcolor{userFillColour}{0.8 0.8 0.8}
\psline[fillcolor=userFillColour,fillstyle=solid](21.25,19.5)
(24,25.5)
(13,25.5)
(13,19.5)(21.25,19.5)
\rput(18,22.5){$\epsilon_A$}
\psbezier{->}(14,25.5)(14,29.23)(22,25.5)(22,31.5)
\psbezier{->}(22,25.5)(22,29.23)(14,25.5)(14,31.5)
\psline{->}(7,15.5)(7,31.5)
\rput(3,18.5){$A$}
\rput(26,35.5){$A$}
\psline{->}(118,10.5)(118,54.5)
\rput(124,11.5){$A$}
\rput(30.5,27.5){$=$}
\psline(44.5,46.43)(44.5,49)
\rput{0}(44.5,46){\psellipse[linestyle=none,fillstyle=solid](0,0)(1,-1)}
\psbezier{<-}(45.5,46)(49.5,45)(49.5,43)(49.5,41)
\psbezier{<-}(43.5,46)(39.5,45)(39.5,43)(39.5,41)
\psline{->}(39.5,12)(39.5,41)
\rput{0}(44.5,49){\psellipse[linestyle=none,fillstyle=solid](0,0)(1,-1)}
\psline(54.5,10)(54.5,12.5)
\rput{0}(54.5,13.5){\psellipse[linestyle=none,fillstyle=solid](0,0)(1,-1)}
\psbezier{<-}(59.5,19.5)(59.5,13.5)(54.5,13.5)(54.5,13.5)
\psbezier{<-}(49.5,19.5)(49.5,13.5)(54.5,13.5)(54.5,13.5)
\rput(64,40){$A$}
\rput{0}(54.5,10.5){\psellipse[linestyle=none,fillstyle=solid](0,0)(1,-1)}
\psline{->}(59.5,34)(59.5,54)
\psbezier{->}(49.5,25)(49.5,33.5)(59.5,27.5)(59.5,34)
\psbezier{->}(59.5,25.5)(59.5,33)(49.5,28)(49.5,34)
\rput(44.5,12.5){$A$}
\newrgbcolor{userFillColour}{0.8 0.8 0.8}
\psline[fillcolor=userFillColour,fillstyle=solid](61.56,25.5)
(64.56,19.5)
(55.56,19.5)
(55.56,25.5)(61.5,25.5)
\newrgbcolor{userFillColour}{0.8 0.8 0.8}
\psline[fillcolor=userFillColour,fillstyle=solid](55.5,40.5)
(52.5,34.5)
(46.5,34.5)
(46.5,40.5)(55.5,40.5)
\rput(69,27){$=$}
\psline(82.5,43.93)(82.5,46.5)
\rput{0}(82.5,43.5){\psellipse[linestyle=none,fillstyle=solid](0,0)(1,-1)}
\psbezier{<-}(83.5,43.5)(87.5,42.5)(87.5,40.5)(87.5,38.5)
\psbezier{<-}(81.5,43.5)(77.5,42.5)(77.5,40.5)(77.5,38.5)
\psline{->}(77.5,11)(77.5,38.5)
\rput{0}(82.5,46.5){\psellipse[linestyle=none,fillstyle=solid](0,0)(1,-1)}
\psline(92.5,15)(92.5,17.5)
\rput{0}(92.5,18.5){\psellipse[linestyle=none,fillstyle=solid](0,0)(1,-1)}
\psbezier{<-}(97.5,24.5)(97.5,18.5)(92.5,18.5)(92.5,18.5)
\psbezier{<-}(87.5,24.5)(87.5,18.5)(92.5,18.5)(92.5,18.5)
\rput(102,43){$A$}
\rput{0}(92.5,15.5){\psellipse[linestyle=none,fillstyle=solid](0,0)(1,-1)}
\psline{->}(97,37)(97,54)
\rput(108,27.5){$=$}
\psbezier{->}(87.5,24.5)(87.5,33)(97,30.5)(97,37)
\psbezier{->}(97.5,24.5)(97.5,32)(87.5,32)(87.5,38)
\rput(83,12){$A$}
\psline{->}(49.5,19)(49.5,25)
\rput(50,37){$d_A$}
\rput(59.5,22){$d_A$}
\end{pspicture}
}
\vspace{-1.2cm}
\end{proof}
\vspace{-5mm}
\begin{definition}
A \emph{$\dagger$-compact category with bases} is a $\dagger$-SMC such that every object $A$ comes with $\dagger$-dual Frobenius structure 
\[
(A,\delta_A:A\to A\otimes A,\gamma_A:A\to I,d_A:A\to A^*)\,,
\]
and where the $\dagger$-dual Frobenius structures on object $A$ and its dual $A^*$ are connected by the fact that $d_{A^*}=d_A^\dagger$ and that $d_A$ is a \emph{permutation}.
\end{definition}

\begin{lemma}
A $\dagger$-compact category with bases is `indeed' a $\dagger$-compact category.
\end{lemma}

\begin{proof}  For  each object $A$, let $\epsilon_A := \gamma_A\circ \delta_A^\dagger\circ (1_A \otimes d^\dagger_A)$; according to Thm.~\ref{thm:dagDFdagCdagF} this entails that $(A,\epsilon_A)$ is a $\dagger$-compact structure. Moreover, $\epsilon_{A^*} = \epsilon_A\circ \sigma_{A^*,A}$ since\vspace{3mm}

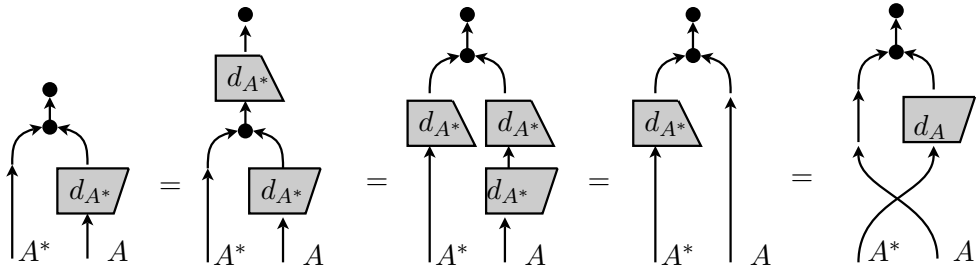
\centerline{
\ifx\JPicScale\undefined\def\JPicScale{1}\fi
\psset{unit=\JPicScale mm}
\psset{linewidth=0.3,dotsep=1,hatchwidth=0.3,hatchsep=1.5,shadowsize=1,dimen=middle}
\psset{dotsize=0.7 2.5,dotscale=1 1,fillcolor=black}
\psset{arrowsize=1 2,arrowlength=1,arrowinset=0.25,tbarsize=0.7 5,bracketlength=0.15,rbracketlength=0.15}
\begin{pspicture}(0,0)(134,37)
\psline{->}(11,20.93)(11,24.5)
\rput{0}(11,20.5){\psellipse[linestyle=none,fillstyle=solid](0,0)(1,-1)}
\psbezier{<-}(12,20.5)(16,19.5)(16,17.5)(16,15.5)
\psbezier{<-}(10,20.5)(6,19.5)(6,17.5)(6,15.5)
\psline{->}(6,3)(6,15.5)
\rput{0}(11,25.5){\psellipse[linestyle=none,fillstyle=solid](0,0)(1,-1)}
\rput(20,4){$A$}
\psline{->}(16,3)(16,9)
\rput(27,12.5){$=$}
\rput(9,4){$A^*$}
\psline{->}(37,20.43)(37,24)
\rput{0}(37,20){\psellipse[linestyle=none,fillstyle=solid](0,0)(1,-1)}
\psbezier{<-}(38,20)(42,19)(42,17)(42,15)
\psbezier{<-}(36,20)(32,19)(32,17)(32,15)
\psline{->}(32,2.5)(32,15)
\rput{0}(37,35.5){\psellipse[linestyle=none,fillstyle=solid](0,0)(1,-1)}
\rput(46,3.5){$A$}
\psline{->}(42,2.5)(42,8.5)
\rput(35,3.5){$A^*$}
\psline{->}(37,30.5)(37,34.07)
\rput(54.5,13){$=$}
\newrgbcolor{userFillColour}{0.8 0.8 0.8}
\psline[fillcolor=userFillColour,fillstyle=solid](75,24)
(78,18)
(69,18)
(69,24)(75,24)
\newrgbcolor{userFillColour}{0.8 0.8 0.8}
\psline[fillcolor=userFillColour,fillstyle=solid](78.5,15)
(76.5,9)
(69,9)
(69,15)(78.5,15)
\psline{->}(66.5,30.43)(66.5,34.5)
\rput{0}(66.5,30){\psellipse[linestyle=none,fillstyle=solid](0,0)(1,-1)}
\psbezier{<-}(67.5,30)(71.5,29)(71.5,27)(71.5,25)
\psbezier{<-}(65.5,30)(61.5,29)(61.5,27)(61.5,25)
\psline{->}(61.5,2.5)(61.5,18)
\rput{0}(66.5,35.5){\psellipse[linestyle=none,fillstyle=solid](0,0)(1,-1)}
\rput(75.5,3.5){$A$}
\psline{->}(71.5,2.5)(71.5,8.5)
\rput(72,12){$d_{A^*}$}
\rput(64.5,3.5){$A^*$}
\rput(73.5,21){$d_{A^*}$}
\rput(84,13){$=$}
\newrgbcolor{userFillColour}{0.8 0.8 0.8}
\psline[fillcolor=userFillColour,fillstyle=solid](64.56,24)
(67.56,18)
(58.56,18)
(58.56,24)(64.5,24)
\newrgbcolor{userFillColour}{0.8 0.8 0.8}
\psline[fillcolor=userFillColour,fillstyle=solid](39.06,30)
(42.06,24)
(33.06,24)
(33.06,30)(39,30)
\newrgbcolor{userFillColour}{0.8 0.8 0.8}
\psline[fillcolor=userFillColour,fillstyle=solid](47,15)
(45,9)
(37.5,9)
(37.5,15)(47,15)
\newrgbcolor{userFillColour}{0.8 0.8 0.8}
\psline[fillcolor=userFillColour,fillstyle=solid](21.5,15)
(19.5,9)
(12,9)
(12,15)(21.5,15)
\rput(16.5,12){$d_{A^*}$}
\rput(37.5,27){$d_{A^*}$}
\rput(42,12){$d_{A^*}$}
\rput(63,21){$d_{A^*}$}
\psline{->}(72,15)(72,18)
\psline{->}(96.5,30.43)(96.5,34.5)
\rput{0}(96.5,30){\psellipse[linestyle=none,fillstyle=solid](0,0)(1,-1)}
\psbezier{<-}(97.5,30)(101.5,29)(101.5,27)(101.5,25)
\psbezier{<-}(95.5,30)(91.5,29)(91.5,27)(91.5,25)
\psline{->}(91.5,2.5)(91.5,18)
\rput{0}(96.5,35.5){\psellipse[linestyle=none,fillstyle=solid](0,0)(1,-1)}
\rput(105.5,3.5){$A$}
\psline{->}(101.5,2.5)(101.5,24.5)
\rput(94.5,3.5){$A^*$}
\rput(111,13.5){$=$}
\newrgbcolor{userFillColour}{0.8 0.8 0.8}
\psline[fillcolor=userFillColour,fillstyle=solid](94.56,24)
(97.56,18)
(88.56,18)
(88.56,24)(94.5,24)
\rput(93,21){$d_{A^*}$}
\psline{->}(123.5,30.93)(123.5,35)
\rput{0}(123.5,30.5){\psellipse[linestyle=none,fillstyle=solid](0,0)(1,-1)}
\psbezier{<-}(124.5,30.5)(128.5,29.5)(128.5,27.5)(128.5,25.5)
\psbezier{<-}(122.5,30.5)(118.5,29.5)(118.5,27.5)(118.5,25.5)
\psline{->}(118.5,18.5)(118.5,25.5)
\rput{0}(123.5,36){\psellipse[linestyle=none,fillstyle=solid](0,0)(1,-1)}
\rput(132.5,4){$A$}
\rput(122,4){$A^*$}
\newrgbcolor{userFillColour}{0.8 0.8 0.8}
\psline[fillcolor=userFillColour,fillstyle=solid](134,24.5)
(132,18.5)
(124.5,18.5)
(124.5,24.5)(134,24.5)
\rput(128,20.5){$d_A$}
\psbezier{->}(129,3)(129,10.5)(118.5,12)(118.5,18)
\psbezier{->}(118.5,2.5)(118.5,11)(128.5,11.5)(128.5,18)
\end{pspicture}
}

\noindent
where the two first steps use the fact that $d_{A^*}$ is a function and the third one uses unitarity of $d_{A^*}$.
\end{proof}

The graphical language for $\dagger$-compact categories can now be extended to $\dagger$-compact categories with bases: 

\noindent
- First, the dualiser depicts as:\vspace{2mm}

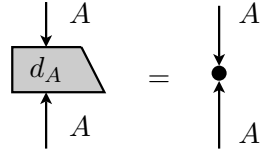
\centerline{
\ifx\JPicScale\undefined\def\JPicScale{1}\fi
\psset{unit=\JPicScale mm}
\psset{linewidth=0.3,dotsep=1,hatchwidth=0.3,hatchsep=1.5,shadowsize=1,dimen=middle}
\psset{dotsize=0.7 2.5,dotscale=1 1,fillcolor=black}
\psset{arrowsize=1 2,arrowlength=1,arrowinset=0.25,tbarsize=0.7 5,bracketlength=0.15,rbracketlength=0.15}
\begin{pspicture}(0,0)(33,19.5)
\psline{<-}(6,13.5)(6,19.5)
\psline{->}(6,0)(6,7.5)
\rput(10.5,3){$A$}
\rput(10.5,18){$A$}
\psline{<-}(29,11)(29,19)
\rput{0}(29,10){\psellipse[linestyle=none,fillstyle=solid](0,0)(1,-1)}
\psline{->}(29,0)(29,9)
\rput(33,2){$A$}
\rput(33,18){$A$}
\newrgbcolor{userFillColour}{0.8 0.8 0.8}
\psline[fillcolor=userFillColour,fillstyle=solid](10.69,13.5)
(13.75,7.5)
(1.5,7.5)
(1.5,13.5)(10.69,13.5)
\rput(6,10.5){$d_A$}
\rput(21,9){$=$}
\end{pspicture}
} 

\noindent
- The $\dagger$-dual Frobenius structure `factorises' the $\dagger$-compact structure:\vspace{2mm}

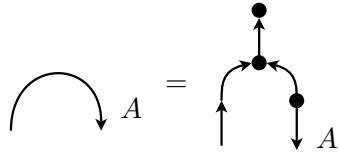
\centerline{
\ifx\JPicScale\undefined\def\JPicScale{1}\fi
\psset{unit=\JPicScale mm}
\psset{linewidth=0.3,dotsep=1,hatchwidth=0.3,hatchsep=1.5,shadowsize=1,dimen=middle}
\psset{dotsize=0.7 2.5,dotscale=1 1,fillcolor=black}
\psset{arrowsize=1 2,arrowlength=1,arrowinset=0.25,tbarsize=0.7 5,bracketlength=0.15,rbracketlength=0.15}
\begin{pspicture}(0,0)(43,21)
\psbezier{->}(1,4)(1,14)(13,14)(13,4)
\psline{->}(34,13.43)(34,19)
\rput{0}(34,13){\psellipse[linestyle=none,fillstyle=solid](0,0)(1,-1)}
\psbezier{<-}(35,13)(39,12)(39,10)(39,8)
\psbezier{<-}(33,13)(29,12)(29,10)(29,8)
\rput{0}(34,20){\psellipse[linestyle=none,fillstyle=solid](0,0)(1,-1)}
\rput{0}(39,8){\psellipse[linestyle=none,fillstyle=solid](0,0)(1,-1)}
\psline{<-}(39,1.43)(39,7)
\psline{->}(29,2)(29,8)
\rput(17,7){$A$}
\rput(43,3){$A$}
\rput(23,10){$=$}
\end{pspicture}
} 

\noindent
- Finally, the coherence conditions are those of the $\dagger$-Frobenius structures and:\vspace{2mm}

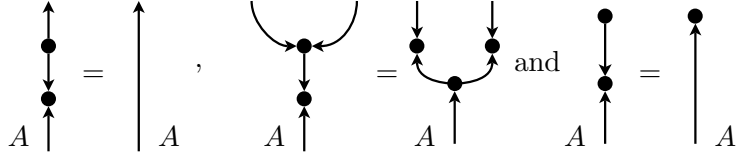
\centerline{
\ifx\JPicScale\undefined\def\JPicScale{1}\fi
\psset{unit=\JPicScale mm}
\psset{linewidth=0.3,dotsep=1,hatchwidth=0.3,hatchsep=1.5,shadowsize=1,dimen=middle}
\psset{dotsize=0.7 2.5,dotscale=1 1,fillcolor=black}
\psset{arrowsize=1 2,arrowlength=1,arrowinset=0.25,tbarsize=0.7 5,bracketlength=0.15,rbracketlength=0.15}
\begin{pspicture}(0,0)(95,20)
\psline{->}(5,0)(5,6)
\rput{0}(59,9){\psellipse[linestyle=none,fillstyle=solid](0,0)(1,-1)}
\psbezier{<-}(64,13)(64,9)(59,9)(59,9)
\rput{0}(64,14){\psellipse[linestyle=none,fillstyle=solid](0,0)(1,-1)}
\rput{0}(5,7){\psellipse[linestyle=none,fillstyle=solid](0,0)(1,-1)}
\psline{<-}(5,8)(5,13)
\rput{0}(5,14){\psellipse[linestyle=none,fillstyle=solid](0,0)(1,-1)}
\psline{->}(5,15)(5,20)
\psline{->}(17,0)(17,20)
\rput(21,2){$A$}
\rput(1,2){$A$}
\rput(11,10){$=$}
\psbezier{<-}(54,13)(54,9)(59,9)(59,9)
\rput{0}(54,14){\psellipse[linestyle=none,fillstyle=solid](0,0)(1,-1)}
\psline{->}(59,1)(59,8)
\psline{<-}(54,15)(54,20)
\psline{<-}(64,15)(64,20)
\rput{0}(39,14){\psellipse[linestyle=none,fillstyle=solid](0,0)(1,-1)}
\rput{0}(39,7){\psellipse[linestyle=none,fillstyle=solid](0,0)(1,-1)}
\psline{<-}(39,8)(39,13)
\psline{->}(39,0)(39,6)
\psbezier{->}(32,20)(32,17)(35,14)(38,14)
\psbezier{->}(46,20)(46,17)(44,14)(40,14)
\rput(50,10){$=$}
\rput(25,11){,}
\rput(70,12){and}
\psline{->}(79,1)(79,8)
\rput{0}(79,9){\psellipse[linestyle=none,fillstyle=solid](0,0)(1,-1)}
\psline[arrowscale=1.1 1]{<-}(79,10)(79,17)
\rput{0}(79,18){\psellipse[linestyle=none,fillstyle=solid](0,0)(1,-1)}
\psline{->}(91,1)(91,17)
\rput{0}(91,18){\psellipse[linestyle=none,fillstyle=solid](0,0)(1,-1)}
\rput(85,10){$=$}
\rput(35,2){$A$}
\rput(55,2){$A$}
\rput(75,2){$A$}
\rput(95,2){$A$}
\end{pspicture}
} 

\begin{definition}
A \em strict $\dagger$-compact category with bases \em is a $\dagger$-compact category with bases which is such that for any objects $A$ and $B$ we have  
\[
d_{A\otimes B} = (d_B \otimes d_A)\circ \sigma_{A,B}
\]
and also
\[
\delta_{A\otimes B}= (1_A\otimes \sigma_{A,B}\otimes 1_B)\circ (\delta_A\otimes \delta_B)
\qquad\mbox{\rm and}\qquad
\gamma_{A\otimes B}=\gamma_A \otimes \gamma_B\,.
\]
Graphically, these are

~\\
\centerline{
\ifx\JPicScale\undefined\def\JPicScale{1}\fi
\psset{unit=\JPicScale mm}
\psset{linewidth=0.3,dotsep=1,hatchwidth=0.3,hatchsep=1.5,shadowsize=1,dimen=middle}
\psset{dotsize=0.7 2.5,dotscale=1 1,fillcolor=black}
\psset{arrowsize=1 2,arrowlength=1,arrowinset=0.25,tbarsize=0.7 5,bracketlength=0.15,rbracketlength=0.15}
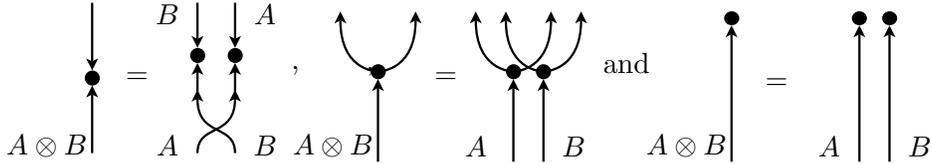
\begin{pspicture}(0,0)(119,22)
\psline{<-}(23,16)(23,22)
\rput{0}(23,15){\psellipse[linestyle=none,fillstyle=solid](0,0)(1,-1)}
\psline{->}(23,10)(23,14)
\rput(32,20.5){$A$}
\rput(19,3){$A$}
\psline{<-}(28,16)(28,22)
\rput{0}(28,15){\psellipse[linestyle=none,fillstyle=solid](0,0)(1,-1)}
\psline{->}(28,10)(28,14)
\psbezier{->}(23,2)(23,6)(28,4)(28,10.43)
\psbezier{->}(28,2)(28,6)(23,4)(23,10.43)
\rput(32,3){$B$}
\rput(19,20.5){$B$}
\psline{->}(65,0.85)(65,11.85)
\rput{0}(65,12.85){\psellipse[linestyle=none,fillstyle=solid](0,0)(1,-1)}
\psbezier{<-}(70,20.85)(70,12.85)(65,12.85)(65,12.85)
\psbezier{<-}(60,20.85)(60,12.85)(65,12.85)(65,12.85)
\rput(60,2.85){$A$}
\psline{->}(69,0.85)(69,11.85)
\rput{0}(69,12.85){\psellipse[linestyle=none,fillstyle=solid](0,0)(1,-1)}
\psbezier{<-}(74,20.85)(74,12.85)(69,12.85)(69,12.85)
\psbezier{<-}(64,20.85)(64,12.85)(69,12.85)(69,12.85)
\rput(73,2.85){$B$}
\psline{->}(111,0.85)(111,19)
\rput{0}(111,19.93){\psellipse[linestyle=none,fillstyle=solid](0,0)(1,-0.93)}
\psline{->}(115,0.85)(115,19)
\rput{0}(115,19.93){\psellipse[linestyle=none,fillstyle=solid](0,0)(1,-0.93)}
\rput(107,2.85){$A$}
\rput(119,2.85){$B$}
\psline{->}(94,0.85)(94,19)
\rput{0}(94,19.93){\psellipse[linestyle=none,fillstyle=solid](0,0)(1,-0.93)}
\rput(88,3){$A\otimes B$}
\rput(100,11.2){$=$}
\rput(80,14){and}
\rput(46,12.85){,}
\rput(41,-0.15){}
\psline{->}(47,0.85)(47,11.85)
\rput{0}(47,12.85){\psellipse[linestyle=none,fillstyle=solid](0,0)(1,-1)}
\psbezier{<-}(52,20.85)(52,12.85)(47,12.85)(47,12.85)
\psbezier{<-}(42,20.85)(42,12.85)(47,12.85)(47,12.85)
\rput(41,3){$A\otimes B$}
\rput(56,12){$=$}
\psline{<-}(9,13)(9,22)
\rput{0}(9,12){\psellipse[linestyle=none,fillstyle=solid](0,0)(1,-1)}
\psline{->}(9,2)(9,11)
\rput(3,3){$A\otimes B$}
\rput(15,12){$=$}
\rput(36,13){,}
\end{pspicture}
}
\end{definition}

\begin{lemma}
A strict $\dagger$-compact category with bases is `indeed'  strict.
\end{lemma}

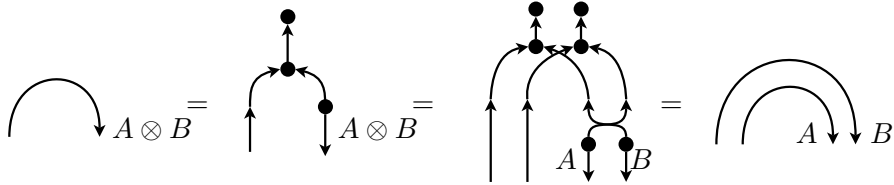
\begin{proof}
Since\vspace{2mm}

\centerline{
\ifx\JPicScale\undefined\def\JPicScale{1}\fi
\psset{unit=\JPicScale mm}
\psset{linewidth=0.3,dotsep=1,hatchwidth=0.3,hatchsep=1.5,shadowsize=1,dimen=middle}
\psset{dotsize=0.7 2.5,dotscale=1 1,fillcolor=black}
\psset{arrowsize=1 2,arrowlength=1,arrowinset=0.25,tbarsize=0.7 5,bracketlength=0.15,rbracketlength=0.15}
\begin{pspicture}(0,0)(118,25)
\psbezier{->}(2,7)(2,17)(14,17)(14,7)
\psline{->}(39,16.43)(39,22)
\rput{0}(39,16){\psellipse[linestyle=none,fillstyle=solid](0,0)(1,-1)}
\psbezier{<-}(40,16)(44,15)(44,13)(44,11)
\psbezier{<-}(38,16)(34,15)(34,13)(34,11)
\rput{0}(39,23){\psellipse[linestyle=none,fillstyle=solid](0,0)(1,-1)}
\rput{0}(44,11){\psellipse[linestyle=none,fillstyle=solid](0,0)(1,-1)}
\psline{<-}(44,4.43)(44,10)
\psline{->}(34,5)(34,11)
\rput(21,8){$A\otimes B$}
\rput(51,8){$A\otimes B$}
\rput(27,11){$=$}
\psline{<-}(72,23)(72,20)
\rput{0}(72,19){\psellipse[linestyle=none,fillstyle=solid](0,0)(1,1)}
\psbezier{<-}(71,19)(66,18.12)(66,12)(66,12)
\rput(76,4){$A$}
\psline{<-}(78,23)(78,20)
\rput{0}(78,19){\psellipse[linestyle=none,fillstyle=solid](0,0)(1,1)}
\psbezier{<-}(79,19)(84,18)(84,12)(84,12)
\psbezier{<-}(77,19)(70,16.37)(70.88,12)(70.88,12)
\rput(86,4){$B$}
\psbezier{->}(79,7)(79,10.11)(84,7)(84,12)
\psbezier{->}(84,7)(84,10.11)(79,7)(79,12)
\rput{0}(72,24){\psellipse[linestyle=none,fillstyle=solid](0,0)(1,1)}
\rput{0}(78,24){\psellipse[linestyle=none,fillstyle=solid](0,0)(1,1)}
\rput{0}(79,6){\psellipse[linestyle=none,fillstyle=solid](0,0)(1,1)}
\rput{0}(84,6){\psellipse[linestyle=none,fillstyle=solid](0,0)(1,1)}
\psbezier{<-}(73,19)(79,17)(79,12)(79,12)
\psline{<-}(66,12)(66,1)
\psline{<-}(70.88,12)(70.88,1)
\psline{->}(79,5)(79,1)
\psline{->}(84,5)(84,1)
\rput(57,11){$=$}
\psbezier{->}(96,6)(96,20.5)(114.5,21)(114.5,6)
\psbezier{->}(99.5,6)(99.5,16)(111.5,16)(111.5,6)
\rput(108,7.5){$A$}
\rput(118,7.5){$B$}
\rput(90,11){$=$}
\end{pspicture}
} 

\noindent
we indeed obtain strictness as in Defn.~\ref{def:strict}.
\end{proof}

\subsection{Spider theorem for basis structures}  

The spider theorem for $\dagger$-Frobenius structures of \cite{CPaquette} also holds in $\dagger$-compact categories with bases provided we allow the spider's legs to be directed.  Hence a spider now takes the form\vspace{2mm}

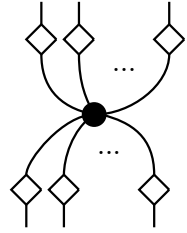
\centerline{
\ifx\JPicScale\undefined\def\JPicScale{1}\fi
\psset{unit=\JPicScale mm}
\psset{linewidth=0.3,dotsep=1,hatchwidth=0.3,hatchsep=1.5,shadowsize=1,dimen=middle}
\psset{dotsize=0.7 2.5,dotscale=1 1,fillcolor=black}
\psset{arrowsize=1 2,arrowlength=1,arrowinset=0.25,tbarsize=0.7 5,bracketlength=0.15,rbracketlength=0.15}
\begin{pspicture}(0,0)(24,31)
\psbezier(12,16)(7,17)(5,20)(5,24)
\psline(7,26)
(5,24)
(3,26)
(5,28)(7,26)
\psbezier(12,16)(11,18)(10,20)(10,24)
\psbezier(8,8)(8,10)(9,14)(12,16)
\psbezier(20,8)(20,13)(17,15)(13,16)
\psbezier(12,16)(17,16)(22,20)(22,24)
\psline(12,26)
(10,24)
(8,26)
(10,28)(12,26)
\psline(24,26)
(22,24)
(20,26)
(22,28)(24,26)
\psline(22,6)
(20,4)
(18,6)
(20,8)(22,6)
\psbezier(3,8)(3,10)(8,16)(12,16)
\psline(5,6)
(3,4)
(1,6)
(3,8)(5,6)
\psline(10,6)
(8,4)
(6,6)
(8,8)(10,6)
\rput(16,22){$...$}
\rput(14,11){$...$}
\rput{0}(12,16){\psellipse[fillstyle=solid](0,0)(1.5,-1.5)}
\psline(5,28)(5,31)
\psline(10,28)(10,31)
\psline(22,28)(22,31)
\psline(3,1)(3,4)
\psline(8,1)(8,4)
\psline(20,1)(20,4)
\end{pspicture}
}  

\noindent
where the diamonds either represent an up- or a down-arrow.  There are six special cases, namely\vspace{2mm}

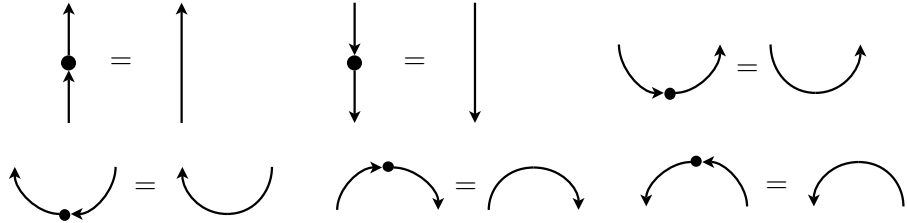
\centerline{
\ifx\JPicScale\undefined\def\JPicScale{1}\fi
\psset{unit=\JPicScale mm}
\psset{linewidth=0.3,dotsep=1,hatchwidth=0.3,hatchsep=1.5,shadowsize=1,dimen=middle}
\psset{dotsize=0.7 2.5,dotscale=1 1,fillcolor=black}
\psset{arrowsize=1 2,arrowlength=1,arrowinset=0.25,tbarsize=0.7 5,bracketlength=0.15,rbracketlength=0.15}
\begin{pspicture}(0,0)(108.24,18)
\psline{<-}(3,9)(3,2)
\rput{0}(3,10){\psellipse[linestyle=none,fillstyle=solid](0,0)(1,-1)}
\psline{<-}(3,18)(3,11)
\psline{->}(41,9)(41,2)
\rput{0}(41,10){\psellipse[linestyle=none,fillstyle=solid](0,0)(1,-1)}
\psline{->}(41,18)(41,11)
\rput(49,10){$=$}
\psline{->}(57,18)(57,2)
\rput{90}(82.94,5.86){\psellipse[linestyle=none,fillstyle=solid](0,0)(0.81,-0.75)}
\rput(93.24,9.06){$=$}
\psbezier{->}(76.14,12.44)(76.14,9.22)(79.13,6)(82.11,6)
\psbezier{<-}(89.58,12.44)(89.58,9.22)(86.59,6)(83.61,6)
\psbezier(96.3,12.44)(96.3,9.22)(98.54,6)(102.27,6)
\psbezier{<-}(108.24,12.44)(108.24,9.22)(106,6)(102.27,6)
\rput(10,10){$=$}
\psline{<-}(18,18)(18,2)
\end{pspicture}
}\vspace{2mm}  

\centerline{
\ifx\JPicScale\undefined\def\JPicScale{1}\fi
\psset{unit=\JPicScale mm}
\psset{linewidth=0.3,dotsep=1,hatchwidth=0.3,hatchsep=1.5,shadowsize=1,dimen=middle}
\psset{dotsize=0.7 2.5,dotscale=1 1,fillcolor=black}
\psset{arrowsize=1 2,arrowlength=1,arrowinset=0.25,tbarsize=0.7 5,bracketlength=0.15,rbracketlength=0.15}
\begin{pspicture}(0,0)(119.72,9.78)
\rput{90}(8.2,1.87){\psellipse[linestyle=none,fillstyle=solid](0,0)(0.78,0.74)}
\rput(18.93,5.51){$=$}
\psbezier{->}(14.98,8.26)(14.98,5.13)(12,2)(9.02,2)
\psbezier{<-}(1.58,8.26)(1.58,5.13)(4.56,2)(7.53,2)
\psbezier(35.77,8.26)(35.77,5.13)(33.53,2)(29.81,2)
\psbezier{<-}(23.86,8.26)(23.86,5.13)(26.09,2)(29.81,2)
\rput{0}(51.23,8.3){\psellipse[linestyle=none,fillstyle=solid](0,0)(0.75,0.7)}
\rput(61.53,5.5){$=$}
\psbezier{->}(44.44,2.54)(44.44,5.36)(47.42,8.18)(50.41,8.18)
\psbezier{<-}(57.87,2.54)(57.87,5.36)(54.89,8.18)(51.9,8.18)
\psbezier(64.59,2.54)(64.59,5.36)(66.83,8.18)(70.56,8.18)
\psbezier{<-}(76.53,2.54)(76.53,5.36)(74.3,8.18)(70.56,8.18)
\rput{0}(92.16,9.04){\psellipse[linestyle=none,fillstyle=solid](0,0)(0.74,-0.74)}
\rput(102.89,5.6){$=$}
\psbezier{->}(98.93,3)(98.93,5.96)(95.95,8.91)(92.98,8.91)
\psbezier{<-}(85.53,3)(85.53,5.96)(88.51,8.91)(91.49,8.91)
\psbezier(119.72,3)(119.72,5.96)(117.49,8.91)(113.77,8.91)
\psbezier{<-}(107.81,3)(107.81,5.96)(110.05,8.91)(113.77,8.91)
\end{pspicture}
}  

\noindent
in which we can drop the dot. However, in the six cases\vspace{2mm}

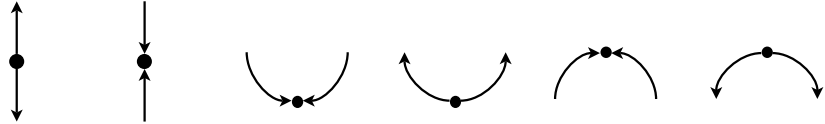
\centerline{
\ifx\JPicScale\undefined\def\JPicScale{1}\fi
\psset{unit=\JPicScale mm}
\psset{linewidth=0.3,dotsep=1,hatchwidth=0.3,hatchsep=1.5,shadowsize=1,dimen=middle}
\psset{dotsize=0.7 2.5,dotscale=1 1,fillcolor=black}
\psset{arrowsize=1 2,arrowlength=1,arrowinset=0.25,tbarsize=0.7 5,bracketlength=0.15,rbracketlength=0.15}
\begin{pspicture}(0,0)(111.44,18)
\psline{<-}(22,9)(22,2)
\rput{0}(22,10){\psellipse[linestyle=none,fillstyle=solid](0,0)(1,-1)}
\psline{->}(22,18)(22,11)
\psline{->}(5,9)(5,2)
\rput{0}(5,10){\psellipse[linestyle=none,fillstyle=solid](0,0)(1,-1)}
\psline{<-}(5,18)(5,11)
\rput{90}(42.36,4.65){\psellipse[linestyle=none,fillstyle=solid](0,0)(0.81,-0.75)}
\psbezier{->}(35.57,11.22)(35.57,8)(38.55,4.78)(41.54,4.78)
\psbezier{->}(49,11.22)(49,8)(46.02,4.78)(43.03,4.78)
\rput{90}(63.36,4.65){\psellipse[linestyle=none,fillstyle=solid](0,0)(0.81,-0.74)}
\psbezier{<-}(56.56,11.22)(56.56,8)(59.55,4.78)(62.54,4.78)
\psbezier{<-}(70,11.22)(70,8)(67.01,4.78)(64.03,4.78)
\rput{90}(83.36,11.24){\psellipse[linestyle=none,fillstyle=solid](0,0)(0.76,0.75)}
\psbezier{->}(76.57,5)(76.57,8.05)(79.55,11.11)(82.54,11.11)
\psbezier{->}(90,5)(90,8.05)(87.02,11.11)(84.03,11.11)
\rput{90}(104.79,11.24){\psellipse[linestyle=none,fillstyle=solid](0,0)(0.76,0.74)}
\psbezier{<-}(98,5)(98,8.05)(100.99,11.11)(103.98,11.11)
\psbezier{<-}(111.44,5)(111.44,8.05)(108.45,11.11)(105.47,11.11)
\end{pspicture}
}  

the presence of the dot is essential.

\begin{theorem}[oriented spider]
Let $A$ and $A^*$ be objects in a $\dagger$-compact category with bases.  Then, in the graphical representation, each `connected' diagram $\Xi$ obtained  from $\dagger$-SMC structure and the $\dagger$-dual Frobenius structure both on  $A$ and $A^*$  is equal to a `spider with directed legs', of which the inputs/outputs have the same orientation as the inputs/outputs of $\Xi$.  
\end{theorem}

Hence, every such  `connected' diagram only depends on its number of inputs, it's number of outputs, and the directions of the arrows at these inputs and outputs. 

\section{Protocols}

\subsection{The quantum teleportation protocol}\label{sec:telep} 

The quantum teleportation protocol \cite{BBC+93a} involves three qubits $a,b,c$ and two parties Alice and Bob. At the start of the protocol: 
\begin{itemize}
\item[-] The pair $(a,b)$ is in state $\ket {00}+\ket {11}$, and qubit $a$  is in Alice's possession  while qubit $b$ is in Bob's possession.
\item[-] Qubit $c$  is in an unknown state $\ket \phi$ and in Alice's possession.
\end{itemize}
To realise the protocol, the following steps are taken:
\begin{enumerate}
\item Alice performs a Bell basis measurement on her pair of qubits $(c,a)$, 
\item Alice sends the classical outcome $x$ of this measurement to Bob, and,
\item Bob applies a particular -- on  $x$ depending -- unitary operation $U_x$. 
\end{enumerate}
As a result  qubit $b$ will now be the unknown state $\ket \phi$. 
The unitary transformation $U_x$ applied by Bob is one of the Pauli operators, and the Bell measurement applied by Alice is composed of four projectors $\{P_x\}_{x=0..3}$, such that for any $x$, 
\[
P_x = (U_x\otimes 1_\HH)\circ (\delta_{std}\circ \gamma_{std}^\dagger) \circ (\gamma_{std}\circ \delta_{std}^\dagger) \circ (U^\dagger_x\otimes 1_\HH)
\]
Thus, the protocol of teleportation can be described in the graphical language of $\dagger$-SMC with $\dagger$-Frobenius structures:\vspace{2mm}

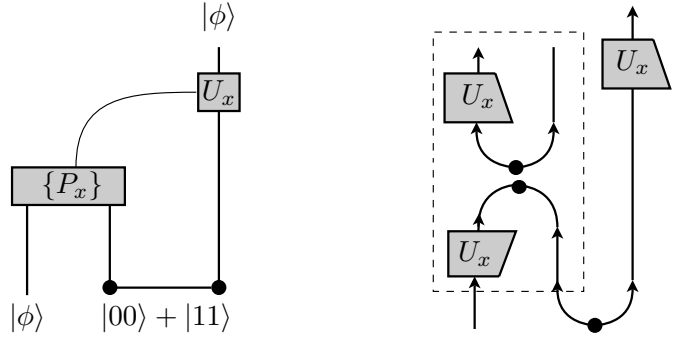
\centerline{
\ifx\JPicScale\undefined\def\JPicScale{1}\fi
\psset{unit=\JPicScale mm}
\psset{linewidth=0.3,dotsep=1,hatchwidth=0.3,hatchsep=1.5,shadowsize=1,dimen=middle}
\psset{dotsize=0.7 2.5,dotscale=1 1,fillcolor=black}
\psset{arrowsize=1 2,arrowlength=1,arrowinset=0.25,tbarsize=0.7 5,bracketlength=0.15,rbracketlength=0.15}
\begin{pspicture}(0,0)(91.5,44)
\psline(31.5,35)(31.5,38.5)
\psline(17,19)(17,6.5)
\psline(6,18)(6,5.5)
\newrgbcolor{userFillColour}{0.8 0.8 0.8}
\pspolygon[fillcolor=userFillColour,fillstyle=solid](4,17.5)(19,17.5)(19,22.5)(4,22.5)
\rput(12,20){$\{P_x\}$}
\rput(6,2.5){$|\phi\rangle$}
\rput{0}(17,6.5){\psellipse[linestyle=none,fillstyle=solid](0,0)(1,-1)}
\psline(17,6.5)(30.5,6.5)
\rput{0}(31.5,6.5){\psellipse[linestyle=none,fillstyle=solid](0,0)(1,-1)}
\psline(31.5,6.5)(31.5,30)
\newrgbcolor{userFillColour}{0.8 0.8 0.8}
\pspolygon[fillcolor=userFillColour,fillstyle=solid](28.75,35)(34.25,35)(34.25,30)(28.75,30)
\rput(31.5,32.5){$U_x$}
\rput(24.5,2.5){$|00\rangle+|11\rangle$}
\rput(31.5,42.5){$|\phi\rangle$}
\psbezier[linewidth=0.15](12.5,22.5)(12.5,32)(19.5,32.5)(28.5,32.5)
\newrgbcolor{userFillColour}{0.8 0.8 0.8}
\psline[fillcolor=userFillColour,fillstyle=solid](68.75,8)
(71,14)
(62,14)
(62,8)(68.75,8)
\rput{0}(71,22.5){\psellipse[linestyle=none,fillstyle=solid](0,0)(1,-1)}
\psbezier{<-}(76,28.5)(76,22.5)(71,22.5)(71,22.5)
\psbezier{<-}(65.75,28.5)(65.75,22.5)(71,22.5)(71,22.5)
\rput{0}(71.4,19.9){\psellipse[linestyle=none,fillstyle=solid](0,0)(1,0.93)}
\psbezier(76.5,14.53)(76.5,20.1)(71.25,20.1)(71.25,20.1)
\psbezier(66,14.63)(66,20.2)(71.25,20.2)(71.25,20.2)
\psline{<-}(65.5,8)(65.5,1)
\psline{<-}(76.5,14.6)(76.5,7)
\rput{0}(81.5,1.5){\psellipse[linestyle=none,fillstyle=solid](0,0)(1,-1)}
\psbezier{<-}(86.5,7.5)(86.5,1.5)(81.5,1.5)(81.5,1.5)
\psbezier{<-}(76.5,7.5)(76.5,1.5)(81.5,1.5)(81.5,1.5)
\psline(86.5,33)(86.5,7.5)
\newrgbcolor{userFillColour}{0.8 0.8 0.8}
\psline[fillcolor=userFillColour,fillstyle=solid](89.25,39.5)
(91.5,33)
(82.5,33)
(82.5,39.5)(89.25,39.5)
\psline(76,38.5)(76,28.5)
\psline{<-}(66,38.5)(66,32.56)
\psline{<-}(86.5,44)(86.5,39.5)
\newrgbcolor{userFillColour}{0.8 0.8 0.8}
\psline[fillcolor=userFillColour,fillstyle=solid](68.25,35)
(70.5,28.5)
(61.5,28.5)
(61.5,35)(68.25,35)
\pspolygon[linewidth=0.15,linestyle=dashed,dash=1 1](60,40.5)(80,40.5)(80,6)(60,6)
\rput(86.5,36.5){$U_x$}
\rput(66,32){$U_x$}
\rput(65.5,11){$U_x$}
\psline{->}(66,14)(66.1,16.4)
\end{pspicture}
}

\noindent where we used following diagrammatic notation:\vspace{2mm}

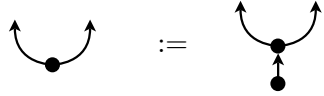
\centerline{
\ifx\JPicScale\undefined\def\JPicScale{1}\fi
\psset{unit=\JPicScale mm}
\psset{linewidth=0.3,dotsep=1,hatchwidth=0.3,hatchsep=1.5,shadowsize=1,dimen=middle}
\psset{dotsize=0.7 2.5,dotscale=1 1,fillcolor=black}
\psset{arrowsize=1 2,arrowlength=1,arrowinset=0.25,tbarsize=0.7 5,bracketlength=0.15,rbracketlength=0.15}
\begin{pspicture}(0,0)(40,15)
\rput{0}(5,6.5){\psellipse[linestyle=none,fillstyle=solid](0,0)(1,-1)}
\psbezier{<-}(10,12.5)(10,6.5)(5,6.5)(5,6.5)
\psbezier{<-}(0,12.5)(0,6.5)(5,6.5)(5,6.5)
\psline{->}(35,4)(35,8)
\rput{0}(35,9){\psellipse[linestyle=none,fillstyle=solid](0,0)(1,1)}
\psbezier{<-}(30,15)(30,9)(35,9)(35,9)
\psbezier{<-}(40,15)(40,9)(35,9)(35,9)
\rput{0}(35,4){\psellipse[linestyle=none,fillstyle=solid](0,0)(1,1)}
\rput(21,9){$:=$}
\end{pspicture}
}\vspace{-3mm}

\noindent
Note here in particular, as compared to the presentation in \cite{AC04}, that all arrows point in the direction of the actual physical flow of time.  It are the `dots' (= dualisers) which enable it.  Lemma \ref{lem:dagFSdagCS} provides a both intuitive and rigorous proof of correctness for teleportation:\vspace{2mm}

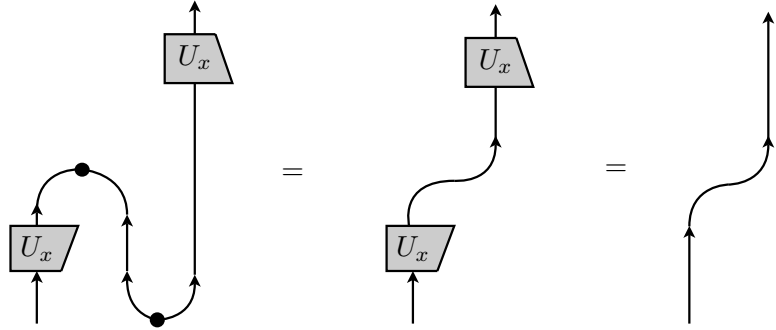
\centerline{
\ifx\JPicScale\undefined\def\JPicScale{1}\fi
\psset{unit=\JPicScale mm}
\psset{linewidth=0.3,dotsep=1,hatchwidth=0.3,hatchsep=1.5,shadowsize=1,dimen=middle}
\psset{dotsize=0.7 2.5,dotscale=1 1,fillcolor=black}
\psset{arrowsize=1 2,arrowlength=1,arrowinset=0.25,tbarsize=0.7 5,bracketlength=0.15,rbracketlength=0.15}
\begin{pspicture}(0,0)(103.25,45)
\newrgbcolor{userFillColour}{0.8 0.8 0.8}
\psline[fillcolor=userFillColour,fillstyle=solid](9.25,9)
(11.5,15)
(2.5,15)
(2.5,9)(9.25,9)
\rput{0}(12,22.5){\psellipse[linestyle=none,fillstyle=solid](0,0)(1,0.93)}
\psbezier(18,16.5)(18,22.07)(12,22.5)(12,22.5)
\psbezier(6,16.93)(6,22.5)(11.25,22.5)(11.25,22.5)
\psline{<-}(6,9)(6,2)
\psline{<-}(18,16.5)(18,9)
\rput{0}(22,2.5){\psellipse[linestyle=none,fillstyle=solid](0,0)(1,-1)}
\psbezier{<-}(27,8.5)(27,2.5)(22,2.5)(22,2.5)
\psbezier{<-}(18,9)(18,3)(22,2.5)(22,2.5)
\psline(27,34)(27,8.5)
\newrgbcolor{userFillColour}{0.8 0.8 0.8}
\psline[fillcolor=userFillColour,fillstyle=solid](29.75,40.5)
(32,34)
(23,34)
(23,40.5)(29.75,40.5)
\psline{<-}(27,45)(27,40.5)
\rput(27,37.5){$U_x$}
\rput(6,12){$U_x$}
\newrgbcolor{userFillColour}{0.8 0.8 0.8}
\psline[fillcolor=userFillColour,fillstyle=solid](59.25,9)
(61.5,15)
(52.5,15)
(52.5,9)(59.25,9)
\psbezier{<-}(67,27)(67,20.5)(61.5,21)(61.5,21)
\psbezier(61.5,21)(54,21)(55.5,15)(55.5,15)
\psline{<-}(56,9)(56,2)
\psline(67,33.5)(67,26.5)
\newrgbcolor{userFillColour}{0.8 0.8 0.8}
\psline[fillcolor=userFillColour,fillstyle=solid](69.75,40)
(72,33.5)
(63,33.5)
(63,40)(69.75,40)
\psline{<-}(67,44.5)(67,40)
\rput(67,37){$U_x$}
\rput(56,12){$U_x$}
\rput(40,22){$=$}
\psbezier{<-}(103.25,27)(103.25,20.5)(98,20.5)(98,20.5)
\psbezier(92.75,14.93)(92.75,20.5)(98,20.5)(98,20.5)
\psline{<-}(92.75,15)(92.75,2)
\psline{<-}(103.25,43.5)(103.25,26.5)
\rput(83,22.5){$=$}
\psline{->}(6,15)(6,18)
\end{pspicture}
}\vspace{-1.5mm}

This diagrammatic proof very much resembles `yanking a rope' as it was the case in the original diagrammatic proof in \cite{AC04} for correctness of the teleportation protocol, with as only difference the `annihilation' of the two dots involved (see \S\ref{InfoFlow} below).  Formally however, our proof is not based on $\dagger$-compact structures but on $\dagger$-Frobenius structures. This presentation enables easy comparison of the teleportation protocol with the following one.

\subsection{The quantum state transfer protocol}\label{sec:st}

The state transfer involves only two qubits $(a,b)$. At the start of the protocol:
\begin{itemize}
\item[-] Qubit $a$ is an unknown state $\ket \phi$, and, 
\item[-] Qubit  $b$ is in state $\ket 0 + \ket 1$. 
\end{itemize}
\noindent To realise the protocol, the following steps are taken:
\begin{enumerate}
\item Qubits $a$ and $b$ are measured according to the parity measurement, 
\item Qubit $a$ is measured in the diagonal basis, and, 
\item A unitary operation $U_{xy}$ is applied on qubit $b$, depending on the classical outcomes $x,y$ of the previous two measurements. 
\end{enumerate}
As a result  qubit $b$ will now be the unknown state $\ket \phi$. A parity measurement is a partial measurement. The state of the measured $2$-qubit system is not projected on a vector, but on a plane: either on the \emph{even plane} spanned by  $\ket{00}$ and $\ket{11}$ or on the \emph{odd plane} spanned by $\ket{01}$ and $ \ket{10}$. The projectors are, for $x\in \{0,1\}$, 
$$\pi_x = (1_\HH\otimes f_x)\circ \delta_{std}\circ \delta_{std}^\dagger \circ (1_\HH\otimes f_x^\dagger)$$ where $f_x$ is a \emph{permutation}. A diagonal basis measurement is a 1-qubit measurement described by the following projectors, for $y\in \{0,1\}$ and $g_y$ unitary \emph{phase maps}, 
\[
P'_y = g_y\circ \gamma^\dagger_{std}\circ \gamma_{std} \circ g_y^\dagger\,.
\] 
In the graphical language of $\dagger$-SMC with $\dagger$-Frobenius structures we obtain: \vspace{2mm}

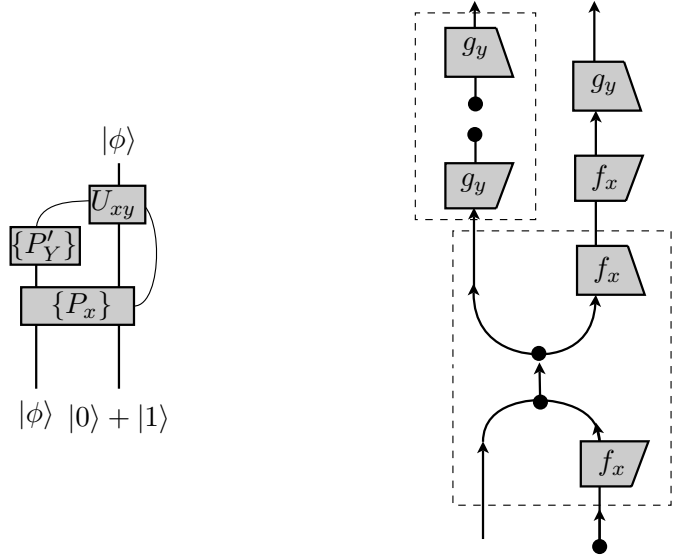
\centerline{
\ifx\JPicScale\undefined\def\JPicScale{1}\fi
\psset{unit=\JPicScale mm}
\psset{linewidth=0.3,dotsep=1,hatchwidth=0.3,hatchsep=1.5,shadowsize=1,dimen=middle}
\psset{dotsize=0.7 2.5,dotscale=1 1,fillcolor=black}
\psset{arrowsize=1 2,arrowlength=1,arrowinset=0.25,tbarsize=0.7 5,bracketlength=0.15,rbracketlength=0.15}
\begin{pspicture}(0,0)(92,72.6)
\psbezier[linewidth=0.15](20.6,32)(24.6,32)(24.6,46)(20.6,46)
\psline{<-}(65.9,72.6)(65.9,63.8)
\psline{<-}(81.8,72.6)(81.8,62)
\psline{<-}(82,58)(82,49.3)
\psline(82,48.8)(82,38.2)
\psline(18.6,51)(18.6,46)
\psbezier[linewidth=0.15](7.6,42)(7.6,46)(12.6,46)(15.6,46)
\psline(7.6,38)(7.6,33)
\psline(18.6,43)(18.6,34)
\psline(18.6,31)(18.6,21)
\psline(7.6,30.29)(7.6,21)
\newrgbcolor{userFillColour}{0.8 0.8 0.8}
\pspolygon[fillcolor=userFillColour,fillstyle=solid](5.6,29.5)(20.6,29.5)(20.6,34.5)(5.6,34.5)
\rput(13.6,32){$\{P_x\}$}
\rput(7.6,17.5){$|\phi\rangle$}
\newrgbcolor{userFillColour}{0.8 0.8 0.8}
\pspolygon[fillcolor=userFillColour,fillstyle=solid](14.7,48)(22.1,48)(22.1,43)(14.7,43)
\rput(17.85,45.5){$U_{xy}$}
\rput(18.6,17){$|0\rangle+|1\rangle$}
\rput(18.6,54){$|\phi\rangle$}
\psbezier{<-}(82,33.5)(81.5,25)(74.07,25.57)(74.07,25.57)
\psbezier{<-}(65.75,35)(65.75,25.57)(74.07,25.57)(74.07,25.57)
\psbezier(82.5,13.93)(82.5,19.5)(74.39,19.5)(74.39,19.5)
\psbezier(67,13.93)(67,19.5)(74.39,19.5)(74.39,19.5)
\psline{<-}(67,14)(67,1)
\psline(82.5,14)(82.5,0)
\psline(66,54)(66,49.56)
\pspolygon[linewidth=0.15,linestyle=dashed,dash=1 1](58,71)(74,71)(74,43.5)(58,43.5)
\newrgbcolor{userFillColour}{0.8 0.8 0.8}
\pspolygon[fillcolor=userFillColour,fillstyle=solid](4.2,42.5)(13.4,42.5)(13.4,37.5)(4.2,37.5)
\rput(8.6,40){$\{P'_Y\}$}
\psline(66,64.94)(66,59)
\newrgbcolor{userFillColour}{0.8 0.8 0.8}
\psline[fillcolor=userFillColour,fillstyle=solid](68.75,69)
(71,62.5)
(62,62.5)
(62,69)(68.75,69)
\newrgbcolor{userFillColour}{0.8 0.8 0.8}
\psline[fillcolor=userFillColour,fillstyle=solid](86.25,40)
(88.5,33.5)
(79.5,33.5)
(79.5,40)(86.25,40)
\newrgbcolor{userFillColour}{0.8 0.8 0.8}
\psline[fillcolor=userFillColour,fillstyle=solid](86,46)
(88.25,52)
(79.25,52)
(79.25,46)(86,46)
\newrgbcolor{userFillColour}{0.8 0.8 0.8}
\psline[fillcolor=userFillColour,fillstyle=solid](85.75,64.5)
(88,58)
(79,58)
(79,64.5)(85.75,64.5)
\pspolygon[linewidth=0.15,linestyle=dashed,dash=1 1](63,42)(92,42)(92,5.5)(63,5.5)
\newrgbcolor{userFillColour}{0.8 0.8 0.8}
\psline[fillcolor=userFillColour,fillstyle=solid](86.75,8)
(89,14)
(80,14)
(80,8)(86.75,8)
\rput{0}(74.6,19.2){\psellipse[linestyle=none,fillstyle=solid](0,0)(1,-1)}
\rput{0}(74.4,25.7){\psellipse[linestyle=none,fillstyle=solid](0,0)(1,-1)}
\psline{<-}(74.5,24.5)(74.6,19.2)
\psline{<-}(65.8,45.1)(65.8,34.5)
\newrgbcolor{userFillColour}{0.8 0.8 0.8}
\psline[fillcolor=userFillColour,fillstyle=solid](68.75,45)
(71,51)
(62,51)
(62,45)(68.75,45)
\rput{0}(65.9,54.8){\psellipse[linestyle=none,fillstyle=solid](0,0)(1,-1)}
\rput{0}(66,58.9){\psellipse[linestyle=none,fillstyle=solid](0,0)(1,-1)}
\rput(83.5,36.8){$f_x$}
\rput(83.2,49.1){$f_x$}
\rput(84.1,11.1){$f_x$}
\rput(83.1,61.4){$g_y$}
\rput(66,66.3){$g_y$}
\rput(65.8,48.2){$g_y$}
\rput{0}(82.5,0){\psellipse[linestyle=none,fillstyle=solid](0,0)(1,-1)}
\psline{<-}(82.5,5)(82.5,0)
\psline{->}(82.5,14)(82,16.5)
\end{pspicture}
}\vspace{1mm}

\noindent
The $\dagger$-Frobenius structure and the properties of the permutations and phase morphisms provide a diagrammatic proof of the state transfer:\vspace{2mm}

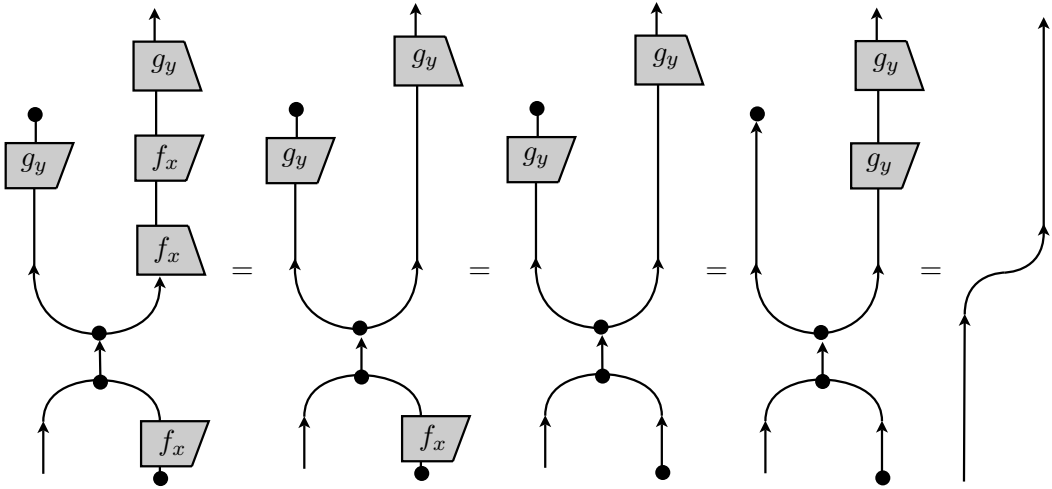
\centerline{
\ifx\JPicScale\undefined\def\JPicScale{1}\fi
\psset{unit=\JPicScale mm}
\psset{linewidth=0.3,dotsep=1,hatchwidth=0.3,hatchsep=1.5,shadowsize=1,dimen=middle}
\psset{dotsize=0.7 2.5,dotscale=1 1,fillcolor=black}
\psset{arrowsize=1 2,arrowlength=1,arrowinset=0.25,tbarsize=0.7 5,bracketlength=0.15,rbracketlength=0.15}
\begin{pspicture}(0,0)(139.5,64.5)
\psline{<-}(21.3,63.7)(21.3,56.73)
\psline(21.5,54.63)(21.5,44.03)
\psline(21.5,43.53)(21.5,32.93)
\psbezier{<-}(22,28)(22,20)(13.57,20.3)(13.57,20.3)
\psbezier{<-}(5.25,29.73)(5.25,20.3)(13.57,20.3)(13.57,20.3)
\psbezier(22,8.66)(22,14.23)(13.89,14.23)(13.89,14.23)
\psbezier(6.5,8.66)(6.5,14.23)(13.89,14.23)(13.89,14.23)
\psline{<-}(6.5,8.73)(6.5,1.73)
\psline(22,8.73)(22,1.73)
\psline(5.5,48.73)(5.5,44.29)
\newrgbcolor{userFillColour}{0.8 0.8 0.8}
\psline[fillcolor=userFillColour,fillstyle=solid](25.75,34.73)
(28,28.23)
(19,28.23)
(19,34.73)(25.75,34.73)
\newrgbcolor{userFillColour}{0.8 0.8 0.8}
\psline[fillcolor=userFillColour,fillstyle=solid](25.5,40.73)
(27.75,46.73)
(18.75,46.73)
(18.75,40.73)(25.5,40.73)
\newrgbcolor{userFillColour}{0.8 0.8 0.8}
\psline[fillcolor=userFillColour,fillstyle=solid](25.25,59.23)
(27.5,52.73)
(18.5,52.73)
(18.5,59.23)(25.25,59.23)
\newrgbcolor{userFillColour}{0.8 0.8 0.8}
\psline[fillcolor=userFillColour,fillstyle=solid](26.25,2.73)
(28.5,8.73)
(19.5,8.73)
(19.5,2.73)(26.25,2.73)
\rput{0}(14.1,13.93){\psellipse[linestyle=none,fillstyle=solid](0,0)(1,-1)}
\rput{0}(13.9,20.43){\psellipse[linestyle=none,fillstyle=solid](0,0)(1,-1)}
\psline{<-}(14,19.5)(14.1,13.93)
\psline(5.3,39.83)(5.3,29.23)
\newrgbcolor{userFillColour}{0.8 0.8 0.8}
\psline[fillcolor=userFillColour,fillstyle=solid](8.25,39.73)
(10.5,45.73)
(1.5,45.73)
(1.5,39.73)(8.25,39.73)
\rput{0}(5.4,49.53){\psellipse[linestyle=none,fillstyle=solid](0,0)(1,-1)}
\rput(23,31.53){$f_x$}
\rput(22.7,43.83){$f_x$}
\rput(23.6,5.83){$f_x$}
\rput(22.6,56.13){$g_y$}
\rput(5.3,42.93){$g_y$}
\rput(33,28.5){$=$}
\rput{0}(22.1,1.1){\psellipse[linestyle=none,fillstyle=solid](0,0)(1,-1)}
\psline{<-}(56,64.37)(56,57.4)
\psline(56.2,55.3)(56.2,29.9)
\psbezier{<-}(56.2,30.4)(56.2,20.97)(48.27,20.97)(48.27,20.97)
\psbezier{<-}(39.95,30.4)(39.95,20.97)(48.27,20.97)(48.27,20.97)
\psbezier(56.7,9.33)(56.7,14.9)(48.59,14.9)(48.59,14.9)
\psbezier(41.2,9.33)(41.2,14.9)(48.59,14.9)(48.59,14.9)
\psline{<-}(41.2,9.4)(41.2,2.4)
\psline(56.7,9.4)(56.7,2.4)
\psline(40.2,49.4)(40.2,44.96)
\newrgbcolor{userFillColour}{0.8 0.8 0.8}
\psline[fillcolor=userFillColour,fillstyle=solid](59.95,59.9)
(62.2,53.4)
(53.2,53.4)
(53.2,59.9)(59.95,59.9)
\newrgbcolor{userFillColour}{0.8 0.8 0.8}
\psline[fillcolor=userFillColour,fillstyle=solid](60.95,3.4)
(63.2,9.4)
(54.2,9.4)
(54.2,3.4)(60.95,3.4)
\rput{0}(48.8,14.6){\psellipse[linestyle=none,fillstyle=solid](0,0)(1,-1)}
\rput{0}(48.6,21.1){\psellipse[linestyle=none,fillstyle=solid](0,0)(1,-1)}
\psline{<-}(48.8,19.9)(48.8,14.6)
\psline(40,40.5)(40,29.9)
\newrgbcolor{userFillColour}{0.8 0.8 0.8}
\psline[fillcolor=userFillColour,fillstyle=solid](42.95,40.4)
(45.2,46.4)
(36.2,46.4)
(36.2,40.4)(42.95,40.4)
\rput{0}(40.1,50.2){\psellipse[linestyle=none,fillstyle=solid](0,0)(1,-1)}
\rput(58.3,6.5){$f_x$}
\rput(57.3,56.8){$g_y$}
\rput(40,43.6){$g_y$}
\rput{0}(56.8,1.77){\psellipse[linestyle=none,fillstyle=solid](0,0)(1,-1)}
\rput(64.5,28.5){$=$}
\psline{<-}(88.03,64.5)(88.03,57.53)
\psline(88.23,55.43)(88.23,30.03)
\psbezier{<-}(88.23,30.53)(88.23,21.1)(80.3,21.1)(80.3,21.1)
\psbezier{<-}(71.98,30.53)(71.98,21.1)(80.3,21.1)(80.3,21.1)
\psbezier(88.73,9.46)(88.73,15.03)(80.61,15.03)(80.61,15.03)
\psbezier(73.23,9.46)(73.23,15.03)(80.61,15.03)(80.61,15.03)
\psline{<-}(73.23,9.53)(73.23,2.53)
\psline{<-}(88.73,9.53)(88.73,2.53)
\psline(72.23,49.53)(72.23,45.09)
\newrgbcolor{userFillColour}{0.8 0.8 0.8}
\psline[fillcolor=userFillColour,fillstyle=solid](91.98,60.03)
(94.23,53.53)
(85.23,53.53)
(85.23,60.03)(91.98,60.03)
\rput{0}(80.83,14.73){\psellipse[linestyle=none,fillstyle=solid](0,0)(1,-1)}
\rput{0}(80.63,21.23){\psellipse[linestyle=none,fillstyle=solid](0,0)(1,-1)}
\psline{<-}(80.8,20.2)(80.83,14.73)
\psline(72,40.6)(72,30)
\newrgbcolor{userFillColour}{0.8 0.8 0.8}
\psline[fillcolor=userFillColour,fillstyle=solid](74.98,40.53)
(77.23,46.53)
(68.23,46.53)
(68.23,40.53)(74.98,40.53)
\rput{0}(72.13,50.33){\psellipse[linestyle=none,fillstyle=solid](0,0)(1,-1)}
\rput(89.33,56.93){$g_y$}
\rput(72.03,43.73){$g_y$}
\rput{0}(88.83,1.9){\psellipse[linestyle=none,fillstyle=solid](0,0)(1,-1)}
\rput(96,28.5){$=$}
\psline{<-}(117.3,63.77)(117.3,56.8)
\psline(117.5,54.7)(117.5,29.3)
\psbezier{<-}(117.5,29.8)(117.5,20.37)(109.57,20.37)(109.57,20.37)
\psbezier{<-}(101.25,29.8)(101.25,20.37)(109.57,20.37)(109.57,20.37)
\psbezier(118,8.73)(118,14.3)(109.89,14.3)(109.89,14.3)
\psbezier(102.5,8.73)(102.5,14.3)(109.89,14.3)(109.89,14.3)
\psline{<-}(102.5,8.8)(102.5,1.8)
\psline{<-}(118,8.8)(118,1.8)
\newrgbcolor{userFillColour}{0.8 0.8 0.8}
\psline[fillcolor=userFillColour,fillstyle=solid](121.25,59.3)
(123.5,52.8)
(114.5,52.8)
(114.5,59.3)(121.25,59.3)
\rput{0}(110.1,14){\psellipse[linestyle=none,fillstyle=solid](0,0)(1,-1)}
\rput{0}(109.9,20.5){\psellipse[linestyle=none,fillstyle=solid](0,0)(1,-1)}
\psline{<-}(110.1,19.3)(110.1,14)
\psline{<-}(101.3,48.6)(101.3,29.3)
\rput{0}(101.4,49.6){\psellipse[linestyle=none,fillstyle=solid](0,0)(1,-1)}
\rput(118.6,56.2){$g_y$}
\rput{0}(118.1,1.17){\psellipse[linestyle=none,fillstyle=solid](0,0)(1,-1)}
\newrgbcolor{userFillColour}{0.8 0.8 0.8}
\psline[fillcolor=userFillColour,fillstyle=solid](120.65,39.7)
(122.9,45.7)
(113.9,45.7)
(113.9,39.7)(120.65,39.7)
\rput(117.7,42.9){$g_y$}
\rput(124.5,28.5){$=$}
\psbezier{<-}(139.5,35)(139.5,28.5)(134.25,28.5)(134.25,28.5)
\psbezier(129,22.93)(129,28.5)(134.25,28.5)(134.25,28.5)
\psline{<-}(129,23)(128.9,0.7)
\psline{<-}(139.5,62.5)(139.5,34.5)
\end{pspicture}
}

\subsection{Unifying state transfer and teleportation}

In sections \ref{sec:telep} and \ref{sec:st}, the graphical calculus for $\dagger$-SMC with $\dagger$-Frobenius structures has been used to give a diagrammatic representation and proof of both teleportation and state transfer.  State transfer was initially introduced to optimise the resources of measurement-only quantum computation \cite{Per05a}: while teleportation requires three qubits state transfer requires only two qubits.  
The diagrammatic representation of these two protocols leads to a better understanding of the foundational structures of measurement-only quantum computation: one can transform teleportation into state transfer and vice versa just by applying the Frobenius equation as it is illustrated in the following diagram\vspace{2mm}

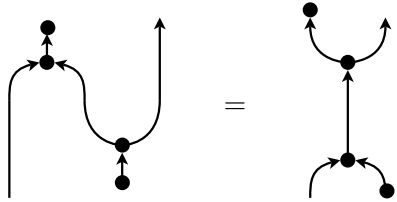
\centerline{
\ifx\JPicScale\undefined\def\JPicScale{1}\fi
\psset{unit=\JPicScale mm}
\psset{linewidth=0.3,dotsep=1,hatchwidth=0.3,hatchsep=1.5,shadowsize=1,dimen=middle}
\psset{dotsize=0.7 2.5,dotscale=1 1,fillcolor=black}
\psset{arrowsize=1 2,arrowlength=1,arrowinset=0.25,tbarsize=0.7 5,bracketlength=0.15,rbracketlength=0.15}
\begin{pspicture}(0,0)(56.2,28)
\psline{->}(50,7)(50,19)
\rput{0}(50,20){\psellipse[linestyle=none,fillstyle=solid](0,0)(1,-1)}
\psbezier{<-}(55,26)(55,20)(50,20)(50,20)
\psbezier{<-}(45,26)(45,20)(50,20)(50,20)
\rput{0}(50,7){\psellipse[linestyle=none,fillstyle=solid](0,0)(1,-1)}
\psbezier{<-}(51,7)(55,6)(55,4)(55,2)
\psbezier{<-}(49,7)(45,6)(45,4)(45,2)
\psline{->}(20,4)(20,8)
\rput{0}(20,9){\psellipse[linestyle=none,fillstyle=solid](0,0)(1,1)}
\psbezier(15,15)(15,9)(20,9)(20,9)
\psbezier(25,15)(25,9)(20,9)(20,9)
\psline{->}(10,20.43)(10,23.8)
\rput{0}(10,20){\psellipse[linestyle=none,fillstyle=solid](0,0)(1,1)}
\psbezier{<-}(9,20)(5,19)(5,17)(5,15)
\psbezier{<-}(11,20)(15,19)(15,17)(15,15)
\psline{->}(25,15)(25,26)
\psline(5,2)(5,15)
\rput(35,14){$=$}
\rput{0}(10.1,24.6){\psellipse[linestyle=none,fillstyle=solid](0,0)(1,1)}
\rput{0}(20,4){\psellipse[linestyle=none,fillstyle=solid](0,0)(1,1)}
\rput{0}(45,27){\psellipse[linestyle=none,fillstyle=solid](0,0)(1,1)}
\rput{0}(55.2,2.9){\psellipse[linestyle=none,fillstyle=solid](0,0)(1,1)}
\end{pspicture}
}

\noindent
where, for reasons of clarity, all unitary transformations are taken to be identity.  

While state transfer requires less ancillary qubits than teleportation, it needs more \em structural resources \em in the sense that the unitary transformations $f_x$ and $g_y$ have to be permutations and phase morphisms respectively.  Also, while for teleportation we could rely on compact structure only, as in \cite{AC04}, for state transfer, even in post-selected form, the use of  Frobenius structures (= a \em base\em) is \em essential\em.

\subsection{Flow of information}\label{InfoFlow}

Our diagrammatic proofs of teleportation and state transfer rely on $\dagger$-Frobenius structures in $\dagger$-SMCs rather than on $\dagger$-compact structure.  However, compactness allows both Joyal, Street and Verity's construction of the trace \cite{JSV} as well as Selinger's CPM construction \cite{Sel05}. 
One can use $\dagger$-compact categories with bases in order to take advantage of both axiomatisations.  
In such a categorical framework the diagrammatic proofs of teleportation due to  Abramsky and Coecke based on compact structure can be converted into the one presented in this paper using basis structures, and vice versa. 

\begin{lemma}
In a $\dagger$-compact category with bases, for any object $A$ we have
\[
(d^\dagger_A\otimes 1_A)\circ \epsilon_{A^*}^\dagger = \delta_A \circ \gamma_A^\dagger = (1_A\otimes d^\dagger_A)\circ \epsilon_{A}^\dagger
\]
that is, graphically,\vspace{5mm}

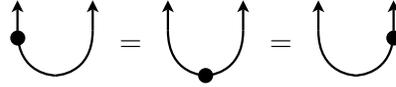
\centerline{
\ifx\JPicScale\undefined\def\JPicScale{1}\fi
\psset{unit=\JPicScale mm}
\psset{linewidth=0.3,dotsep=1,hatchwidth=0.3,hatchsep=1.5,shadowsize=1,dimen=middle}
\psset{dotsize=0.7 2.5,dotscale=1 1,fillcolor=black}
\psset{arrowsize=1 2,arrowlength=1,arrowinset=0.25,tbarsize=0.7 5,bracketlength=0.15,rbracketlength=0.15}
\begin{pspicture}(0,0)(61,20)
\rput{0}(35,10){\psellipse[linestyle=none,fillstyle=solid](0,0)(1,-1)}
\psbezier(40,16)(40,10)(35,10)(35,10)
\psbezier(30,16)(30,10)(35,10)(35,10)
\psline{->}(40,16)(40,20)
\psline{->}(30,16)(30,20)
\rput{0}(10,15){\psellipse[linestyle=none,fillstyle=solid](0,0)(1,-1)}
\psbezier(20,16)(20,10)(15,10)(15,10)
\psbezier(10,16)(10,10)(15,10)(15,10)
\psline{->}(20,16)(20,20)
\psline{->}(10,16)(10,20)
\rput{0}(60,15){\psellipse[linestyle=none,fillstyle=solid](0,0)(1,-1)}
\psbezier(60,16)(60,10)(55,10)(55,10)
\psbezier(50,16)(50,10)(55,10)(55,10)
\psline{->}(60,16)(60,20)
\psline{->}(50,16)(50,20)
\rput(25,14){$=$}
\rput(45,14){$=$}
\end{pspicture}
}\vspace{-8mm}
\end{lemma}

Note that while equality of the left and the right picture reflects a fact derivable in compact categories for arbitrary morphisms -- involving the transposed $(-)^*$ -- the picture in the middle can only be given meaning for basis structure. We have:\vspace{3mm}

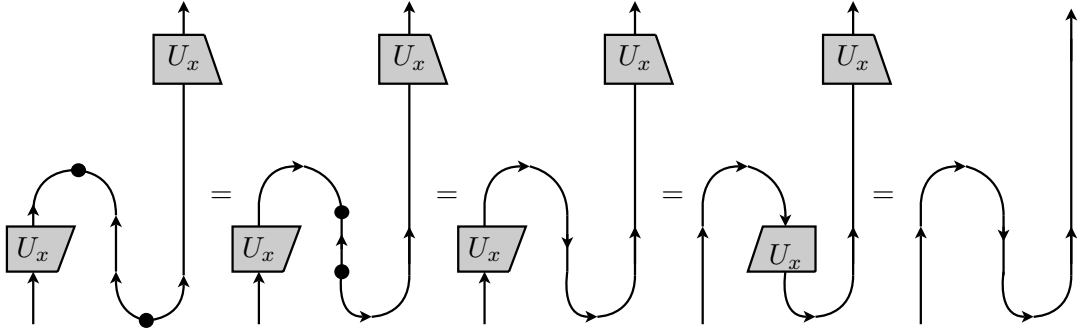
\centerline{
\ifx\JPicScale\undefined\def\JPicScale{1}\fi
\psset{unit=\JPicScale mm}
\psset{linewidth=0.3,dotsep=1,hatchwidth=0.3,hatchsep=1.5,shadowsize=1,dimen=middle}
\psset{dotsize=0.7 2.5,dotscale=1 1,fillcolor=black}
\psset{arrowsize=1 2,arrowlength=1,arrowinset=0.25,tbarsize=0.7 5,bracketlength=0.15,rbracketlength=0.15}
\begin{pspicture}(0,0)(141,43)
\newrgbcolor{userFillColour}{0.8 0.8 0.8}
\psline[fillcolor=userFillColour,fillstyle=solid](6.25,7)
(8.5,13)
(-0.5,13)
(-0.5,7)(6.25,7)
\rput{0}(9,20.5){\psellipse[linestyle=none,fillstyle=solid](0,0)(1,0.93)}
\psbezier(14,14.5)(14,21)(8,20.5)(8,20.5)
\psbezier(3,14.93)(3,20.5)(8.25,20.5)(8.25,20.5)
\psline{<-}(3,7)(3,0)
\psline{<-}(14,14.5)(14,7)
\rput{0}(18,0.5){\psellipse[linestyle=none,fillstyle=solid](0,0)(1,-1)}
\psbezier{<-}(23,6.5)(23,0.5)(18,0.5)(18,0.5)
\psbezier{<-}(14,7)(14,1)(18,0.5)(18,0.5)
\psline(23,32)(23,6.5)
\newrgbcolor{userFillColour}{0.8 0.8 0.8}
\psline[fillcolor=userFillColour,fillstyle=solid](25.75,38.5)
(28,32)
(19,32)
(19,38.5)(25.75,38.5)
\psline{<-}(23,43)(23,38.5)
\rput(23,35.5){$U_x$}
\rput(3,10){$U_x$}
\rput(58,17){$=$}
\psline{->}(3,13)(3,16)
\rput(28,17){$=$}
\newrgbcolor{userFillColour}{0.8 0.8 0.8}
\psline[fillcolor=userFillColour,fillstyle=solid](36.25,7)
(38.5,13)
(29.5,13)
(29.5,7)(36.25,7)
\rput{0}(44,14.93){\psellipse[linestyle=none,fillstyle=solid](0,0)(1,0.93)}
\psbezier(44,14.5)(44,20.07)(39,21)(39,21)
\psbezier{<-}(39,21)(33,21)(33,16)(33,16)
\psline{<-}(33,7)(33,0)
\psline{<-}(44,12)(44,7)
\rput{0}(44,7){\psellipse[linestyle=none,fillstyle=solid](0,0)(1,-1)}
\psbezier(53,6.5)(53,1)(48,1)(48,1)
\psbezier{<-}(48,1)(43,1)(44,7)(44,7)
\psline(53,32)(53,6.5)
\newrgbcolor{userFillColour}{0.8 0.8 0.8}
\psline[fillcolor=userFillColour,fillstyle=solid](55.75,38.5)
(58,32)
(49,32)
(49,38.5)(55.75,38.5)
\psline{<-}(53,43)(53,38.5)
\rput(53,35.5){$U_x$}
\rput(33,10){$U_x$}
\psline(33,13)(33,16)
\psline(44,14.5)(44,10)
\psline{<-}(53,13)(53,8)
\rput(88,17){$=$}
\newrgbcolor{userFillColour}{0.8 0.8 0.8}
\psline[fillcolor=userFillColour,fillstyle=solid](66.25,7)
(68.5,13)
(59.5,13)
(59.5,7)(66.25,7)
\psbezier(74,14.5)(74,20.07)(69,21)(69,21)
\psbezier{<-}(69,21)(63,21)(63,16)(63,16)
\psline{<-}(63,7)(63,0)
\psline(74,12)(74,7)
\psbezier(83,6.5)(83,1)(78,1)(78,1)
\psbezier{<-}(78,1)(73,1)(74,7)(74,7)
\psline(83,32)(83,6.5)
\newrgbcolor{userFillColour}{0.8 0.8 0.8}
\psline[fillcolor=userFillColour,fillstyle=solid](85.75,38.5)
(88,32)
(79,32)
(79,38.5)(85.75,38.5)
\psline{<-}(83,43)(83,38.5)
\rput(83,35.5){$U_x$}
\rput(63,10){$U_x$}
\psline(63,13)(63,16)
\psline{->}(74,14.5)(74,10)
\psline{<-}(83,13)(83,8)
\newrgbcolor{userFillColour}{0.8 0.8 0.8}
\psline[fillcolor=userFillColour,fillstyle=solid](107,7)
(107,13)
(100,13)
(98,7)(107,7)
\psbezier(103,14.43)(103,20)(98,21)(98,21)
\psbezier{<-}(98,21)(92,21)(92,16)(92,16)
\psline{<-}(92,13)(92,0)
\psbezier(112,6.5)(112,1)(107,1)(107,1)
\psbezier{<-}(107,1)(102,1)(103,7)(103,7)
\psline(112,32)(112,6.5)
\newrgbcolor{userFillColour}{0.8 0.8 0.8}
\psline[fillcolor=userFillColour,fillstyle=solid](114.75,38.5)
(117,32)
(108,32)
(108,38.5)(114.75,38.5)
\psline{<-}(112,43)(112,38.5)
\rput(112,35.5){$U_x$}
\rput(103,9){$U_x$}
\psline(92,13)(92,16)
\psline{->}(103,14.5)(103,13)
\psline{<-}(112,13)(112,8)
\rput(116,17){$=$}
\psbezier(132,14.5)(132,20.07)(127,21)(127,21)
\psbezier{<-}(127,21)(121,21)(121,16)(121,16)
\psline{<-}(121,13)(121,0)
\psbezier(141,6.5)(141,1)(136,1)(136,1)
\psbezier{<-}(136,1)(131,1)(132,7)(132,7)
\psline(141,38)(141,6.5)
\psline{<-}(141,42)(141,35)
\psline(121,13)(121,16)
\psline{->}(132,14.5)(132,11)
\psline{<-}(141,13)(141,8)
\psline(132,13)(132,6.5)
\end{pspicture}
}\vspace{2mm}

\noindent
Note in particular that in\vspace{2mm}

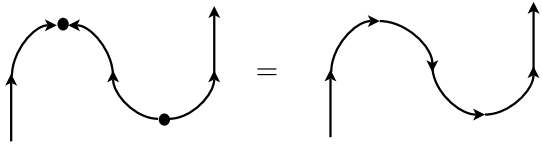
\centerline{
\ifx\JPicScale\undefined\def\JPicScale{1}\fi
\psset{unit=\JPicScale mm}
\psset{linewidth=0.3,dotsep=1,hatchwidth=0.3,hatchsep=1.5,shadowsize=1,dimen=middle}
\psset{dotsize=0.7 2.5,dotscale=1 1,fillcolor=black}
\psset{arrowsize=1 2,arrowlength=1,arrowinset=0.25,tbarsize=0.7 5,bracketlength=0.15,rbracketlength=0.15}
\begin{pspicture}(0,0)(72.45,19.56)
\rput{90}(9.89,16.59){\psellipse[linestyle=none,fillstyle=solid](0,0)(0.85,0.75)}
\psbezier{->}(3.1,9.65)(3.1,13.05)(6.09,16.44)(9.07,16.44)
\psbezier{->}(16.54,9.65)(16.54,13.05)(13.55,16.44)(10.56,16.44)
\rput{90}(23.34,3.86){\psellipse[linestyle=none,fillstyle=solid](0,0)(0.81,-0.75)}
\psbezier{<-}(16.55,10.44)(16.55,7.22)(19.54,4)(22.52,4)
\psbezier{<-}(29.99,10.44)(29.99,7.22)(27,4)(24.01,4)
\psline{->}(3,1)(3,10)
\psline{->}(30,10)(30,19)
\rput(37,10){$=$}
\psbezier{->}(45.55,10.21)(45.55,13.61)(48.77,17)(52,17)
\psbezier{<-}(58.99,10.21)(58.99,13.61)(55.49,17)(52,17)
\psbezier{->}(59,11)(59,7.78)(62.5,4.56)(66,4.56)
\psbezier{<-}(72.44,11)(72.44,7.78)(69.22,4.56)(66,4.56)
\psline{->}(45.45,1.56)(45.45,10.56)
\psline{->}(72.45,10.56)(72.45,19.56)
\end{pspicture}
}  

\noindent
the directions of the arrows in picture on the left can be interpreted as 
\begin{itemize}
\item \em the physical flow of information\em,  
\end{itemize}
since they respect causal ordering, while the directions of the arrows in picture on the right, which do not respect causal ordering, can be interpreted as 
\begin{itemize}
\item \em the logical flow of information\em.
\end{itemize}
This logical flow of information guides the unknown input state to where it will end up at the end of the protocol.  Having both compact structure and Frobenius structure available enables interchange between these two complementary views.

\section{Conclusion}

We introduced non-self-dual basis structures.  Previous accounts introduced abstract bases by factoring a given \em necessarily self-dual \em $\dagger$-compact structure in a \em copying-deleting pair\em:
\vskip 2mm
\centerline{
\ifx\JPicScale\undefined\def\JPicScale{1}\fi
\psset{unit=\JPicScale mm}
\psset{linewidth=0.3,dotsep=1,hatchwidth=0.3,hatchsep=1.5,shadowsize=1,dimen=middle}
\psset{dotsize=0.7 2.5,dotscale=1 1,fillcolor=black}
\psset{arrowsize=1 2,arrowlength=1,arrowinset=0.25,tbarsize=0.7 5,bracketlength=0.15,rbracketlength=0.15}
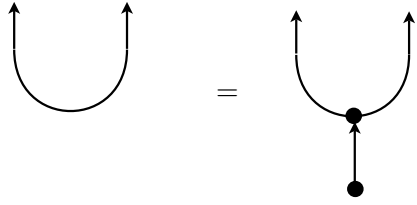
\begin{pspicture}(0,0)(53.12,26.25)
\psline{->}(15.62,20)(15.62,26.25)
\psline{->}(38.12,19.38)(38.12,25.14)
\psbezier(53.12,19.2)(53.12,8.3)(38.12,8.3)(38.12,19.2)
\psline{->}(53.12,19.24)(53.12,25)
\rput{0}(45.8,11.09){\psellipse[fillstyle=solid](0,0)(0.95,-0.94)}
\psline{<-}(45.9,10.3)(45.9,2.1)
\rput{0}(46,1.34){\psellipse[fillstyle=solid](0,0)(0.95,-0.93)}
\rput(29,13.9){$=$}
\psline{->}(0.62,20)(0.62,26.25)
\psbezier(15.62,19.9)(15.62,9)(0.62,9)(0.62,19.9)
\end{pspicture}
} 
\noindent
Our account introduced abstract bases by factoring an arbirtary (not necessarily self-dual) $\dagger$-compact structure in a  $\dagger$-compact structure in a \em copying-deleting-dualiser triple\em:
\vskip 2mm
\centerline{
\ifx\JPicScale\undefined\def\JPicScale{1}\fi
\psset{unit=\JPicScale mm}
\psset{linewidth=0.3,dotsep=1,hatchwidth=0.3,hatchsep=1.5,shadowsize=1,dimen=middle}
\psset{dotsize=0.7 2.5,dotscale=1 1,fillcolor=black}
\psset{arrowsize=1 2,arrowlength=1,arrowinset=0.25,tbarsize=0.7 5,bracketlength=0.15,rbracketlength=0.15}
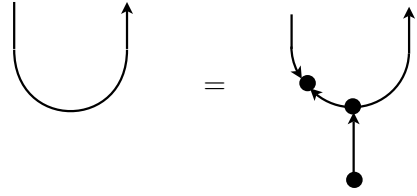
\begin{pspicture}(0,0)(53.12,27.4)
\psline{->}(15.62,20)(15.62,26.25)
\psline(37.5,20)(37.5,24.6)
\psline{->}(53.12,19.38)(53.12,25.62)
\rput{0}(45.65,12.36){\psellipse[fillstyle=solid](0,0)(0.95,-0.93)}
\psline{<-}(45.75,11.57)(45.75,3.36)
\rput{0}(45.85,2.6){\psellipse[fillstyle=solid](0,0)(0.95,-0.93)}
\rput(27.3,14.8){$=$}
\psline(0.62,20)(0.62,26.25)
\psbezier(15.62,19.9)(15.62,9)(0.62,9)(0.62,19.9)
\rput{0}(45.61,19.6){\psellipticarc[]{->}(0,0)(7.51,-7.35){1.68}{138.75}}
\rput{0}(39.65,15.44){\psellipse[fillstyle=solid](0,0)(0.95,-0.93)}
\rput{90}(44.35,20.3){\psellipticarc[]{<-}(0,0)(7.1,-6.85){-127.13}{-90}}
\end{pspicture}
} 
\noindent
In the light of the correspondence between copying-deleting pairs and orthonormal bases in ${\bf FdHilb}$ which was established in \cite{CPV} it are the latter which, given a $\dagger$-compact structure, produce all orthonormal bases as its factorisations.  Indeed, if, for example, we consider the $\dagger$-compact structure
\[
\epsilon:\mathbb{C}\to{\cal H}_2\otimes{\cal H}_2::1\mapsto |00\rangle+|11\rangle\,
\]
then its factorisations produce the $Z$-basis a the $X$-basis, respectively as
\[
\qquad\delta_Z:{\cal H}_2\to{\cal H}_2\otimes{\cal H}_2::\left\{\begin{array}{l}
|0\rangle\mapsto |00\rangle\\
|1\rangle\mapsto |11\rangle
\end{array}\right.
\qquad\qquad\gamma_Z^\dagger:\mathbb{C}\to{\cal H}_2::1\mapsto |0\rangle+|1\rangle
\]
\[
\qquad\delta_X::\left\{\begin{array}{l}
|0\rangle+|1\rangle\mapsto {1\over \sqrt{2}}(|0\rangle+|1\rangle)\otimes(|0\rangle+|1\rangle)\\
|0\rangle-|1\rangle\mapsto {1\over \sqrt{2}}(|0\rangle-|1\rangle)\otimes(|0\rangle-|1\rangle)
\end{array}\right.
\qquad\qquad\qquad\gamma_X^\dagger::1\mapsto \sqrt{2}|0\rangle\qquad\qquad
\]\\ \noindent
but not the $Y$-basis.  We verify this now explicitly. Consider
\[
\delta_Y:{\cal H}_2\to{\cal H}_2\otimes{\cal H}_2::\left\{\begin{array}{l}
|0\rangle+i|1\rangle\mapsto {1\over \sqrt{2}}(|0\rangle+i|1\rangle)\otimes(|0\rangle+i|1\rangle)\\
|0\rangle-i|1\rangle\mapsto {1\over \sqrt{2}}(|0\rangle-i|1\rangle)\otimes(|0\rangle-i|1\rangle)
\end{array}\right.
\]
which copies the $Y$-basis together with the uniform deleting operation
\[
\gamma_Y^\dagger::1\mapsto {1\over \sqrt{2}}(|0\rangle+i|1\rangle)+{1\over \sqrt{2}}(|0\rangle-i|1\rangle)=
\sqrt{2}|0\rangle\,,
\]
which is the only one that yields a $\dagger$-Frobenius structure together with $\delta_Y$.  
The induced $\dagger$-compact structure is now
\[
\epsilon':\mathbb{C}\to{\cal H}_2\otimes{\cal H}_2::1\mapsto |00\rangle-|11\rangle
\]
so not equal to $\epsilon$.\footnote{The general form of a $\dagger$-Frobenius structure representing  the $Y$-axis is obtained by setting  $\delta_Y'(|0\rangle+i|1\rangle)=\delta_Y(|0\rangle+i|1\rangle)$, $\delta_Y'(|0\rangle-i|1\rangle)=e^{i\theta}\cdot\delta_Y(|0\rangle-i|1\rangle)$, 
$\gamma_Y'(|0\rangle+i|1\rangle)=\gamma_Y(|0\rangle+i|1\rangle)$ and
$\gamma_Y'(|0\rangle-i|1\rangle)=e^{-i\theta}\cdot\gamma_Y(|0\rangle-i|1\rangle)$, for which one straightforwardly verifies that there is no $\theta$ such that $\delta_Y'\circ\gamma_Y'=\epsilon$.}  However, relative to the fixed $\dagger$-compact structure $\epsilon$ we can define a dualiser as
\vskip 2mm
\centerline{
\ifx\JPicScale\undefined\def\JPicScale{1}\fi
\psset{unit=\JPicScale mm}
\psset{linewidth=0.3,dotsep=1,hatchwidth=0.3,hatchsep=1.5,shadowsize=1,dimen=middle}
\psset{dotsize=0.7 2.5,dotscale=1 1,fillcolor=black}
\psset{arrowsize=1 2,arrowlength=1,arrowinset=0.25,tbarsize=0.7 5,bracketlength=0.15,rbracketlength=0.15}
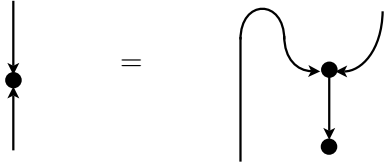
\begin{pspicture}(0,0)(51.2,21.7)
\psline{->}(2.1,1.5)(2.1,10)
\rput(17.8,12.9){$=$}
\psline{<-}(2.1,11.55)(2.1,21.4)
\rput{0}(2.1,10.82){\psellipse[fillstyle=solid](0,0)(0.95,-0.93)}
\psbezier(38.12,16.25)(38.12,21.7)(32.3,21.7)(32.3,16.25)
\psline{<-}(44.1,2.9)(44.1,11.4)
\rput{0}(44.1,12.22){\psellipse[fillstyle=solid](0,0)(0.95,-0.93)}
\rput{0}(44.05,1.96){\psellipse[fillstyle=solid](0,0)(0.95,-0.93)}
\psbezier{->}(38.12,16.7)(38.12,13.6)(39.88,12.1)(42.9,12)
\psbezier{->}(51.2,20)(51.1,14.7)(48.3,12)(44.9,12)
\psline(32.3,0)(32.3,16.6)
\end{pspicture}
} 
\vskip 2mm\noindent
that is, explicitly
\[
d_Y:{\cal H}_2\to{\cal H}_2::\left\{\begin{array}{l}
|0\rangle\mapsto |0\rangle\\
|1\rangle\mapsto -|1\rangle
\end{array}\right.\,,
\]
that is, the $Z$-gate.  Adjoining dualisers only requires a minor refinement of the spider-theorem: the spider's legs are now oriented.  And the dualiser allows to retain planarity in the graphical calculus by comprehending symmetry:
\vskip 2mm
\centerline{
\ifx\JPicScale\undefined\def\JPicScale{1}\fi
\psset{unit=\JPicScale mm}
\psset{linewidth=0.3,dotsep=1,hatchwidth=0.3,hatchsep=1.5,shadowsize=1,dimen=middle}
\psset{dotsize=0.7 2.5,dotscale=1 1,fillcolor=black}
\psset{arrowsize=1 2,arrowlength=1,arrowinset=0.25,tbarsize=0.7 5,bracketlength=0.15,rbracketlength=0.15}
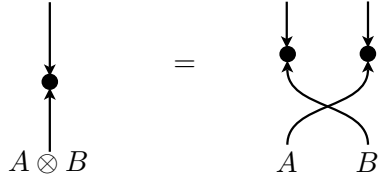
\begin{pspicture}(0,0)(52.65,23.3)
\psline{->}(9.4,3.3)(9.4,11.8)
\rput(27.3,14.8){$=$}
\psline{<-}(9.4,13.35)(9.4,23.2)
\rput{0}(9.4,12.61){\psellipse[fillstyle=solid](0,0)(0.95,-0.93)}
\rput(9.2,1.85){$A\otimes B$}
\psbezier{<-}(40.98,15.44)(40.98,8.9)(51.7,9.83)(51.7,4.24)
\psbezier{<-}(51.7,15.44)(51.7,8.9)(40.98,9.83)(40.98,4.24)
\rput{0}(51.7,16.33){\psellipse[fillstyle=solid](0,0)(0.95,-0.93)}
\rput{0}(41,16.27){\psellipse[fillstyle=solid](0,0)(0.95,-0.93)}
\psline{<-}(40.9,17.1)(40.9,23.27)
\psline{<-}(51.7,17.13)(51.7,23.3)
\rput(40.9,2){$A$}
\rput(51.6,2){$B$}
\end{pspicture}
} 
\vskip 2mm

That the structures introduced in this paper are by no means ad hoc is witnesses by the fact that very similar ones have recently occurred in a variety contexts:  conformal field theory  \cite[and follow-up papers]{Fuchs},  theory of group representations  \cite[and references therein]{Morrison} and  categorical representation of C*-algebras  \cite{VicaryCstar}.  It would be worthwhile to investigate the connections between all of these.

\bibliography{Base_ENTCS_III}

\end{document}